\documentclass[review,sort&compress]{elsarticle}

\usepackage{fullpage} 
\usepackage{afterpage}
\usepackage{enumerate}

\usepackage{nomencl}
\usepackage{amsmath}
\usepackage{mathtools}
\usepackage{amsfonts}
\usepackage{multirow}
\usepackage{subfig}
\DeclareMathOperator*{\argmin}{arg\,min}
\usepackage{graphicx}
\usepackage{booktabs}

\makenomenclature
\usepackage{xpatch}
\xpatchcmd{\thenomenclature}{\section*{\nomname}
}{}{\typeout{Success}}{\typeout{Failure}}
\makenomenclature
\RequirePackage{ifthen}
\printnomenclature[0.8cm]

\renewcommand\nomgroup[1]{
  \item[\itshape
  \ifstrequal{#1}{A}{Symbols}{
  \ifstrequal{#1}{B}{Roman Letters}{
  \ifstrequal{#1}{C}{Greek Letters}{
  \ifstrequal{#1}{D}{Abbreviations}{}}}}
]}

\usepackage{bm}

\newcommand{\region}[1]{\langle #1 \rangle}
\newcommand{\ud}{\,\mathrm{d}}

\graphicspath{{./figs/}}

\usepackage{array}
\newcolumntype{P}[1]{>{\centering\arraybackslash}m{#1}}

\pdfoutput=1

\begin{document}

\begin{frontmatter}
  \title{Learning nonlocal constitutive models with neural networks}

  \author[vt]{Xu-Hui Zhou}
    
  \author[pu]{Jiequn Han\corref{cor}}
  \ead{jiequnh@princeton.edu}
  \cortext[cor]{Corresponding author}

  \author[vt]{Heng Xiao}
  \ead{hengxiao@vt.edu}
  
  \address[vt]{Kevin T. Crofton Department of Aerospace and Ocean Engineering, Virginia Tech, Blacksburg, VA 24060, USA}
  \address[pu]{Department of Mathematics, Princeton University, Princeton, NJ 08544, USA}

  \begin{abstract}
    Constitutive and closure models play important roles in computational mechanics and computational physics in general. Classical constitutive models for solid and fluid materials are typically local, algebraic equations or flow rules describing the dependence of stress on the local strain and/or strain-rate. Closure models such as those describing Reynolds stress in turbulent flows and laminar--turbulent transition can involve transport PDEs (partial differential equations). Such models play similar roles to constitutive relation, but they are often more challenging to develop and calibrate as they describe nonlocal mappings and often contain many submodels. Inspired by the structure of the exact solutions to linear transport PDEs, we propose a neural network representing a region-to-point mapping to describe such nonlocal constitutive models. The range of nonlocal dependence and the convolution structure are derived from the formal solution to transport equations.  The neural network-based nonlocal constitutive model is trained with data.  Numerical experiments demonstrate the predictive capability of the proposed method. Moreover, the proposed network learned the embedded submodel without using data from that level, thanks to its interpretable mathematical structure, which makes it a promising alternative to traditional nonlocal constitutive models. 

  \end{abstract}

  \begin{keyword}
    convolutional neural networks \sep constitutive modeling \sep nonlocal closure model \sep inverse modeling \sep deep learning
  \end{keyword}
\end{frontmatter}

\section{Introduction}
\label{sec:intro}

Computational mechanics, and more generally, computational physics predominantly rely on solving partial differential equations (PDEs) describing the conservation laws for mass, momentum, energy, and species. However, a complementary yet equally critical component to the problem specification is a closure model. Such closure models come in various forms, ranging from local, algebraic constitutive models (e.g., stress-strain relations for linear elastic solids, Newtonian fluids, and viscoelastic fluids) to empirical correlations (e.g., drag force laws in particle-laden flows) and partial differential equations (PDEs, e.g.,  Reynolds stresses transport in turbulent flows~\cite{pope00turbulent}, intermittency transport in laminar--turbulence transition~\cite{menter2015one,coder2014computational}). All these closure models play similar roles in computational mechanics in that they express the unknown quantities in the PDEs with quantities that are known or solved for. However, an important conceptual difference between them is that classical constitutive models describe the macroscopic property of the materials, while the PDE-based models describe the property of the flow and thus can be considered \emph{generalized} constitutive relations. 
In particular, the PDE-based models in generalized constitutive relations tend to be nonlocal because the unresolved physics they aim to represent is nonlocal. This is in stark contrast to the classical local constitutive models for material behaviors, which are usually local. As a result, such PDE-based constitutive models are much more difficult to calibrate due to their high complexities, the presence of submodels with interactions, and the lack of data. In the present work we explore a new approach for this class of nonlocal, PDE-based constitutive models in the context of data-driven modeling.

\subsection{Examples and challenges in nonlocal, PDE-based constitutive models}

An important class of nonlocal constitutive models in transport problems takes the form of convection-diffusion-reaction PDEs. This is in contrast to the classical constitutive models, which are typically linear or nonlinear algebraic relations between fluxes (e.g., heat flux and stress) and the gradient of primary variables (e.g., displacement, velocity, and temperature). Here we use the  Reynolds stress transport models for turbulent flows as an example to motivate the current work, but it is worth noting that such nonlocal, PDE-based constitutive models are representative in a broad range of applications. Turbulent flows are described by the Navier--Stokes equations, which can be written for incompressible flows of constant density $\rho$ and kinematic viscosity $\nu$ as:
\begin{subequations}
\begin{align}
    \nabla \cdot \mathbf{v} & = 0 \\
    \frac{\partial \mathbf{v}}{\partial t} + 
    \mathbf{v} \cdot \nabla \mathbf{v}  - \nu \nabla^2 \mathbf{v} + \frac{1}{\rho} \nabla p & = 0 \; ,
\end{align}
\end{subequations}
where $\mathbf{v}$ and $p$ denote velocity and pressure, respectively, while $t$ indicates temporal coordinate, and $\nabla$ indicates spatial derivatives. Since the flows involve turbulent structures of a wide range of scales for high Reynolds number flows in practice, resolving those scales are prohibitively expensive. Consequently, engineering modeling often relies on descriptions of the mean flow as opposed to the instantaneous flow above. The mean flow equations are obtained by decomposing the flow into mean component and fluctuation components, i.e.,  $\mathbf{v} = \bar{\mathbf{v}} + \mathbf{v}'$ and $p = \bar{p} + p'$ (where $\bar{\mathbf{v}}$ denotes mean velocity, written as $\mathbf{u}$ for brevity hereafter, and $\bar{p}$ is the mean pressure). Taking the ensemble average of the equation above leads to the Reynolds-averaged Navier--Stokes (RANS) equations describing the mean flow:
\begin{subequations}
\label{eq:rans}
\begin{align}
    \nabla \cdot \mathbf{u} & = 0 \\
    \frac{\partial {\mathbf{u}}}{\partial t} + 
    {\mathbf{u}} \cdot \nabla {\mathbf{u}} - \nu \nabla^2 \mathbf{u} +
    \frac{1}{\rho} \nabla \bar{p}  & = \nabla \cdot \bm{\tau} \; ,
    \label{eq:rans-momentum}
\end{align}
\end{subequations}
where the mean velocity $\mathbf{u}$ and mean pressure $\bar{p}$ are the primary variables to be solved for. The RANS equations have a similar form to the Navier--Stokes equation except for the velocity covariance term $\bm{\tau} \equiv - \overline{v'_i v'_j}$, referred to as Reynolds stress tensor~\cite{pope00turbulent}. In this paper we call the Reynolds stress $\bm{\tau}$  as the \emph{closure variable}, for which a closure model is needed for the RANS equations~\eqref{eq:rans}, i.e., the primary equations, to be complete. Such closure models are called \emph{turbulence models}, which include many categories such as the commonly used eddy viscosity models and the more advanced Reynolds stress transport models. 
An important class among them are Reynolds stress transport models, which are based on transport PDEs derived from the original Navier--Stokes equations and take the following form~\cite{launder75progress}:
\begin{equation}
\label{eq:rstm}
  \frac{\partial \bm{\tau}}{\partial t} +  \mathbf{u} \cdot \nabla \bm{\tau} + \nabla \cdot T^{(\tau)} = 
  \mathcal{P}^{(\tau)} - {\mathcal{E}}^{(\tau)} + {\Pi}^{(\tau)} ,
\end{equation}
where $T^{(\tau)}$, $\mathcal{P}^{(\tau)}$, $\mathcal{E}^{(\tau)}$, and~$\Pi^{(\tau)}$ indicate flux (due to diffusion and triple velocity correlation), production (energy extraction from mean flow), turbulence dissipation (destruction due to viscosity), and pressure--strain-rate (energy exchange among different components of the Reynolds stress tensor), respectively. We refer to the Reynolds stress transport equation as the constitutive (closure) PDE for the RANS equation (the primary equation). Unfortunately, the Reynolds stress transport equation itself needs to be closed by using covariance terms of even higher order.
The unclosed terms include velocity triple correlation, pressure--strain-rate, and dissipation, among others. 
In particular, the pressure--strain-rate term plays a critical role and has a large impact on the predictive performances of the Reynolds stress models,  but it is notoriously difficult to model. 
Consequently, this difficulty impairs the robustness of Reynolds stress models and restricts their usage to a small fraction of practical turbulent flows despite their theoretical superiority~\cite{spalart15philosophies,xiao2019quantification}.

\begin{figure}[!htb]
\centering
\subfloat[mapping in nonlocal constitutive models]
{\includegraphics[width=0.46\textwidth]{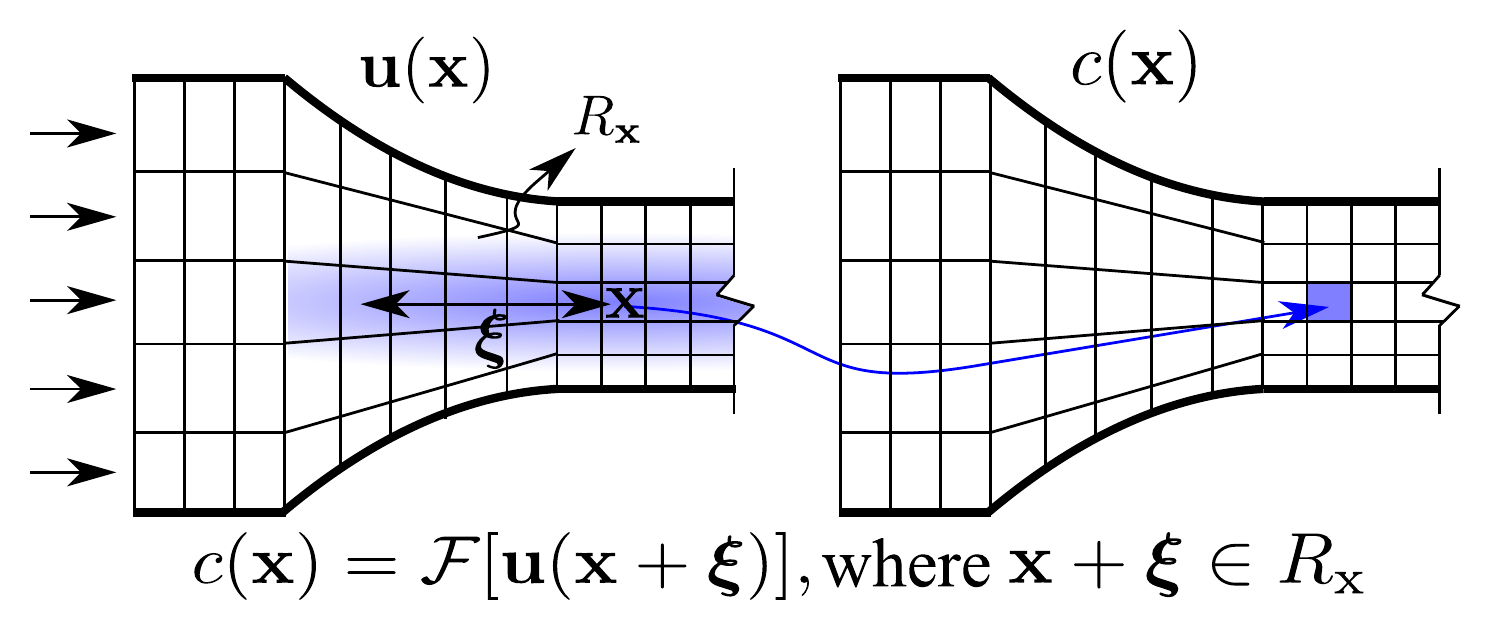}}
\hspace{2em}
 \subfloat[mapping in local constitutive models]
{\includegraphics[width=0.46\textwidth]{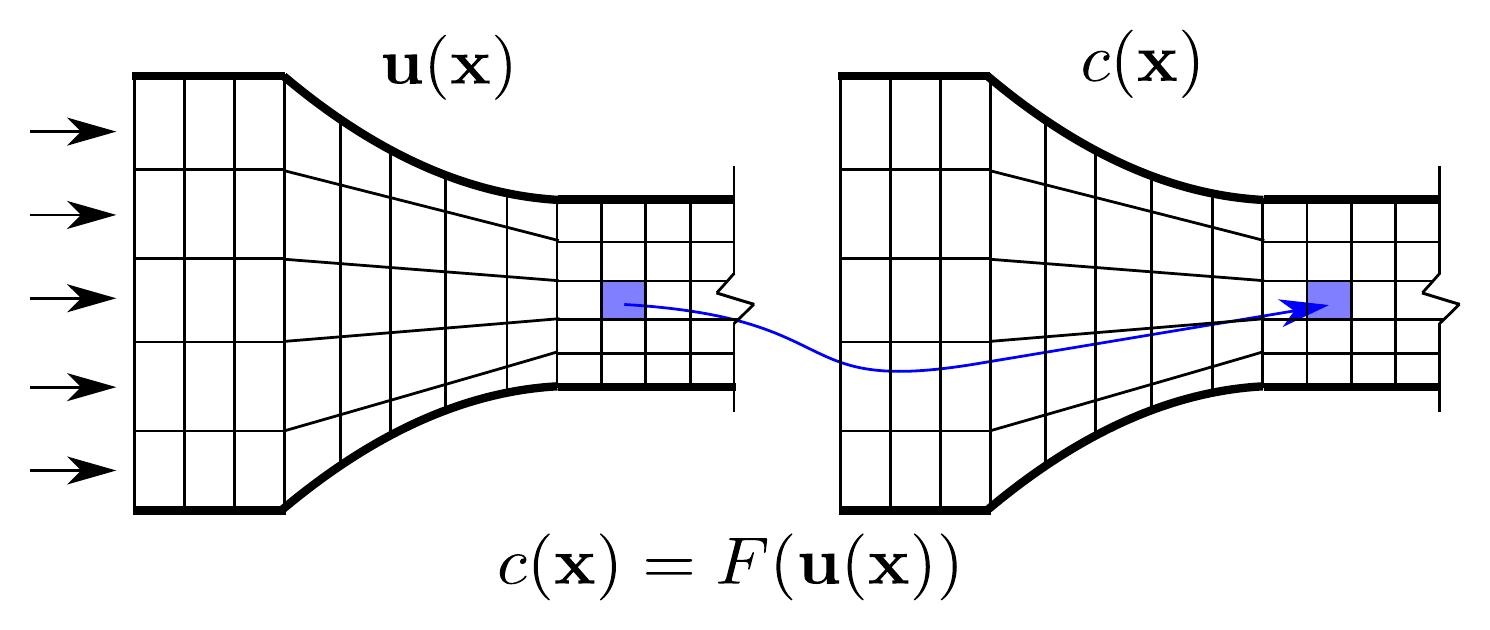}} 
\caption{Schematic comparison between nonlocal and local constitutive models mapping the velocity field $\mathbf{u}(\mathbf{x})$ to a generic closure variable field $c(\mathbf{x})$. (a) Constitutive models with nonlocal mapping 
$c(\mathbf{x}) = \mathcal{F} [\mathbf{u}(\mathbf{x} + \bm{\xi}) ]$, where $\mathbf{x} + \bm{\xi} \in R_{\mathbf{x}}$ and the influence region $R_{\mathbf{x}}$ is indicated in graded shade (inspired by Fig.~2.4.2 in Ref.~\cite{gatski1996simulation}). For example, $c$ can be considered Reynolds stress $\bm{\tau}$ or its anisotropy $\bm{b}$ in Reynolds stress transport models, both of which depend nonlocally on the velocity gradient. (b) Constitutive models involving only local, point-to-point mapping $c(\mathbf{x}) = F(\mathbf{u}(\mathbf{x})$.  For example, the generic closure variable $c$ here can be considered the (normalized) Reynolds stress anisotropy $\bm{b}$ in eddy viscosity models, where it depends only on the local velocity gradient.
}
  \label{fig:consti-model}
\end{figure}

The Reynolds stress transport equation~\eqref{eq:rstm} implies a nonlocal mapping from the primary variable (mean velocity $\mathbf{u}$) to the closure variable (Reynolds stress $\bm{\tau})$.
Following Lumley~\cite{lumley70toward,gatski1996simulation}, such a nonlocal mapping in the turbulent constitutive relation is illustrated in Fig.~\ref{fig:consti-model}a (where the generic closure variable $c$ should be interpreted as $\bm{\tau}$), and the mapping can be written as follows:
\begin{equation}
\label{eq:constitutive-eq}
\bm{\tau}(\mathbf{x}) = \mathcal{F} [\mathbf{u}(\mathbf{x} + \bm{\xi}) ],
\quad \text{with} \quad \mathbf{x} + \bm{\xi} \in R_{\mathbf{x}},
\end{equation}
where $\mathbf{x}$ indicates spatial coordinates, $\bm{\xi}$ is the spatial offset from $\mathbf{x}$, and $R_{\mathbf{x}}$ is the region of dependence (influence region) for the closure variable~$\bm{\tau}(\mathbf{x})$.
In contrast, the widely used eddy viscosity models
can be loosely considered local since they are derived based on weak equilibrium assumption~\cite{rodi1972prediction,pope00turbulent}, where the anisotropy $\bm{b}$ of the Reynolds stress (i.e., its orientation and aspect ratio) depends only on the local strain-rate, i.e., $\bm{b}(\mathbf{x}) = F (\mathbf{u}(\mathbf{x}))$, where $\displaystyle \bm{b} = -\frac{\bm{\tau}}{2\mathsf{k}} + \frac{\mathsf{I}}{3}$ and $\mathsf{I}$ is the identity tensor. This is illustrated in Fig.~\ref{fig:consti-model}b. Note that the magnitude of the Reynolds stress (i.e., turbulent kinetic energy $\mathsf{k}$) in such eddy viscosity models does contain upstream information, and as such the generic closure variable $c$ in Fig.~\ref{fig:consti-model}b should be interpreted as the Reynolds stress anisotropy $\bm{b}$. This fundamental difference contributes to the theoretical superiority of the Reynolds stress transport models over eddy viscosity models mentioned above.

The example of Reynolds stress transport models above highlights a number of intrinsic difficulties with the traditional PDE-based constitutive models.
In particular, they typically contain layers of submodels embedded in the PDEs. 
Moreover, the data needed to develop and individually calibrate such submodels are rarely available. Finally, interactions and coupling among the closure terms are difficult to quantify. 
Some of these difficulties are shared by other turbulence models such as the two-equation turbulence models~\cite{launder74application} and the laminar--turbulent transition models based on transport equations, as well as continuum-based modeling of granular flows~\cite{sun11constitutive}. For example, both classes of the commonly used laminar--turbulent transition models, the intermittency transport model~\cite{langtry2009correlation,menter2015one} and the amplification factor transport (AFT) model~\cite{coder2014computational,coder2019further}, are based on PDEs of the following form:
\begin{equation}
  \frac{\partial \bm{\phi}}{\partial t} +  \mathbf{u} \cdot \nabla \bm{\phi} + \nabla \cdot T^{(\phi)} = 
  \mathcal{P}^{(\phi)} - \mathcal{E}^{(\phi)} \, ,
  \label{eq:transition}
\end{equation}
where $\phi$ indicates the closure variable (intermittency $\gamma$ or amplification factor $\tilde{n}$) in the respective models; $T^{(\phi)}$, ${\mathcal{P}}^{(\phi)}$, and ${\mathcal{E}}^{(\phi)}$ denote diffusion transport, production, and dissipation/destruction (for intermittency transport model only) for the corresponding primary variable $\phi$. In particular, the production term in both models are closed by layers of submodels with empirical calibrations.

\subsection{Neural networks for nonlocal constitutive modeling}

In view of the common difficulty of such PDE-based constitutive models in transport-dominated physics, we propose a neural network that can represent the nonlocal, region-to-point mapping. The motivations are twofold: (1) to resolve the conflict between the need for calibrating many submodels and the lack of data at such levels, and (2) to improve the robustness of the resulted constitutive models. 
Our data-driven approach to nonlocal constitutive modeling draws inspiration from the seminal work of Lumley~\cite{lumley70toward}. He pointed out that turbulence in homogeneous shear flow at high Reynolds number behaves like a viscoelastic solid, based on which he argued that a universal constitutive relation exists for turbulence far removed from boundaries (e.g., wakes, jets, and the outer boundary layer in wall-bounded flows). A turbulence constitutive model is defined as a functional mapping from mean velocities to the Reynolds stresses~\cite{lumley70toward}. The dependence can be nonlocal, with the nonlocality (size of the influence region) determined from the spatial scales of the physics~\cite[see e.g.,][]{hamlington2008reynolds}.
However, by definition, it is only determined by the local structure of the turbulence and should not depend on initial or boundary conditions (except for a level of time and length scales being set therefrom).

In the present work, we study a broader class of nonlocal constitutive models mapping the mean velocity to the closure variable $c$ in the context of data-driven modeling. The details of the nonlocal models and the role of the closure variable $c$ are explained in Section~\ref{sec:problem}.
Such closure models are typically governed by the underlying transport processes. In some cases the exact physics may be written in explicit transport PDEs (e.g., the Reynolds stress transport equations) while in other cases an exact PDE is not available (e.g., intermittency or amplification factor in transition) and is heuristically motivated instead.
For brevity, hereafter the nonlocal mapping in Eq.~\eqref{eq:constitutive-eq} above is written in a more compact form for a general closure variable $c$ as
\begin{equation}
\label{eq:constitutive-eq-brief}
   c(\mathbf{x}) = \mathcal{F} [\region{\mathbf{u}}] \, ,
\end{equation}
where $\region{\mathbf{u}}$ indicates the velocities at all points in the region $R_{\mathbf{x}}$ (the shaded region in Fig.~\ref{fig:consti-model}a).
In these models, the closure variable $c$ at location $\mathbf{x}$ depends on $\mathbf{u}$ in the region $R_{\mathbf{x}}$, which may be spatially varying depending on the local physics.
Typically, the influence diminishes for points further away from $\mathbf{x}$ (i.e., as the offset distance $\|\bm{\xi}\|$ increases), which is indicated by gradually fading shade towards the edge of the shaded region. In contrast, classical constitutive models mostly involve local, point-to-point mapping as shown in Fig.~\ref{fig:consti-model}b.

\begin{figure}[!htb]
\centering
  \includegraphics[width=0.7\textwidth]{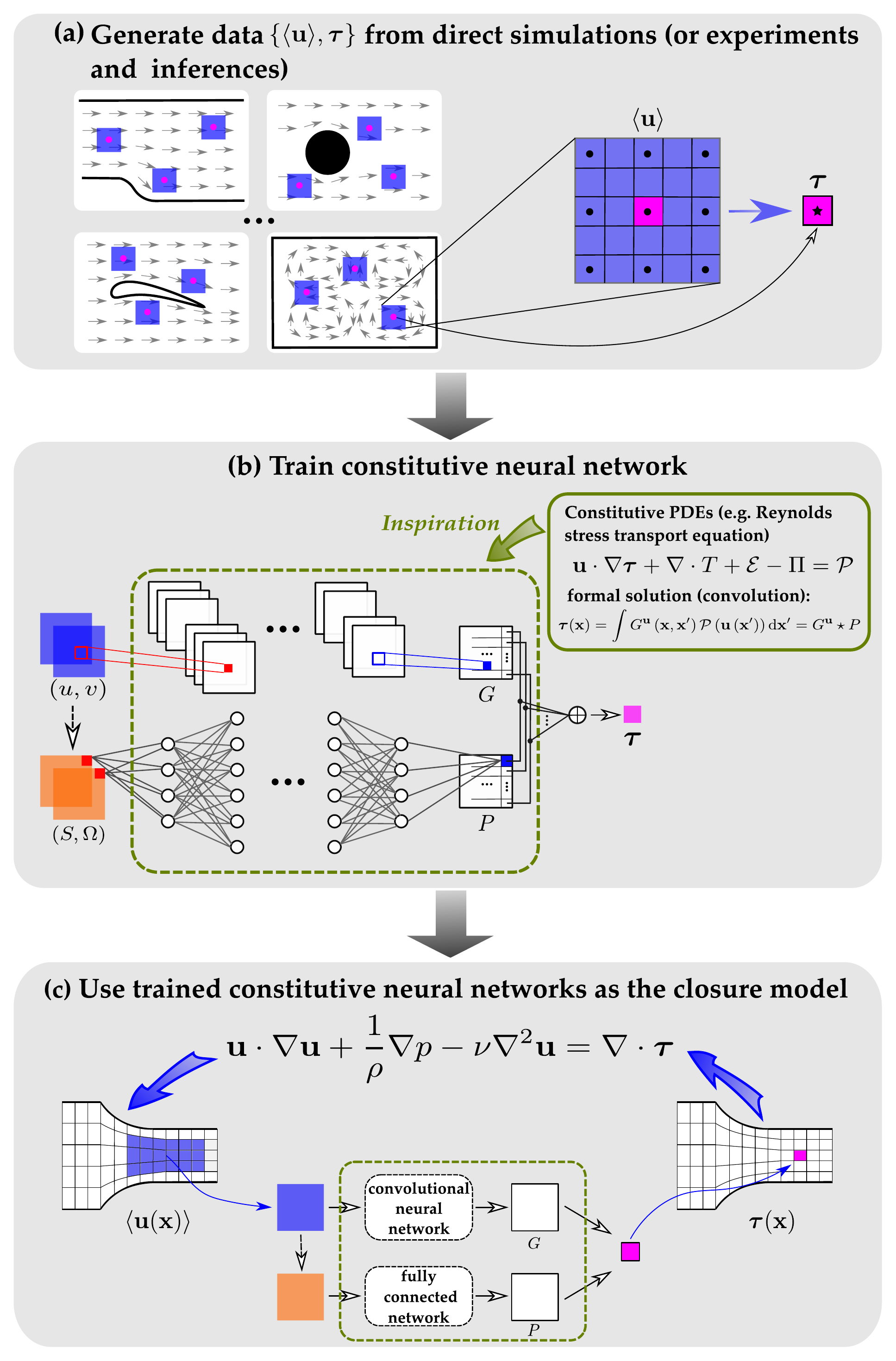}
  \caption{
  \label{fig:workflow}
  Schematic representation of the proposed workflow: (a) generate training data in the form of pairs $(\region{\mathbf{u}}, \bm{\tau})$ consisting of velocity patch $\region{\mathbf{u}}$ and closure variable $\bm{\tau}$, (b) train the neural network with such region-to-point mapping data, and (c) use the trained constitutive neural network as closure for the primary equations (using the RANS equations~\eqref{eq:rans-momentum} as an example).
  The neural network architecture in panel (b) is inspired by the structure of the formal solution to transport PDEs (using Reynolds stress transport equation~\eqref{eq:rstm} as an example), which is written as a convolution $G \star P(\mathbf{u})$ and represents nonlocal constitutive relations. A convolutional network is thus constructed to represent the nonlocal mapping
  from velocity field to the kernel $G$, and a fully connected network is used to model the production term $P$ as a local, pointwise function of $\mathbf{u}$ (or, more precisely, the strain rate $S$ and rotation rate $\Omega$). The framework lacks rotational invariance, and this deficiency is addressed in our follow-on work~\cite{zhou2021frame}.
  }
\end{figure}

We propose representing the region-to-point mapping in nonlocal constitutive models (as depicted in Fig.~\ref{fig:consti-model} and Eq.~\eqref{eq:constitutive-eq}) by using a neural network. The structure of the neural network is chosen to mimic that of the underlying transport PDEs in the classical constitutive models, e.g., Eqs.~\eqref{eq:rstm} and~\eqref{eq:transition} for Reynolds stress models and transition models, respectively. Specifically, the neural network consists of a convolution between a kernel and a production term. The kernel is a nonlocal function of the velocities on the influence region, while the production term is a local function of the velocity field $\mathbf{u}$ and the closure variable $c$. The development and application of such a neural network-based constitutive model thus consists of the following steps (taking turbulence modeling as an example) as illustrated in Fig.~\ref{fig:workflow}: 
\begin{enumerate}[(a)]
    \item perform direct simulation, experiments, or inferences to generate training data by extracting the region-to-point mapping from representative flow configurations and regions therein (Fig.~\ref{fig:workflow}a), 
    \item train the neural network of region-to-point mapping $\region{\mathbf{u}} \mapsto c$ by using the generated data~(Fig.~\ref{fig:workflow}b); once trained, the neural network model replaces the PDEs traditionally used to describe nonlocal constitutive relations (e.g., the Reynolds stress transport equation~\eqref{eq:rstm}, where closure variable $c$ is Reynolds stress $\bm{\tau}$); 
    \item use the trained constitutive neural network $c=\mathcal{F}_\text{NN}[\region{\mathbf{u}}]$ as the closure to the primary PDEs, e.g., the RANS momentum equation in turbulence modeling to solve for mean velocities and pressure (Fig.~\ref{fig:workflow}c).
\end{enumerate}

The constitutive neural network proposed here is distinct from the local, point-to-point data-driven constitutive models~\cite[e.g.,][in turbulence modeling]{ling16reynolds,wang17physics-informed,wu2018physics-informed,schmelzer2020discovery}. Clearly, the purely local models are not able to reflect the nonlocal physics implied in many constitutive models as outlined above. 
Recently, field-to-field mappings or global operators parameterized by machine learning models for approximating solutions of the PDEs~\cite[see, e.g.,][]{guo2016convolutional,long2018pde,raissi2019physics,sun2020surrogate,kim19deep,lu2019deeponet,ribeiro2020deepcfd,li2020neural,li2020fourier,gin2020deepgreen} have emerged as a new computation tool. Such tools can speed up outer-loop applications such as uncertainty quantification~\cite{gao2020bi,yang2020bifidelity,liu2020bi} and design optimization~\cite{li2017review,obiols2020cfdnet,patel2020parametric}, and also hold the promise to serve as nonlocal constitutive modelling.
However, some trained global mapping therein has initial conditions embedded in the network and thus cannot serve as constitutive models.
In contrast to most work cited above, this work focuses on a region-to-point mapping better suiting the needs for representing nonlocal constitutive models instead of finding global PDE solutions with machine learning models.

In this work we will focus on steps (a) and (b) outlined in Figs.~\ref{fig:workflow}a and \ref{fig:workflow}b, respectively, i.e., generating data and training a nonlocal constitutive model, while the integration with the primary solver is left for future work.
Specifically, we demonstrate the concept of the constitutive neural network on a convection-diffusion-reaction equation, which is a hypothetical constitutive model yet a prototype for many actual constitutive models in computational mechanics. Moreover, the velocity field in the convection-diffusion-reaction equation is specified and frozen, i.e., it is not affected by the closure variable $c$.
The objective is to demonstrate that the proposed network architecture representing a nonlocal, region-to-point mapping, is a viable approach to describe nonlocal constitutive relations typically depicted by PDEs. Compared to such \textit{ad hoc} and often heuristically justified closure PDEs, the neural network-based constitutive model is easier to train and has more flexibility and expressive power. We expect this preliminary study shall pave way for full-fledged data-driven turbulence models~\cite{duraisamy2019turbulence} and data-driven constitutive models in other related fields, such as computational mechanics~\cite{kirchdoerfer2016data,bock2019review,huang2020learning,xu2020learning,masi2020thermodynamics} and non-Newtonian fluid dynamics~\cite{lei2020machine}. 
Finally, we point out that the framework lacks rotational invariance, and this deficiency is addressed in our follow-on work~\cite{zhou2021frame}.

\afterpage{\clearpage}

\subsection{Contributions of present work}
The present work aims to demonstrate the merits of utilizing the underlying mathematical structure and physics to guide the design of neural network models. 
Even though the motivation example (linear differential equation) for the design of the network structure is generally invalid for intended practical applications with nonlinearity, the machine learning framework is much more disciplined than a purely black-box model, and it performs very well in the nonlinear problems. 
The same spirit has been used in Refs.~\cite{khoo2019switchnet,fan2019solving} in inverse problems like tomographic reconstruction and wave scattering inversion.
The present work focuses on transport constitutive relations that have a convection-diffusion-reaction PDE structure. Such problems are commonly found in computational physics and mechanics, particularly in fluid dynamics.

Neural networks have been used to solve PDEs~\cite{HanJentzenE2018,sirignano2018dgm,berg2018unified,han2019uniformly,han2019solving} and to build surrogate models for the solutions of a class of parameterized PDEs~\cite{tripathy2018deep,sun2020surrogate}. 
In such works the intent was to \emph{solve} clearly specified PDEs by using the solutions of similar PDEs as training data. In contrast, the present work aims to learn from data a nonlocal mapping implied by an unknown PDE, albeit with its structure assumed to be known (convection-diffusion-reaction type). In physical modeling it is common to know only the structure and not the specific form of a PDE, particularly when modeling unresolved processes. For example, the widely used Spalart--Allmaras turbulence model is based on a transport PDE of a viscosity-like variable $\tilde{\nu}$ without rigorous derivation~\cite{spalart92one}, so are the intermittency and amplification factor transport equations in laminar--turbulent transition modeling~\cite{menter2015one,coder2014computational}. 
Moreover, even though the turbulence dissipation needed for two-equation and Reynolds stress models can be derived exactly, an \textit{ad hoc} form is often used~\cite{wilcox06turbulence,pope00turbulent}.
As such, in this work we aim to learn a region-to-point mapping that embodies the physics described by the underlying (possibly unknown) PDE without requiring data at the submodel levels (e.g., those for the pressure--strain-rate, triple correlation in Reynolds stress model). Given the context of the present work in constitutive modeling,  which by definition shall be free from the boundary and initial conditions, we consider only the physics governed by PDEs. This is consistent with the comment made in Lumley~\cite{lumley70toward}, who, through analogy to St Venant’s principle in solid mechanics, stated that it is unlikely for a turbulent constitutive relation to exist near the boundaries. This insight applies not only to turbulence models but also to many other nonlocal constitutive relations. In practice, turbulence modeling within RANS or LES (large eddy simulation) solvers, the boundary regions near solid walls (fluid-solid interfaces) or free surfaces (liquid-gas interfaces) are handled by separate yet complementary closure models, i.e., wall models~\cite{yang2019predictive} and free surface models \cite{shirani2006turbulence}, respectively.

The rest of the paper is organized as follows. Section~\ref{sec:method} describes the problem the neural network-based constitutive modeling aims to solve and presents the technical details of the proposed framework, including the PDE-inspired neural network architecture, the choice of region size in the nonlocal mapping, and the details of data generation and model training. Section~\ref{sec:result} presents detailed verification of the proposed method, followed by multiple test cases to demonstrate the predictive capability of the constitutive neural network. Finally, Section~\ref{sec:conclude} concludes the paper.

\section{Methodology}
\label{sec:method}

\subsection{Problem statement}
\label{sec:problem}
We consider a steady-state convection-diffusion-reaction PDE describing the transport of an active tracer
\begin{equation}
    \label{eq:cdr3}
    \mathbf{u} \cdot \nabla c - \nabla \cdot \nu \nabla c + \mathcal{E} = \mathcal{P}(c; \mathbf{u})
\end{equation}
where $c$ is the concentration of the tracer, which is the closure variable (analogous to Reynolds stress $\bm{\tau}$ in RANS equations), $\mathbf{u}$ is the given flow field (whose governing equation contains unclosed term involving~$c$), $\nu$ is the diffusion coefficient, $\mathcal{P}$ is a production term depending only locally on the flow velocity and concentration, and $\mathcal{E}$ contains the dissipation term
and any sources that are nonlocal functions of $\mathbf{u}$ and $c$. The active tracer concentration variable $c$ corresponds to unclosed terms such as Reynolds stresses, intermittency, or amplification factor in the examples given in Section~\ref{sec:intro}.  
In this work we consider the scenario where $\nu$ is a specified constant, ${\mathcal{E}} = \zeta c$ with $\zeta > 0$, and~$\mathcal{P}$ is a function of the flow field and concentration:
\begin{equation}
\label{eq:P3}
    \mathcal{P}(c; \mathbf{u}) = f(S) \; g(\Omega) \; (K\,h(c) + h_0)
\end{equation}
where $S$ = $\tilde{S}/\tilde{S}_\text{max}$, $\Omega = \tilde{\Omega}/\tilde{\Omega}_0$, $\tilde{S} = \| \nabla \mathbf{u} + (\nabla \mathbf{u})^\top \|$ and $\tilde{\Omega} = \| \nabla \mathbf{u} - (\nabla \mathbf{u})^\top \|$ are the magnitudes, of the strain rate and rotation rate tensors, respectively, with $\|\cdot\|$ denoting tensor norm; $\tilde{S}_\text{max}$ and $\tilde{\Omega}_0$ are normalization constants. The functions are defined as:
\begin{subequations}
\begin{align}
\label{eq:fs}
    f(S) & = 4\sin{(2\pi S)} + 6 {S}^2 + 5 e^{S}, \\
\label{eq:gomega}
    g(\Omega) & = \frac{1}{1+\Omega^2}, \\
\label{eq:hc}
    h(c) & = 
    \begin{cases}
    e^{{-3c(1-c)}}, & c \in [0, 0.5], \\
    e^{{-10(c-0.5)}} + h_{12}, & c \in [0.5, 0.55], \\
    0.8 \ln c + h_{23}, & c \in [0.55, 1]. 
    \end{cases}
\end{align}
\end{subequations}
Here $\ln$ indicates natural logarithm, while $h_{12} = e^{-0.75}-1$ and $h_{23} =e^{-0.75}+e^{-0.5}-0.8 \ln(0.55)-1$ are constants used to ensure the continuity of the function. $K$ and $h_0$ are constants used to adjust the strength of nonlinear dependence of the production $\mathcal{P}$ on the concentration $c$.

\begin{figure}[!htb]
\centering
\subfloat[Dependence of production on strain]
{\includegraphics[width=0.46\textwidth]{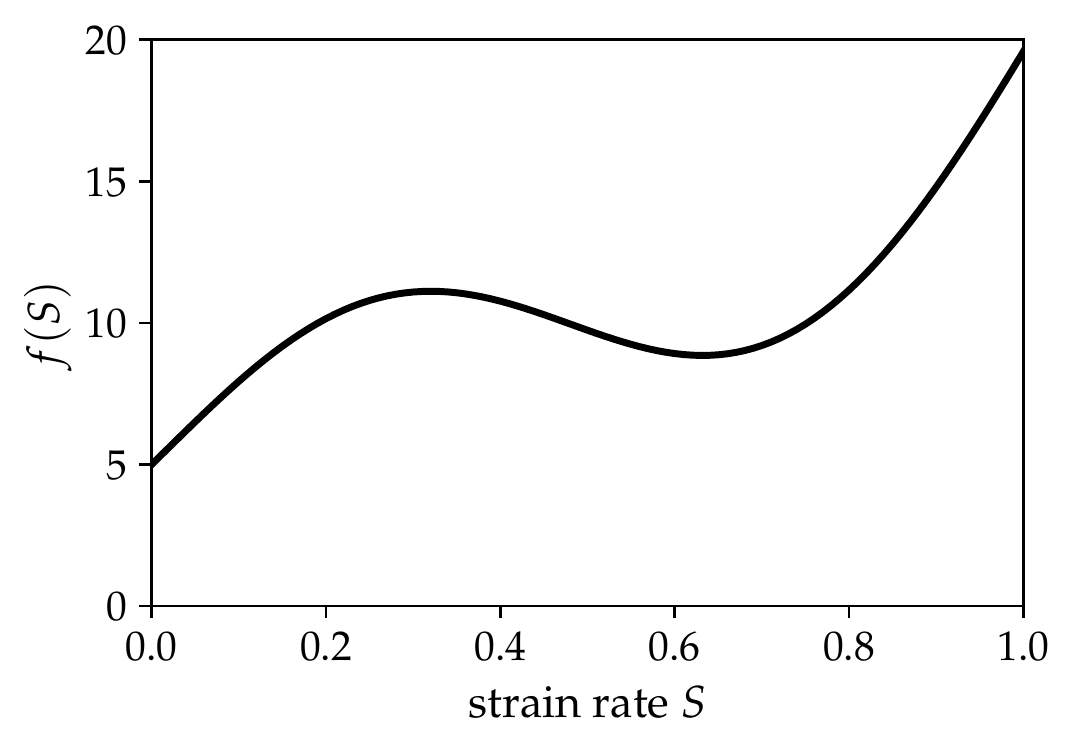}}
\hspace{1em}
\subfloat[Dependence of production on concentration]
{\includegraphics[width=0.46\textwidth]{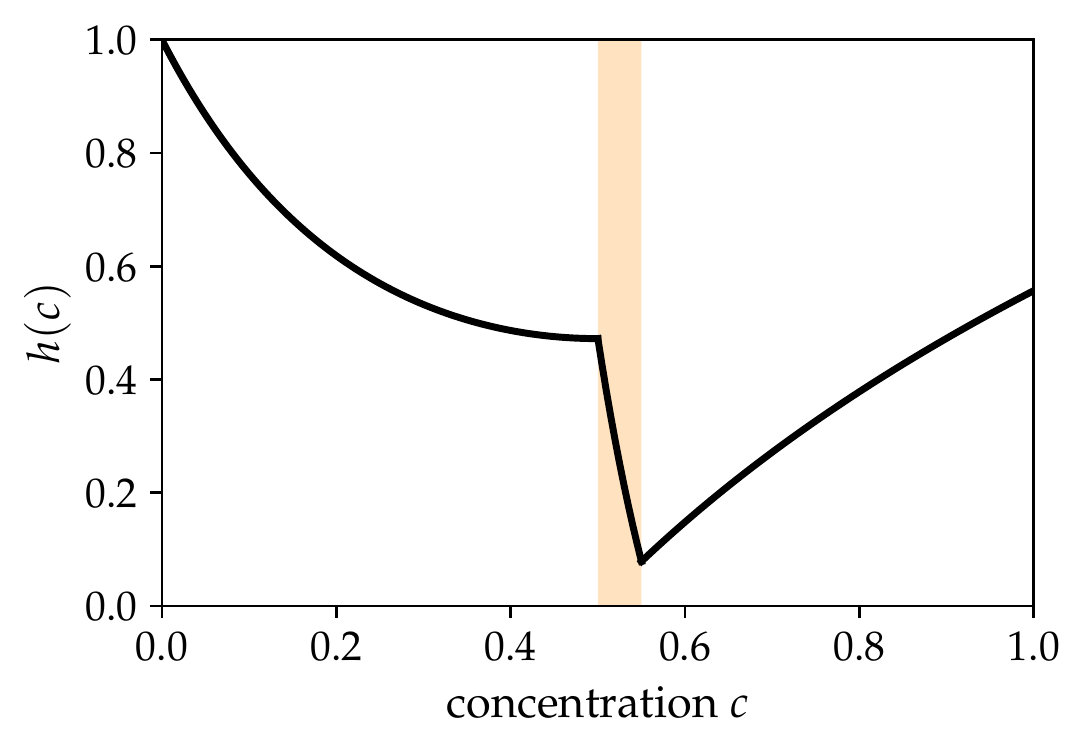}}
  \caption{
  \label{fig:source}
  (a) Dependence of production term $\mathcal{P}$ on normalized strain rate $S$ as described in Eq.~\eqref{eq:fs} and (b) dependence of production term $\mathcal{P}$ on concentration $c$ showing three distinct regimes with a steep gradient in the middle transition regime as described in Eq.~\eqref{eq:hc}. 
  }
\end{figure}

The behaviors of function $f(S)$ and $h(c)$ are presented in Fig.~\ref{fig:source}.
The function $h(c)$ has three distinct regimes, which is representative of the closure submodels in fluid dynamics. 
In particular, $h(c)$ has a transition regime (shaded in Fig.~\ref{fig:source}b) with a steep gradient, which can pose challenges for constitutive modeling. A similar submodel can be found in the particle-laden flow simulations, where
the drag on a spherical particle as a function of Reynolds number can be primarily divided into laminar and turbulent flow regimes, with a transition regime in between~\cite{currie2016fundamental}.

The overall problem statement above is reminiscent of the closure models in fluid dynamics, particularly the laminar-turbulence transition equation in Eq.~\eqref{eq:transition}. Specifically, while the whole ad hoc transport PDE serves as a constitutive model in a mean flow solver, the equation itself contains production and destruction terms that require closure (with several layers of embedded, regime-dependent models as amplified in the setup of Eq.~\eqref{eq:hc}). The Reynolds transport model, on the other hand, has additional difficulties involving inter-component coupling and energy transfer through pressure--strain-rate term $\Pi$ and wall models, which are not considered in present work.  As such, the present work can be seen as a first attempt towards a data-driven Reynolds stress transport model.

\subsection{Proposed network architecture}
\label{sec:method-nn}
The objective of this work is to use a neural network as a surrogate model that represents the nonlocal mapping described by the convection-diffusion-reaction PDEs. The mapping not only embodies the nonlocal dependence due to convection, diffusion, and other nonlocal production terms, but also contains the effects of the submodels, i.e., when the production is a nonlinear function of the flow field $\mathbf{u}$ and active tracer concentration $c$.

Our neural network architecture for nonlocal constitutive modeling (Fig.~\ref{fig:workflow}b) is inspired by the linear simplification of the convection-diffusion-reaction equation. 
To see the motivation, we write the transport equations in the constitutive models in an abstract form:
\begin{equation}
    \label{eq:ce-abstract}
    \mathcal{N}(c; \mathbf{u}) [c] = \mathcal{P}(c; \mathbf{u}),
\end{equation}
where $\mathbf{u}$ is the flow field (determined from the primary solver for which the constitutive model serve as closure), $c$ is the primary variable in the constitutive model of concern, $\mathcal{N}$ denotes the differential operators including convective and diffusive transport terms along with any \emph{nonlocal} production terms, $\mathcal{P}$ is the production function containing \emph{local}, pointwise dependence on $c$ and $\mathbf{u}$. 
Notice that the turbulence models Eq.~\eqref{eq:rstm}, the transport model Eq.~\eqref{eq:transition}, and the prototype model~Eq.~\eqref{eq:cdr3} can all fit into the form of Eq.~\eqref{eq:ce-abstract} above.
The separation of local and nonlocal operators in Eq.~\eqref{eq:ce-abstract} is often justified by the relevant physics. For example, in Reynolds stress transport equation~\eqref{eq:rstm}, the production, dissipation, and velocity triple correlation terms are all local, while the pressure--strain-rate is nonlocal due to the elliptic nature of the pressure equation. 

If Eq.~\eqref{eq:ce-abstract} is linear in $c$, we can denote the linear differential operator associated with the flow field $\mathbf{u}$ by $\mathcal{L}^{\mathbf{u}}$ and assume the production only depends on $\mathbf{u}$, i.e., :
\begin{equation}
\label{eq:linear-abstract}
    \mathcal{L}^{\mathbf{u}}[c] = \mathcal{P}(\mathbf{u}),
\end{equation}
which is defined on an infinite domain~$\mathbb{R}^d$ with vanishing boundary condition, then the solution can be written as~\cite[see, e.g.,][]{Evans2010partial}:
\begin{equation}
\label{eq:green-sol}
    c(\mathbf{x}) = \int_{\mathbb{R}^d} G^{\mathbf{u}}(\mathbf{x}, \mathbf{x}') \; \mathcal{P}(\mathbf{u}(\mathbf{x}'))\ud\mathbf{x}' \equiv G^{\mathbf{u}} \star P.
\end{equation}
Here $G^{\mathbf{u}}$ is the Green's function corresponding to, specifically the inverse operator of, the linear differential operator $\mathcal{L}^{\mathbf{u}}$, $P=\mathcal{P}(\mathbf{u}(\mathbf{x}'))$, and $\star$ denotes the convolution operator.
Motivated by the structure of Eq.~\eqref{eq:green-sol}, we propose the following network architecture to represent the general nonlocal constitutive relation.

\begin{figure}[!htb]
\centering
  \includegraphics[width=0.98\textwidth]{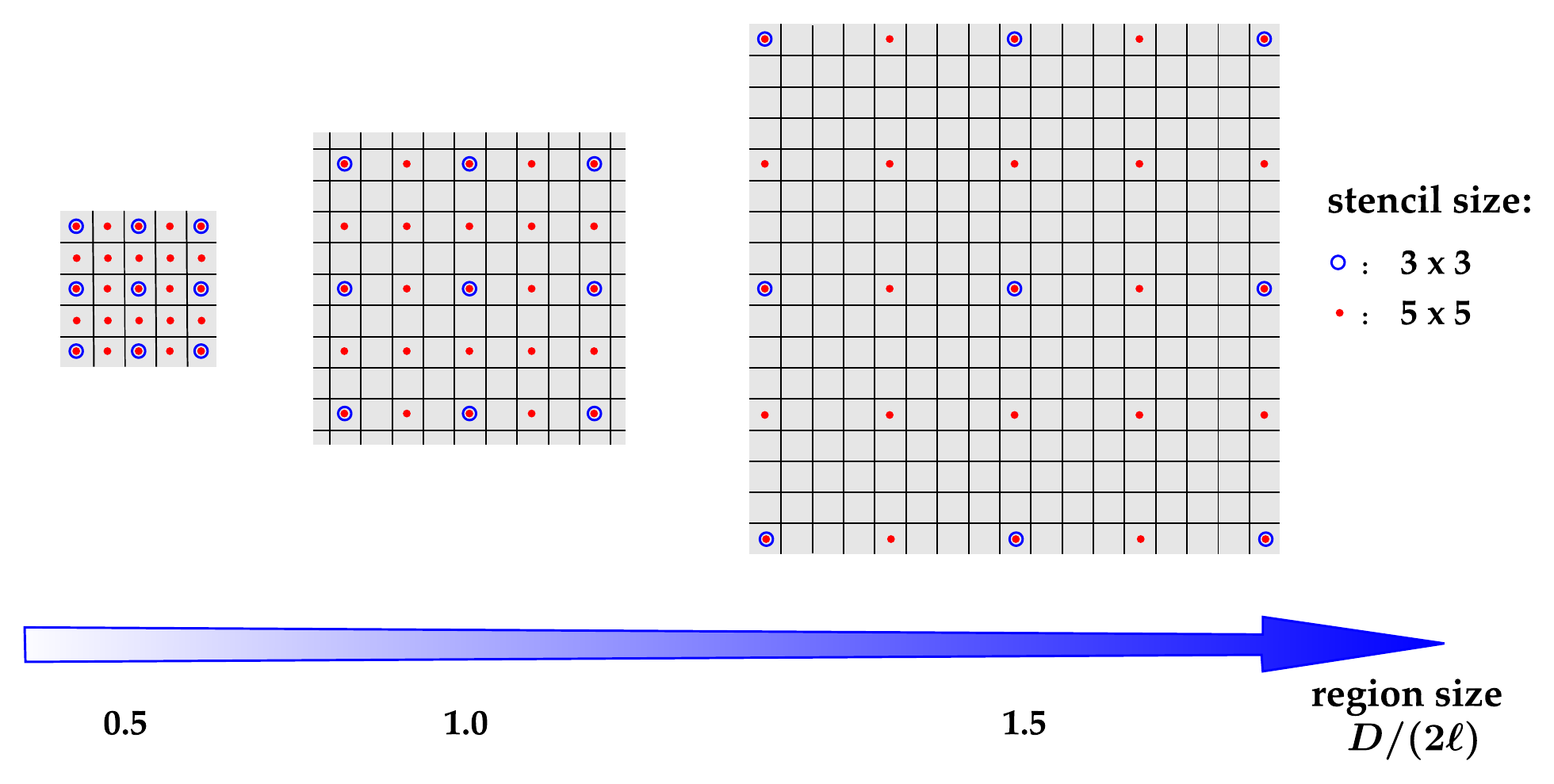}
  \caption{
  \label{fig:stencil-setup}
    Schematic figure on the different choices of stencil parameters $k$, $n_0$ and region size $D$.
    Blue/black circles and red/grey dots denote a $3\times 3~(k=1)$ and $5\times 5~(k=2)$ stencil, respectively. From left to right, with a stencil of size $3\times 3$ ($5\times 5$), by choosing the downsampling parameter $n_0=2, 4, 8$ ($n_0=1,2,4$), the resulting patch sizes are approximately $D/(2 \ell)=0.5, 1.0, 1.5$.
    Here $\ell$ denotes the half-length of the influence region introduced in Eq.~\eqref{eq:stencil-size}, Section~\ref{sec:region_size}.
  }
\end{figure}

Specifically, we intend to learn a nonlocal mapping from a patch of flow field $\region{\mathbf{u}}$ to the concentration variable $c$ at the point of interest $\mathbf{x}$. In this work, we focus on two-dimensional (2D) problems and assume the field is discretized on a regular grid. 
Suppose the point of interest $\mathbf{x}$ corresponds to the grid index $(i,j)$ and we use a stencil of size $(2k+1)\times(2k+1)~(k\in \mathbb{Z}^+)$ to approximate the velocity field in the considered influence region:
\begin{align}
\region{\mathbf{u}}_{i,j}=\{\mathbf{u}_{i+\Delta i,j+\Delta j}| \, \Delta i\in\mathcal{I},\,\Delta j\in\mathcal{J}\},
\end{align}
where 
\begin{align}
\label{eq:index_set}
\mathcal{I}=\mathcal{J}=\{-kn_0, -(k-1)n_0, \cdots, (k-1)n_0, kn_0\}
\end{align}
is the offset index set with stride $n_0\in\mathbb{Z}^{+}$ (positive integer). Here $k$ (along with the discretization grid) implies the size of stencil, and the parameter $n_0$ determines how frequent the data is sampled in the region (patch). If $n_0>1$, the velocity field in the considered region is downsampled. A few examples of combinations of $k$ and $n_0$ are shown in Fig.~\ref{fig:stencil-setup} to illustrate how they determine the region size and the spacing between stencil points.

The proposed architecture consists of two distinct modules as shown in Fig.~\ref{fig:workflow}b: 
\begin{enumerate}[(1)]
    \item a convolutional network to learn the mapping $\region{\mathbf{u}}_{i,j} \mapsto \region{G}_{i,j}$ from the patch of the velocity $\region{\mathbf{u}}_{i,j}$ surrounding a point to the Green's kernel;
    \item a fully connected neural network to learn the local mapping $\mathbf{u}_{i,j} \mapsto P_{i,j}$ from the velocity at point~$(i, j)$ to the production term at the same point. In order to guarantee the universality of the constitutive relation, the mapping represented by the fully connected network is shared by each point.
\end{enumerate}
Finally, with these two modules above, the final tracer concentration $c$ is computed as a discrete convolution of $G$ and $P$ as follows:
\begin{equation}
\label{eq:discrete_conv}
    c_{i,j} = \sum_{\Delta i\in\mathcal{I}, \Delta j\in\mathcal{J}} G_{i+\Delta i,j+\Delta j} \, P_{i-\Delta i, j-\Delta j}.
\end{equation}
This is a discrete counterpart of the formal solution in Eq.~\eqref{eq:green-sol} for the linear convection-diffusion-reaction equation. 
We remark that the structure in Eqs.~\eqref{eq:green-sol} and~\eqref{eq:discrete_conv} can be generalized to three dimensions (3D) straightforwardly such that our learning approach is capable of dealing with 3D problems in the same way.

Note that no data at the submodel level, e.g., the Green's kernel  $G$ or the production term $P$ will be used, since such data are rarely available in our application scenarios.
In a linear problem (Eq.~\eqref{eq:linear-abstract}), the ground truth of both $G$ and $P$ are known with clear interpretation.
It is expected that both mappings, 
$\region{\mathbf{u}}_{i, j} \mapsto \region{G}_{i, j}$ and $\mathbf{u}_{i, j} \mapsto P_{i, j}$, can be learned approximately (up to a multiplication constant among $G$ and $P$) given the data. This will be verified in two preliminary prediction cases in Section~\ref{sec:res-G} (\emph{Case I}) and Section~\ref{sec:res-P} (\emph{Case II}).
In \emph{Case I}, we consider a steady-state convection-diffusion-reaction PDE describing the transport of an active tracer
\begin{equation}
    \label{eq:cdr1}
    \mathbf{u} \cdot \nabla c - \nabla \cdot \nu \nabla c + \mathcal{E} = \mathcal{P}
\end{equation}
where $\mathcal{P}$ is a random production term generated by Gaussian process. We will verify that the Green's kernel can be learned approximately.
In \emph{Case II}, we consider a similar PDE
\begin{equation}
    \label{eq:cdr2}
    \mathbf{u} \cdot \nabla c - \nabla \cdot \nu \nabla c + \mathcal{E} = \mathcal{P}(\mathbf{u})
\end{equation}
where $\mathcal{P}$ is a production term depending only locally on the flow velocity.
We will verify that the submodel on $\mathcal{P}$ can be learned approximately in two different situations:
\begin{equation}
\label{eq:P21}
    \mathcal{P}(\mathbf{u}) = f(S)
\end{equation}
or
\begin{equation}
\label{eq:P22}
    \mathcal{P}(\mathbf{u}) = f(S)g(\Omega).
\end{equation}

For the nonlinear problem (Eq.~\eqref{eq:cdr3}) of our interest, the exact structure in Eq.~\eqref{eq:green-sol} no longer holds, but we expect that the network structure inspired by the linear problem, together with the nonlinearity of the networks, provide sufficient approximation capability for the resulted nonlocal mapping. This will be verified in Section \ref{sec:res-all}. We remarked that while the production term is chosen to be a product of two or three univariate functions for convenience, this fact is never used in the data-driven modeling, and thus all the conclusions should be equally valid for general production models.

\subsection{Form of training data and region size}
\label{sec:region_size}
As introduced above, the nonlocal, region-to-point mapping is trained by data consisting of pairs of $(\region{\mathbf{u}}, c)$ (or $(\region{\mathbf{u}}, \region{P}, c)$ in the special \emph{Case I}). When preparing the training data $\region{\mathbf{u}}$, we need to specify $k$ and $n_0$ in Eq.~\eqref{eq:index_set}, which essentially dictates the size of the region considered and the downsampling level therein.
The necessary size of the region should be determined by the physical influence (depicted in Fig.~\ref{fig:workflow}), which is in turn determined by the underlying PDE, while the downsampling level should be determined by the budget of points to be used. The question of determining the patch size amounts to finding the minimum region size of a truncated domain $R_{\mathbf{x}}$ such that the true solution can be approximated well by the convolution $\hat{c}(\mathbf{x}) \approx \int_{R_{\mathbf{x}}} G(\mathbf{x}, \mathbf{x}') \; P(\mathbf{x}') \ud\mathbf{x}'$  on $R_{\mathbf{x}}$ up to error tolerance $\epsilon$, i.e., 
\begin{equation}
\frac{|\hat{c}(\mathbf{x}) - c(\mathbf{x})|}{|c(\mathbf{x})|}
\le \epsilon \, ,
\end{equation}
where $|\cdot|$ denote absolute value. We have derived the region size in a simplified setting with details provided in \ref{sec:appd_green} and only a summary is provided below.

We consider a steady-state convection-diffusion-reaction equation with constant coefficients. The diffusion and dissipation constants are denoted by $\nu$ and $\zeta$. Suppose we use a square region as the influence region with error tolerance $\epsilon$. The half-length of its side used for both the upstream and downstream direction should be approximately
\begin{equation}
    \label{eq:stencil-size}
    \ell = \bigg|\frac{2\nu\log\epsilon}{\sqrt{\|\mathbf{u}\|^2+4\nu\zeta}-\|\mathbf{u}\|}\bigg|.
\end{equation}
It can be seen that the region size $2 \ell$ increases with larger diffusion coefficient $\nu$ and velocity magnitude $\|\mathbf{u}\|$ but decreases with larger dissipation coefficient $\zeta$. In fact, the diffusion and dissipation coefficients control the local equilibrium. The larger $\zeta$ is or the smaller $\nu$ is, the more local the problem becomes. We remark that in practice an accurate size of the influence region can be spatially inhomogeneous and difficult to obtain.
Our derivation provides a simple formula to apply in general situations and it can serve as a rule-of-thumb guideline for practice.
In the derivation, we have assumed the truncated region is square in order to employ the analytical solution in 1D. Moreover, we have calculated the size conservatively by treating the upstream and downstream regions equally. The difference between the streamwise and lateral velocities is ignored as well.
Although the derivation is based on these simplifications, we will see in Section~\ref{sec:res-G} that the best learning performance is achieved when stencil size parameter $k$ and sampling density parameter $n_0$ are properly chosen such that the underlying region size matches that suggested by the theoretical derivation.

\subsection{Implementation}
The proposed neural network is implemented in Python by using the machine learning library PyTorch~\cite{paszke2019pytorch}. The training data are generated by solving the underlying convection-diffusion-reaction PDEs with FiPy~\cite{FiPy:2009}, a finite volume PDE solver. The code for training data generation and neural network training are both made available in a public GitHub repository~\cite{zhou2020nonlocal}, and thus the results presented here can be straightforwardly reproduced and further developed.

\section{Results}
\label{sec:result}
In this section we evaluate the performance of the proposed method in three cases, with underlying flow fields being uniform flows (constant velocity in the entire domain), plane Poiseuille flows, and Taylor--Green vortexes.
First, in Section~\ref{sec:res-G}  we use the uniform flow with randomly sampled production field $\mathcal{P}(\mathbf{x})$ to verify that the proposed network structure is consistent with the solution structure for linear PDEs and that the proposed algorithm can learn the Green's kernel well. We will also demonstrate that the theoretically derived region size is indeed near-optimal numerically. As a further verification, in Section~\ref{sec:res-P} we learn from data generated from Eq.~\eqref{eq:cdr1} with flow-dependent production $\mathcal{P}(\mathbf{u}(\mathbf{x}))$. We demonstrate that the proposed architecture is capable of learning the underlying submodel for the production term, i.e., its dependence on the flow field. Finally, the setup in Section~\ref{sec:res-all} showcases the predictive capability of the proposed framework for nonlinear PDE in which the production term is a function of both the flow field and concentration as described in Eq.~\eqref{eq:P3}.

In each numerical experiment, we draw 100 random samples of physical coefficients to construct 100 PDEs and solve the PDEs to generate training data. The diffusion coefficient $\nu$ and dissipation coefficient $\zeta$ are constant in different PDEs belonging to the same case. The domain of the flow field is $[0, 1] \times [0, 1]$ with periodic boundary condition and discretized with grid size 0.01.
In total we use 500,000 pairs of $(\region{\mathbf{u}}, c)$ (or $(\region{\mathbf{u}}, \region{P}, c)$ in \emph{Case I}) as our training data and reserve 50,000 data points for testing. The details of network architectures and training parameters are provided in~\ref{sec:appd_parameter}.
We evaluate the prediction performance of the trained neural network by the normalized prediction error, which is defined as:
\begin{equation}
    \text{error} = \frac{\sqrt{\sum_{i=1}^{N_{\text{s}}} {|c_{i, \text{predict}} -
    c_{i, \text{truth}}|}^2}}{\sqrt{\sum_{i=1}^{N_{\text{s}}} {|c_{i, \text{truth}}|}^2}} , 
    \label{eq:error-def}
\end{equation}
where the summation is performed over all of the $N_{\text{s}}$ training or prediction data points.

\subsection{Case I: Verification of learning Green's kernel}
\label{sec:res-G}
In this case, we generate training data by solving the transport PDE \eqref{eq:cdr1} with uniform velocity fields and random production fields $\mathcal{P}(\mathbf{x})$. The diffusion coefficient  and dissipation coefficient in the PDE are set as $\nu=0.05$ and $\zeta=12$, respectively. The velocities are in the $x$-direction and uniformly distributed between 0 and 2.0.
The random production fields $\mathcal{P}(\mathbf{x})$ are generated by sampling from a zero-mean Gaussian process (also called Gaussian random field) with the periodic covariance kernel~\cite{rasmussen2004gaussian,strofer2020enforcing}, i.e., $\mathcal{P} \sim \mathcal{G} \mathcal{P}(0, \mathsf{K}(\mathbf{x},\mathbf{x}'))$ with
\begin{equation}
\mathsf{K}(\mathbf{x}, \mathbf{x}') = \sigma^{2} \exp \left[-\left(2 \frac{\sin ^{2}\left(\left|{x}_{1}-{x}'_{1}\right| \pi / l_p\right)}{l^{2}}+2 \frac{\sin ^{2}\left(\left|{x}_{2}-{x}'_{2}\right| \pi / l_p\right)}{l^{2}}\right)\right] ,
\end{equation}
where the variance is chosen as $\sigma = 1$, the length scale $l = 0.5$ and the periodicity is consistent with the length of the domain with $l_p = 1$. The magnitude of the sampled fields are then shifted and scaled to fall within the range $[0, 5]$.

The objective of this verification is to demonstrate the capability of the proposed neural architecture in recovering the mathematical structure of the analytical solution that is used to motivate it.
Specifically, the training data imply a region-to-point mapping $(\region{\mathbf{u}}_{i,j}, \region{P}_{i,j} ) \mapsto c_{i, j}$, while the underlying exact mapping is $c = G^{\mathbf{u}} \star P$. In general, the Green's kernel depends on the flow field $\mathbf{u}$. In this case,
since the velocity field is uniform, the analytical kernel $G$ becomes constant that can be easily compared with the learning result.

We remark that in this special case, the discretized constant kernel can be interpreted as the inverse of the finite difference approximation of the operator 
$\mathcal{L} [c] \equiv  (\mathbf{u} \cdot \nabla + \nabla \cdot \nu \nabla + \zeta) [c] $ (see \ref{sec:appd_green} for more details). For the region-to-point mapping in this case, by default we use $11\times11$ stencil points and distance between the adjacent stencil points is 0.06 (downsampling parameter $n_0=6$) such that the size of region covered by these points is approximately $0.6\times0.6$, which is close to the suggested region size $2 \ell$ determined by $\nu$, $\zeta$, $\mathbf{u}$ and error tolerance $\epsilon=0.2$ according to Eq.~\eqref{eq:stencil-size}.

\begin{figure}[!htb]
\centering
\includegraphics[width=0.4\textwidth]{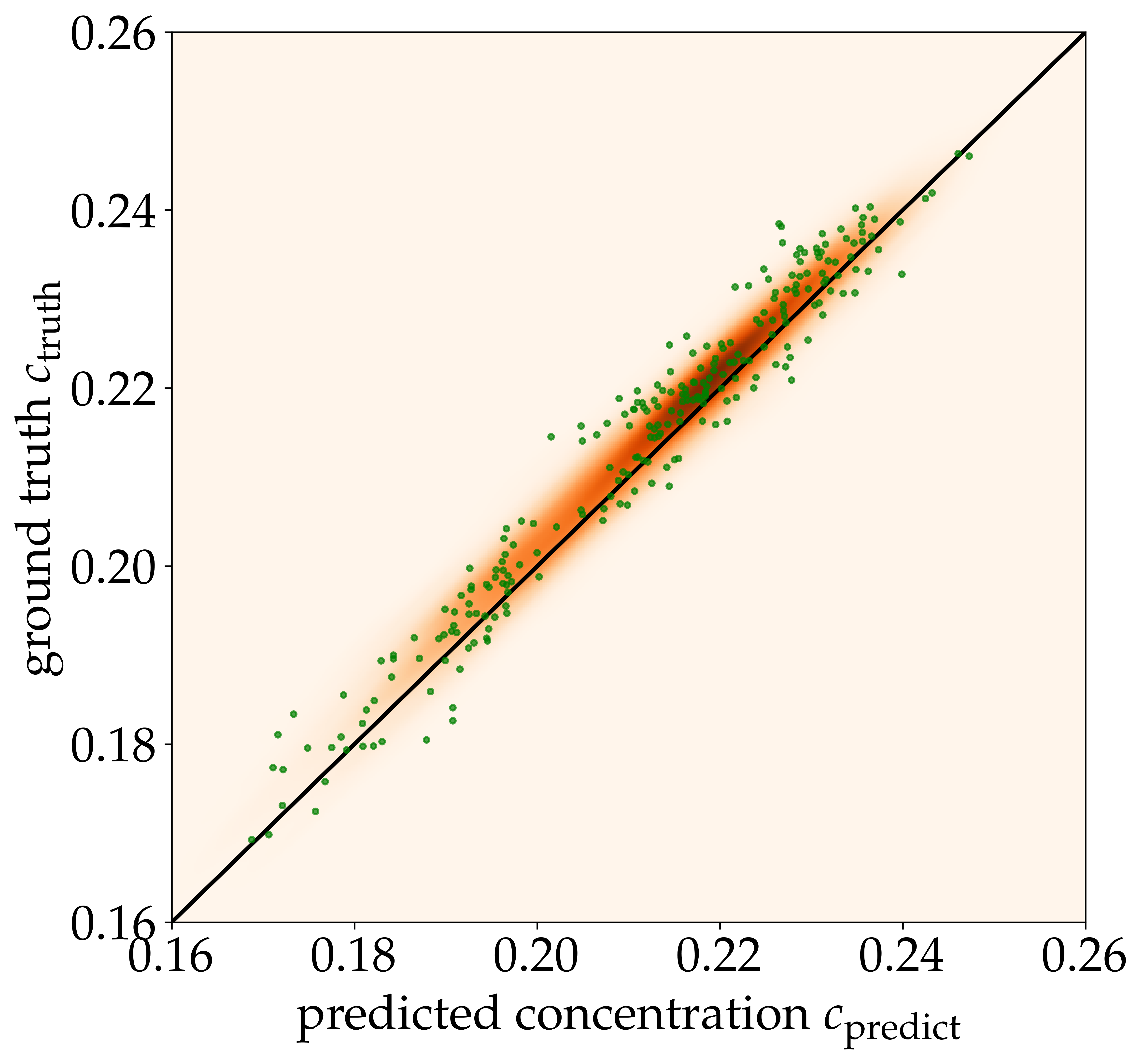}
  \caption{
  \label{fig:vp-case1}
  Prediction error plot for \emph{Case I} (linear PDE with random production field $\mathcal{P}(\mathbf{x})$), showing agreement between the predicted tracer concentrations and their ground truths on five different prediction flows.
  To clarity, only randomly sampled 300 points are shown as markers (filled circles). The shade contour in the background denotes the kernel density estimated from the entire population of $n=50,000$ prediction points, with darker region indicating higher density.
  }
\end{figure}

The prediction error plot in Fig.~\ref{fig:vp-case1} shows the predicted tracer concentration $c_\text{predict}$ obtained from the learned kernel $G$ compared with the corresponding ground truth $c_\text{truth}$.  The normalized prediction error is computed to be $3.63\%$ according to the error metric defined in Eq.~\eqref{eq:error-def}. For clarify, only a randomly sampled 300 points are shown as markers (filled circles). The kernel density estimated from the entire population of 50,000 prediction points is shown as shaded contour background.
If a point falls on the diagonal (indicated as a solid line in Fig.~\ref{fig:vp-case1}), it means that the predicted concentration agrees exactly with the corresponding ground truth. It can be seen from the scatter plot and density contour that all the points are located near the diagonal. Therefore, the concentrations are predicated satisfactorily at all the points. 

\begin{figure}[!htb]
\captionsetup[subfloat]{captionskip=-2pt}
\centering
\subfloat[ground truth of concentrations]
{\includegraphics[height=0.3\textwidth]{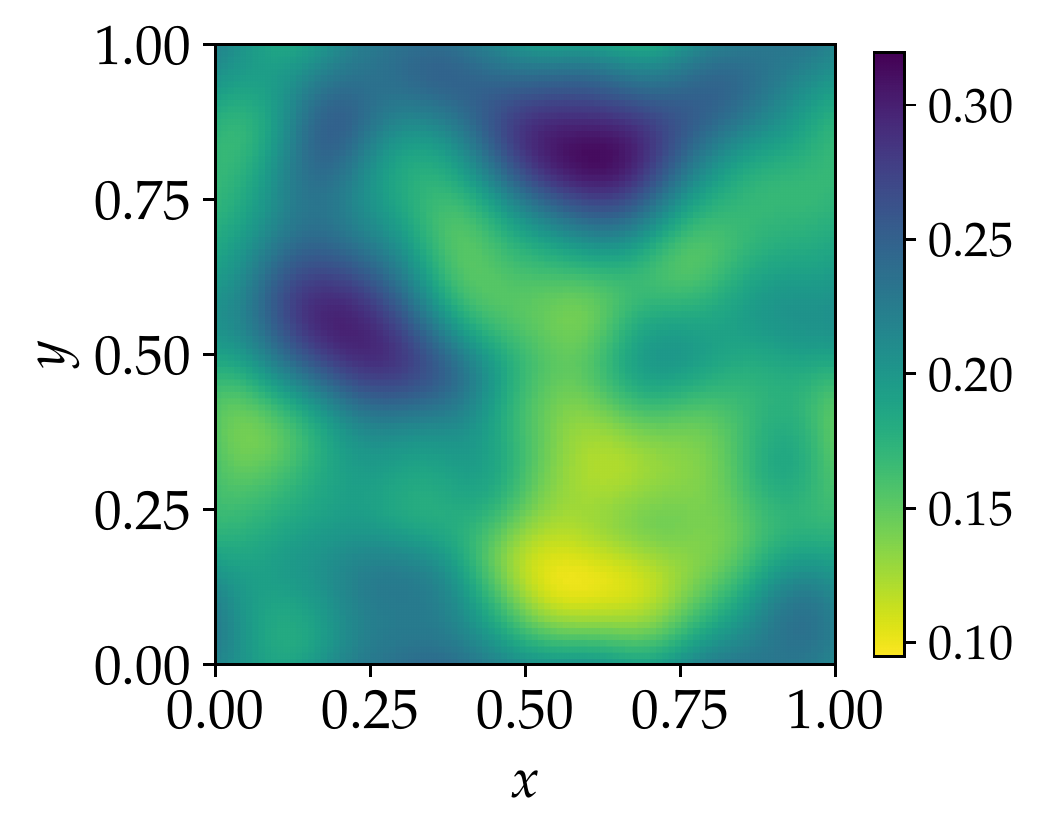}}
\hspace{4em}
\subfloat[predicted concentrations]
{\includegraphics[height=0.3\textwidth]{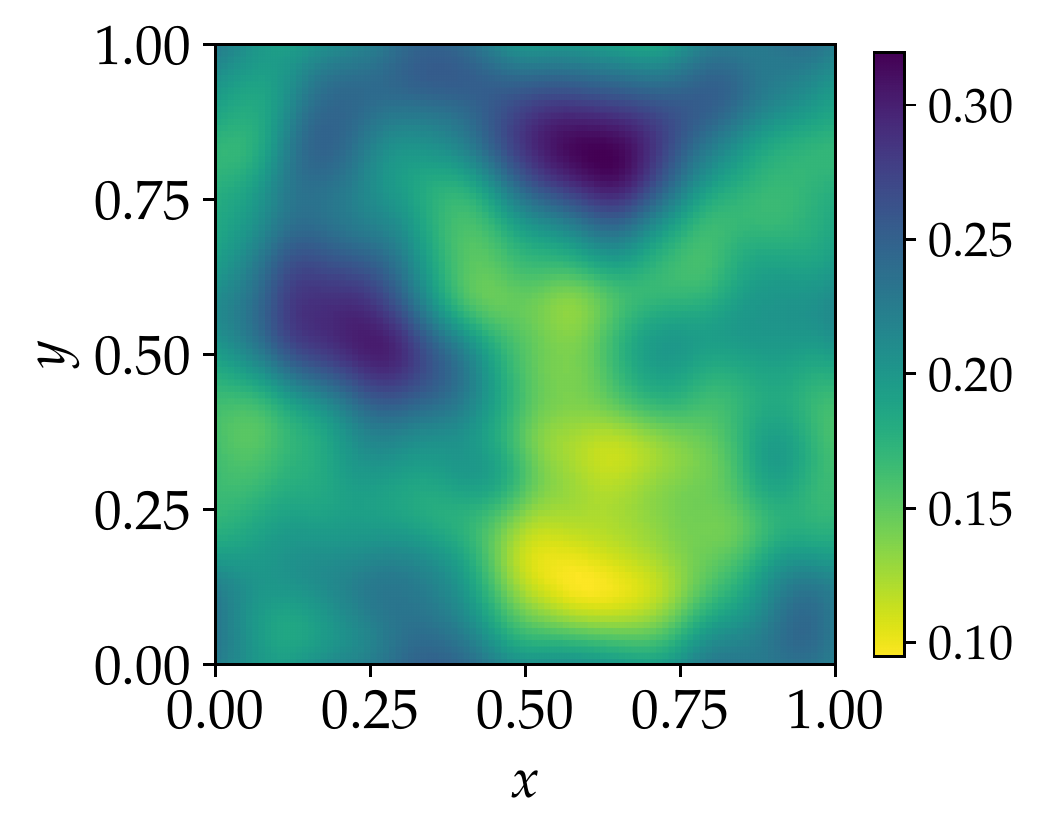}}\\
{\vspace{10pt}
\includegraphics[width=0.5\textwidth]{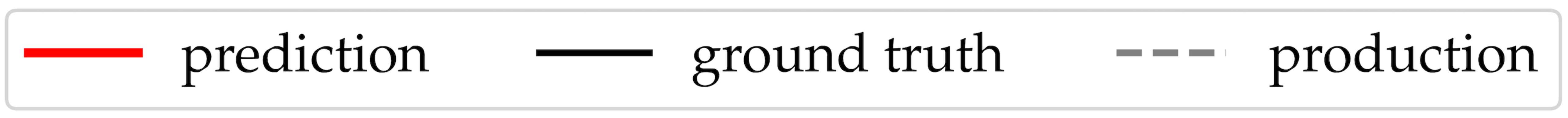}\vspace{-10pt}}\\
\subfloat[predict concentration $c$ at $y=0.3$]
{\includegraphics[width=0.47\textwidth]{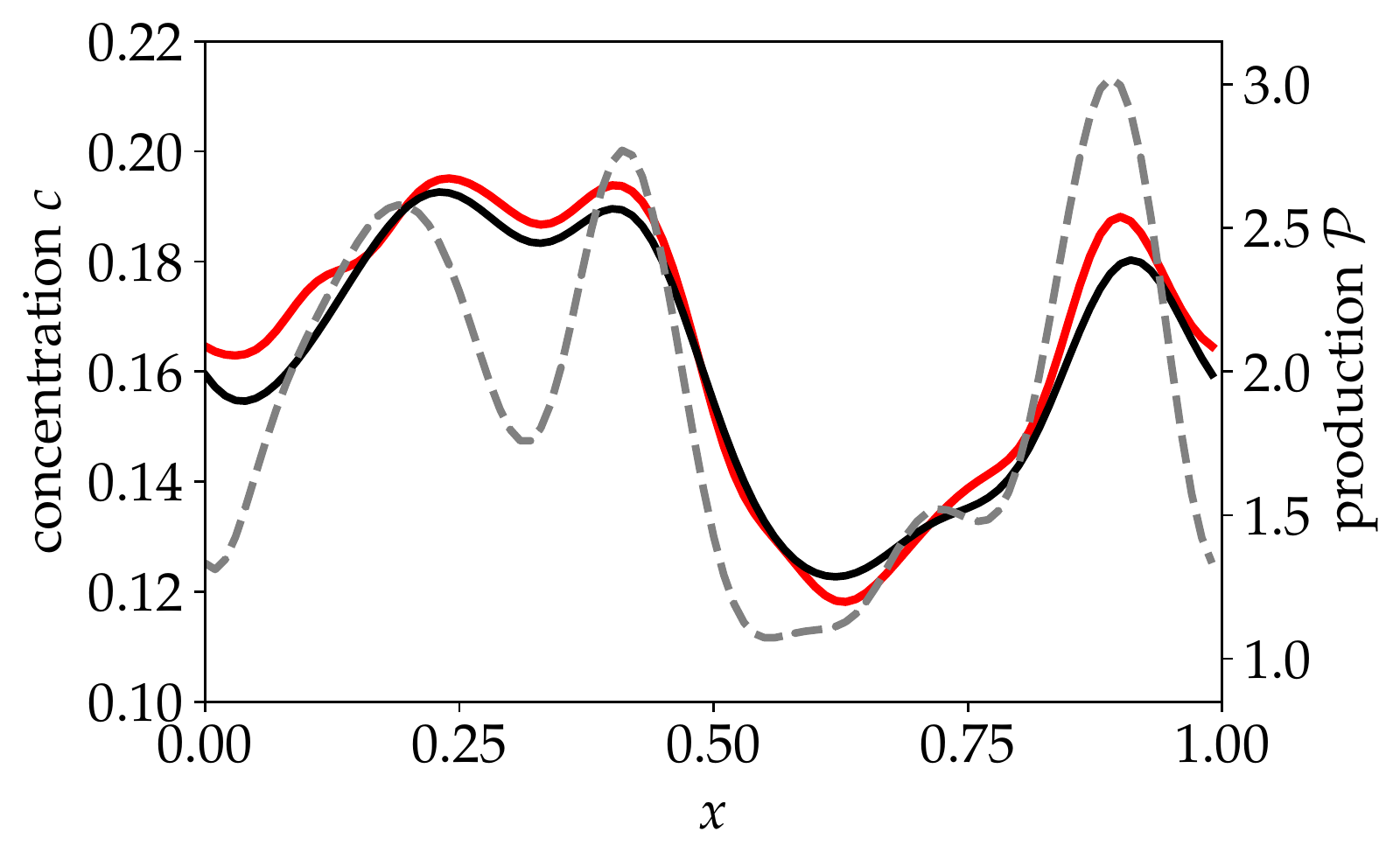}}
\hspace{1em}
\subfloat[predict concentration $c$ at $x=0.4$]
{\includegraphics[width=0.47\textwidth]{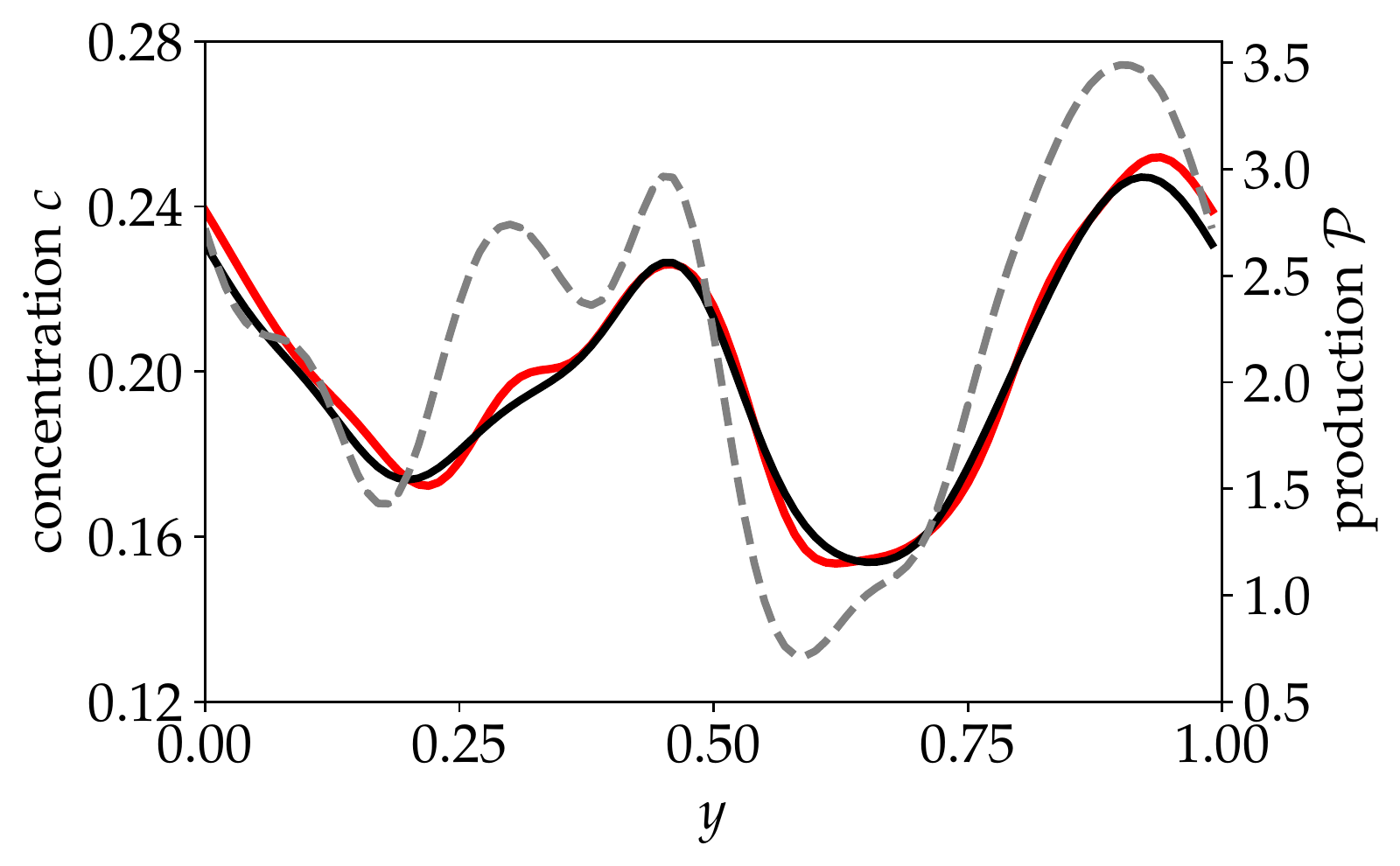}}
  \caption{
  \label{fig:c-compare-case1}
  Comparison of the ground truth of concentration field (panel a) and the corresponding predicted field (panel b) on a prediction flow in \emph{Case I} with a random production field (sampled from the Gaussian process). The concentrations and the corresponding production in two cross-sections are shown in panels (c) and (d).
  }
\end{figure}

A comparison of the predicted tracer concentration field and the ground truth is presented in Fig.~\ref{fig:c-compare-case1}. 
We can see in Figs.~\ref{fig:c-compare-case1}a and~\ref{fig:c-compare-case1}b that the predicted concentration field and the ground truth have good overall agreement.
This can be seen more clearly in the cross section plots in Fig.~\ref{fig:c-compare-case1}c and~\ref{fig:c-compare-case1}d where the predicted and the ground truth concentrations have rather good overall agreement except for some minor discrepancy. 
Note that the dramatic difference between the original production and the resulting concentration reveals that they are related via a non-trivial mapping due to the convection and diffusion effect of the flow. The learning of this mapping from data is the subject of our discussion below.

\begin{figure}[!htb]
\centering
{\includegraphics[width=0.42\textwidth]{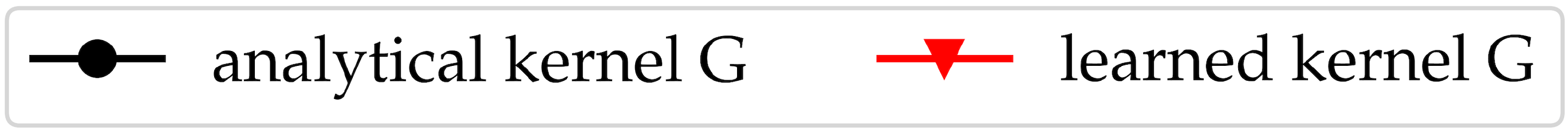}\vspace{-8pt}}\\
\subfloat[$u = 1.06$ m/s]
{\includegraphics[width=0.32\textwidth]{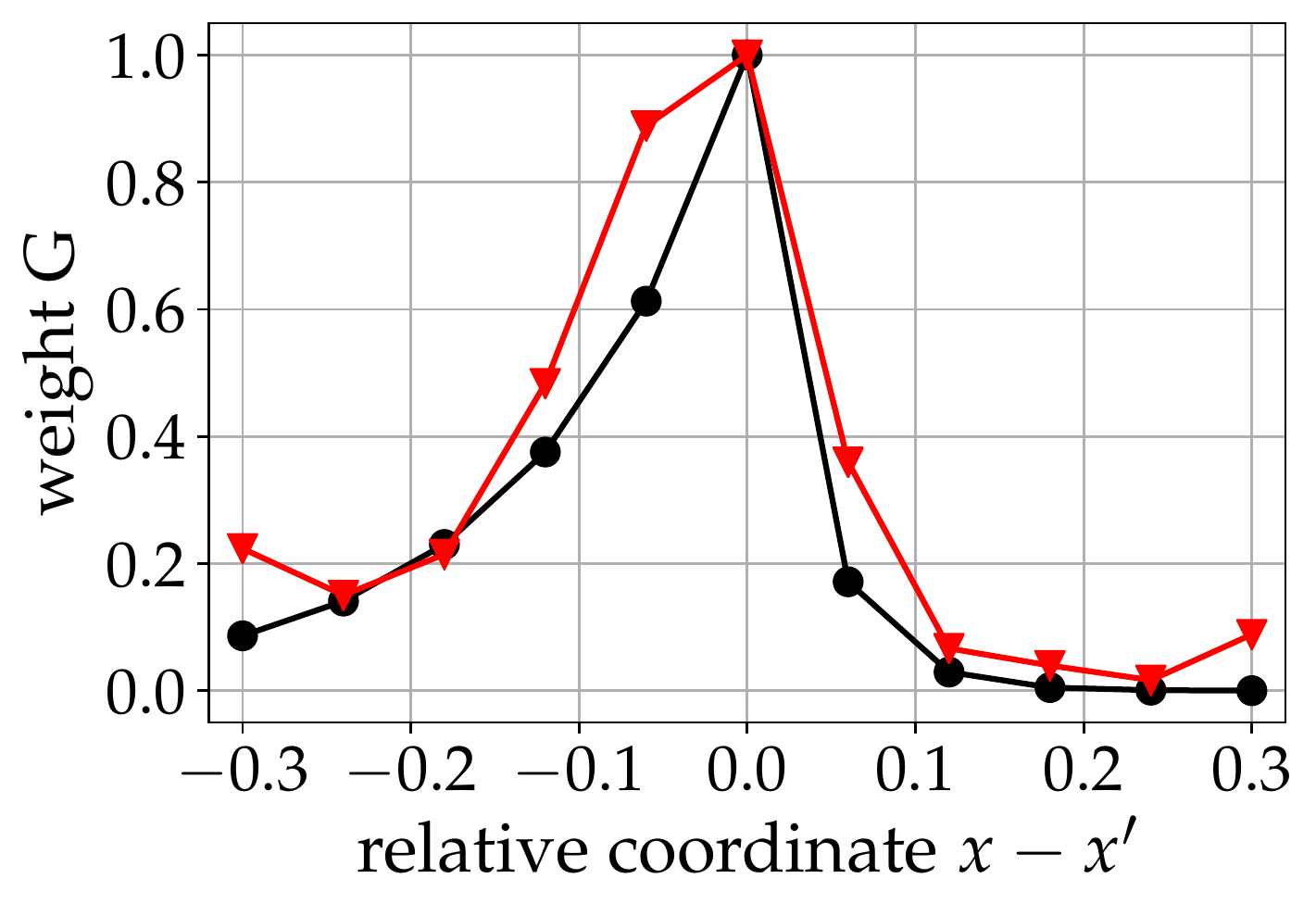}}
\hspace{0.2em}
\subfloat[$u = 1.41$ m/s]
{\includegraphics[width=0.32\textwidth]{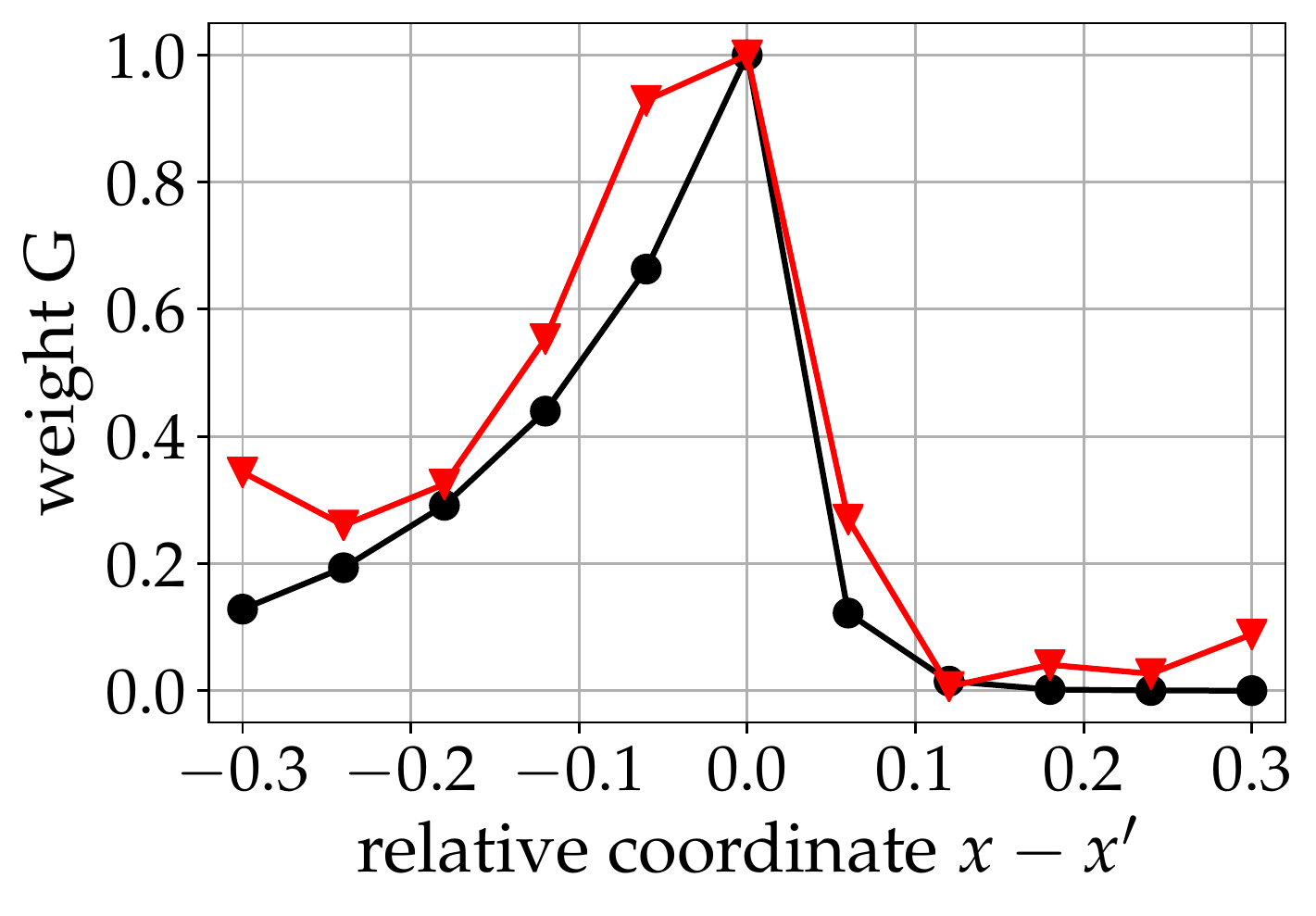}}
\hspace{0.2em}
\subfloat[$u = 0.66$ m/s]
{\includegraphics[width=0.32\textwidth]{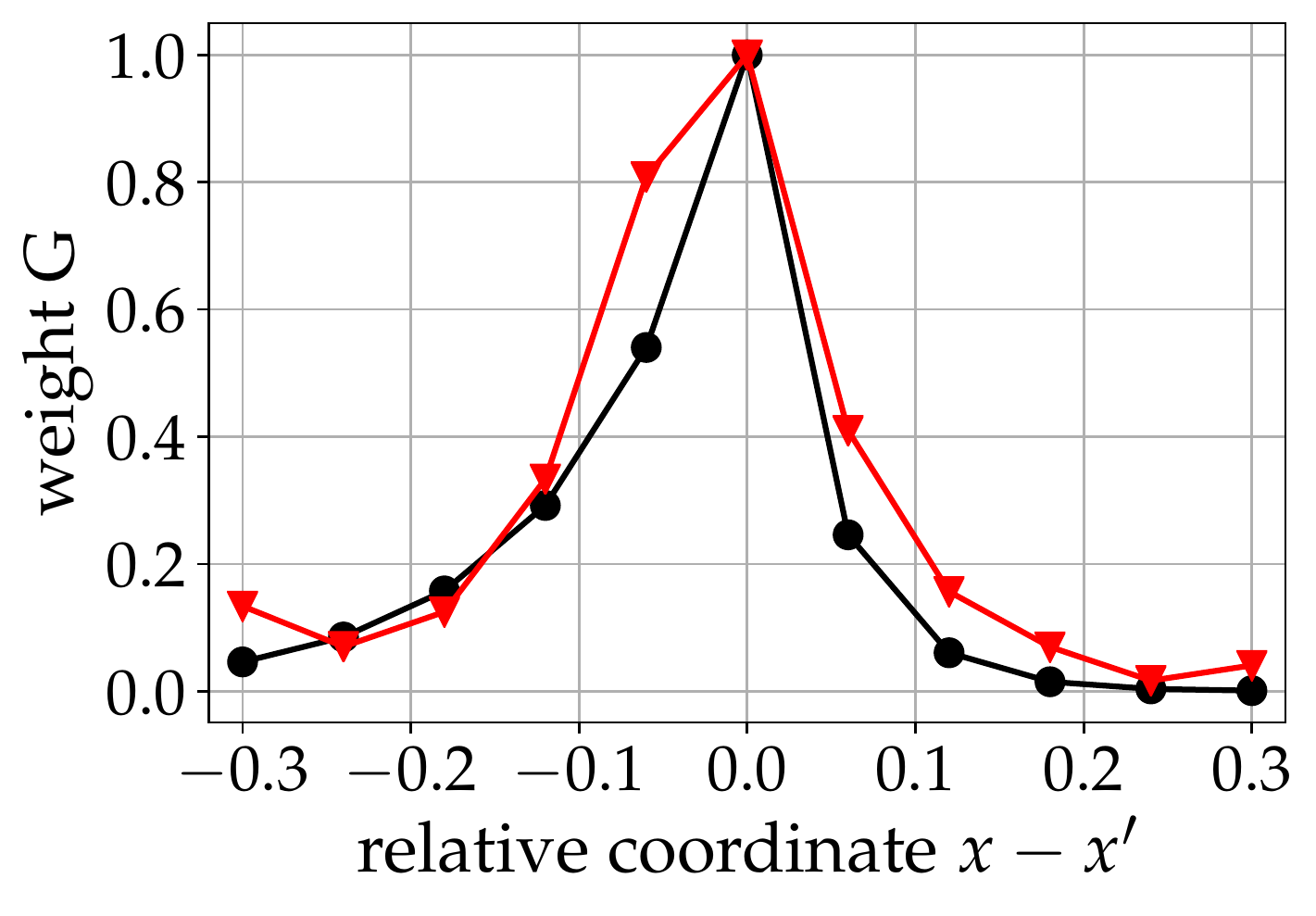}}
  \caption{
  \label{fig:green-kernel}
  Comparison of the learned Green's kernels with the analytical Green's kernels in three uniform flows with different velocities in \emph{Case I}. The maximum weight is normalized to 1.0. The locations of the points in Green's kernel are specified with relative coordinates to the center point in the $x$ direction.
  }
\end{figure}

It is noteworthy that the neural network not only predicts the closure variable $c$ very well, but also achieves it for the right reason: it learned the Green's kernel correctly, as demonstrated in Fig.~\ref{fig:green-kernel}.
In this case, we know the analytic form of the Green's function and we take its discretization with the corresponding grid size as the analytical kernel.
The learned kernel $G$ is of size $11 \times 11$, and we compare it with the analytical kernel. 
Since the velocity $\mathbf{u}$ is in the $x$-direction, the kernel is essentially 1D, depending only on the relative coordinate in the $x$-direction.
Accordingly, we compute the average of the learned kernel along the $y$-direction and obtain the weight in the $x$-direction. The 1D weights of both the learned kernel and theoretical kernel are normalized such that the maximum is 1.0.
We can see that the trend of learned $G$ is quite similar to that of the analytical $G$ in the $x$-direction, explaining the good performance of the concentration prediction. 
Notice that $G$ determines the weight of the production at different points relative to the center point. 
Since the productions in the upstream have a greater influence on the center point than those in the downstream, the weights on the left (upstream) side are larger than the weights of the symmetric positions on the right (downstream) side, and our learned kernel reveals such physics correctly.
Slight differences of the weights are observed in the left-most (upstream) and right-most (downstream) positions, which may be caused by the padding used in our convolutional neural networks.
However, the results are not sensitive to the padding mode that is used (zeros, reflect or replicate padding).
It is also possible that the slight difference is caused by the cutoff of the influence region and downsampling procedure.

\begin{figure}[!htb]
\centering
  \includegraphics[width=0.6\textwidth]{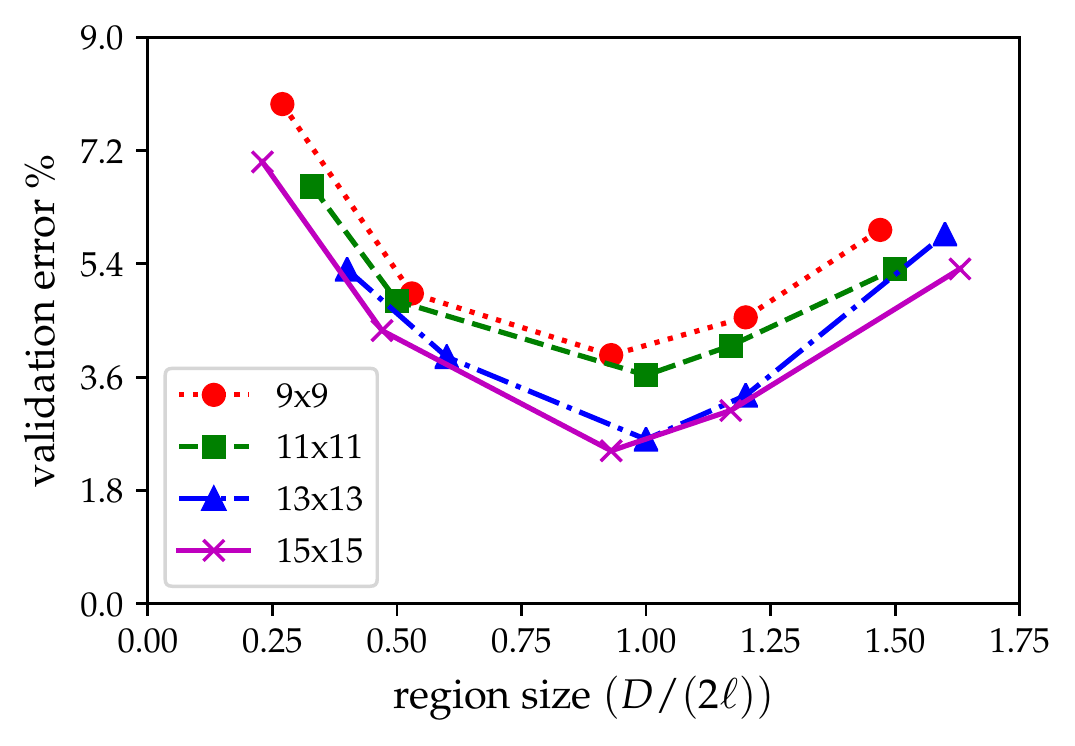}
  \caption{
  \label{fig:optimal-stencil}
   Parametric study of predictive errors at various region sizes and stencil sizes (as defined in Fig.~\ref{fig:stencil-setup}). Points on the same line (same symbol type/color) correspond to cases of the same stencil size (number of points in the patch or region) but with the patch covering different sizes of regions. Points on at the same abscissa (the $D/(2 \ell)$ value) correspond to cases with the same region size but with a varying spatial resolution (stencil size).
}
\end{figure}

We further demonstrate that the region size chosen based on the derivation in Eq.~\eqref{eq:stencil-size} is indeed near-optimal. To this end, we evaluated four different stencil sizes (the number of points $(2k+1)\times(2k+1)$ used in the stencil): $9 \times 9$, $11 \times 11$, $13 \times 13$ and $15 \times 15$.
For each stencil size, the points are uniformly distributed in a square region of size $D$ such that $D/(2 \ell) \in [0.5, 1.5]$. The way how we choose $k,\,n_0$ and $D$ is illustrated in Fig.~\ref{fig:stencil-setup}. A total number of 20 cases with various combinations of region size and stencil size are investigated, and the prediction error of concentration in each case are presented in Fig.~\ref{fig:optimal-stencil}.

From Fig.~\ref{fig:optimal-stencil} we can see that, with each fixed stencil size $(2k+1)\times (2k+1)$, the smallest prediction error is achieved when the region size is close to the theoretical optimum, i.e., $D/(2 \ell)\approx 1$. 
This observation suggests that, given the same number of data points as the computational budget on the input size, the learning is most effective when the information is drawn from the proper physical domain of influence. Otherwise, when the region size $D/(2 \ell)$ departs significantly from the theoretical optimum, e.g., when $D/(2 \ell)$ approach the lower limit 0.5 or the higher limit 1.5, the prediction error increases dramatically. 
This is because when $D/(2 \ell)$ is much smaller than 1, the stencil fails to account for the points whose velocity $u$ has significant influences on the concentration $c$. On the other hand, when the region size $D/(2 \ell)$ is much larger than 1, the information is \emph{diluted} by using many points whose velocity has little influence on the concentration $c$. 
Finally, at a specified region size with a constant $D/(2 \ell)$ value, the error decreases as the stencil size (the number of points) increases. This improved performance comes from the better spatial resolution of a fixed region. Such behavior is expected in most numerical solutions of PDEs.

\subsection{Case II: Verification of learning submodel for PDEs}
\label{sec:res-P}
\begin{figure}[!htb]
\centering
\subfloat[plane Poiseuille flow]
{\includegraphics[height=0.4\textwidth]{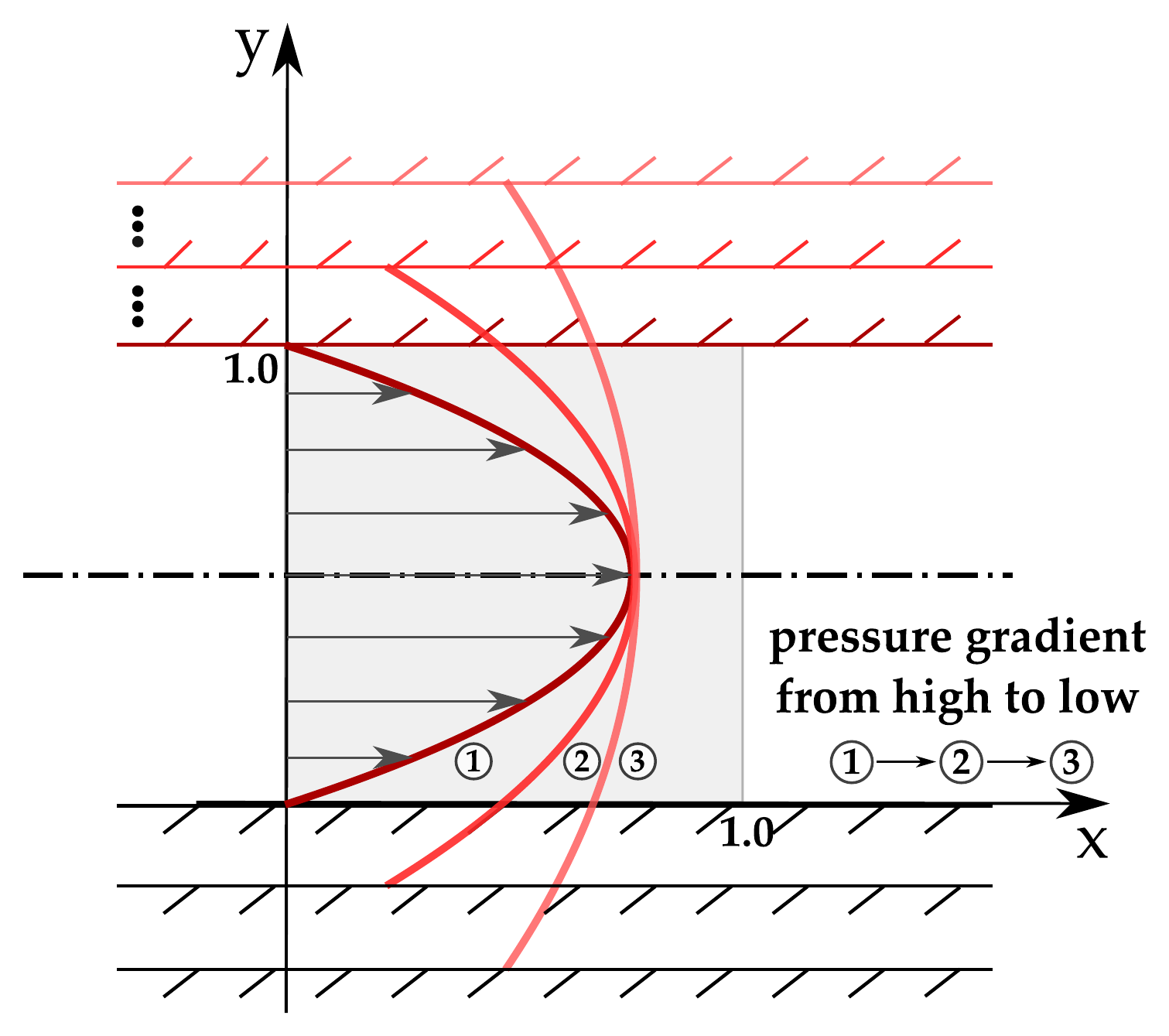}}
\hspace{0.2em}
\subfloat[Taylor--Green vortex]
{\includegraphics[height=0.35\textwidth]{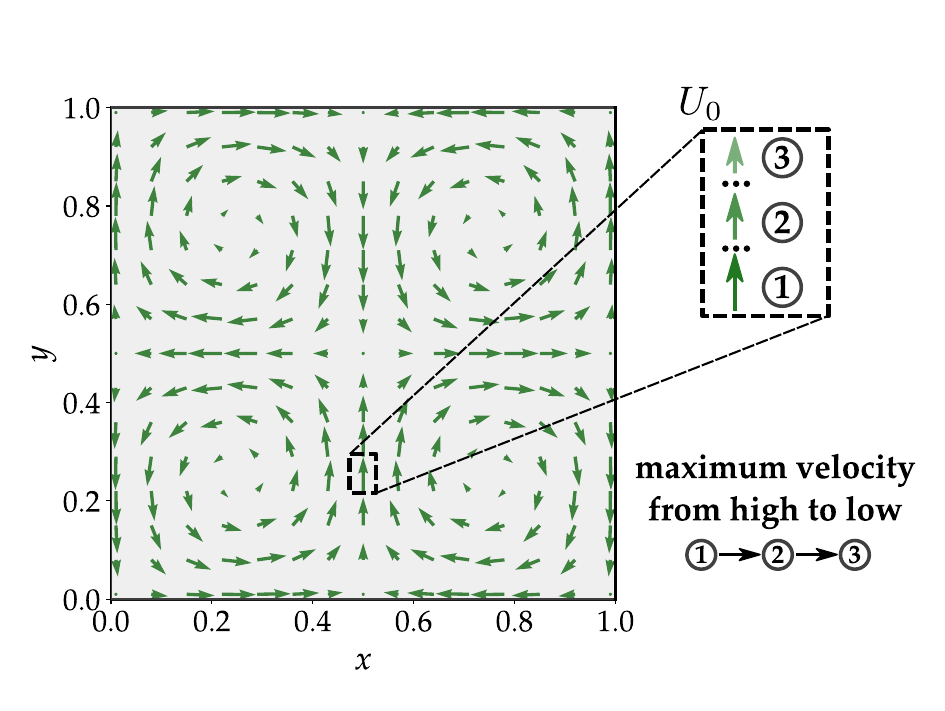}}
  \caption{
  Two families of flows used to generate training data:
  (a) plane Poiseuille flows, whose velocities are parameterized by the $\alpha$ that is proportional to the pressure gradient driving the flow as in Eq.~\eqref{eq:vp}, and (b) Taylor--Green vortex flows, parameterized by the maximum velocity $U_0$ as in   Eq.~\eqref{eq:vt}. For both flows the velocities in the domain $[0, 1] \times [0, 1]$ (shaded) are extracted. For the Poiseuille flows, the wall boundaries are adjusted according to the applied pressure gradient such that the centerline velocity is 1.0, and as such the two walls are located at $y = 0.5 \pm \frac{1}{\sqrt{\alpha}}$.}

    \label{fig:twoflows}
\end{figure}

In this case, we consider two families of flow fields shown in Fig.~\ref{fig:twoflows}. The first is plane Poiseuille flows, whose velocity field $\mathbf{u}=(u,v)$ is 
\begin{equation}
\label{eq:vp}
\begin{cases}
    u(x,y) = 1 - \alpha (y-0.5)^2, \\ 
    v(x,y)=0, 
\end{cases}
\quad (x,y) \in [0, 1]\times[0, 1]
\end{equation}
where $\alpha$ is the pressure gradient parameter (i.e., $\alpha = -\frac{1}{2\mu} \frac{dp}{dx}$, with $\mu$ being the dynamic viscosity). Another family of flows is Taylor--Green vortexes~\cite{taylor1937mechanism}, whose velocity filed $\mathbf{u}=(u,v)$ is
\begin{equation}
\label{eq:vt}
\begin{cases}
    u(x, y)  = U_0 \sin(2\pi x) \cos(2\pi y), \\
    v(x, y) =  -U_0 \cos(2\pi x) \sin(2\pi y),
\end{cases}
\quad (x,y) \in [0, 1]\times[0, 1].
\end{equation} 
where $U_0$ is the maximum velocity. The parameters $\alpha$ and $U_0$ are varied to generate two families of flow fields for training. For both flows, the velocities in the domain $[0, 1] \times [0, 1]$ (shaded in Fig.~\ref{fig:twoflows}a and ~\ref{fig:twoflows}b) are extracted to generate training data.  For plane Poiseuille flows, the velocities are aligned with the $x$-direction. To generate flow fields, we choose 100 standard plane Poiseuille flow cases parameterized by the pressure gradient parameter $\alpha$, which is uniformly sampled in the interval between 1 and 4. The wall boundaries are adjusted according to the applied pressure gradient to ensure that the centerline velocity is $u|_{y=0.5} = 1.0$, and as such the two walls are located at $y = 0.5 \pm \frac{1}{\sqrt{\alpha}}$.
On the other hand, the Taylor--Green vortex flows are parameterized by the maximum velocities $U_0$. Similarly, 100 flows are generated by uniformly sampling $U_0$ in the interval $[0, 1.0]$ to obtain different velocity fields for generating training data.

For the flows above, we consider three scenarios of specified production fields:
\begin{description}
    \item [Case IIa] The production is only a function of the normalized strain rate magnitude $S$, i.e., $\mathcal{P}(\mathbf{u}) = f(S)$, with $f(S)$ defined in Eq.~\eqref{eq:fs} and depicted in Fig.~\ref{fig:source}a,
    \item [Case IIb] The production is a function of normalized strain rate $S$ and rotation rate magnitude $\Omega$, i.e., $\mathcal{P}(\mathbf{u}) = f(S) g(\Omega)$ where $g(\Omega) = 1/(1 + \Omega^2)$  as defined in Eq.~\eqref{eq:gomega}, and
    \item [Case IIc] The production is a function of the strain rate $S$, i.e., $\mathcal{P}(\mathbf{u}) = h(S)$ (see Eq.~\eqref{eq:hc} and Fig.~\ref{fig:source}b). This is a more challenging case that requires a more comprehensive investigation. It is omitted here for brevity (see \ref{sec:appd_learn-h} for details).
\end{description}

Given any combination of a velocity field and a production field generated as detailed above, we then generate the training data ($\region{\mathbf{u}, c}$) by solving the transport PDE~\eqref{eq:cdr2}. The diffusion coefficient in the PDE is set to $\nu=0.02$, and dissipation coefficient is $\zeta=20$. 
The stencil size for the region-to-point mapping in both families of flows is set to be $11\times11$, and the downsampling parameter $n_0$ is 2 such that the region size is approximately  $0.2\times0.2$, which is very close to the suggested region size $2 \ell$ determined by $\nu$, $\zeta$, $\mathbf{u}$ and error tolerance $\epsilon=0.2$ according to Eq.~\eqref{eq:stencil-size}.

The training data define a region-to-point mapping $(\region{\mathbf{u}}_{i,j}) \mapsto c_{i, j}$, while the underlying exact mapping is $c = G^{\mathbf{u}} \star \mathcal{P}(S)$ or $c = G^{\mathbf{u}} \star \mathcal{P}(S, \Omega)$. The production is a function of the strain rate (and rotation rate), so both the Green's kernel and production depend on the flow field $\mathbf{u}$.
Apart from learning the Green's kernel, the constitutive neural network also correctly learns the production term $P$. In fact, here we will demonstrate that the proposed network can even learn the production submodel $\mathcal{P}(S, \Omega)$ itself, i.e.,  the functional dependence of production $P$ on the strain rate magnitude $S$ and rotation rate $\Omega$.

\begin{figure}[!htb]
\centering
\subfloat
{\includegraphics[width=0.30\textwidth]{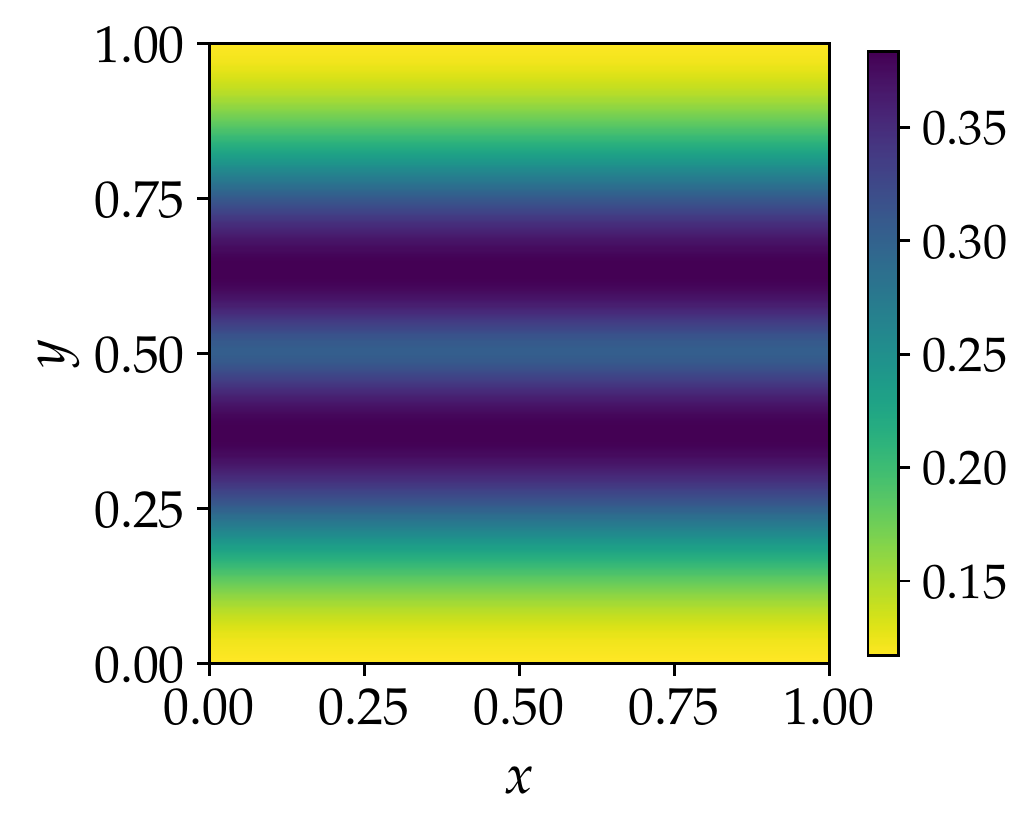}}
\hspace{0.2em}
\subfloat
{\includegraphics[width=0.30\textwidth]{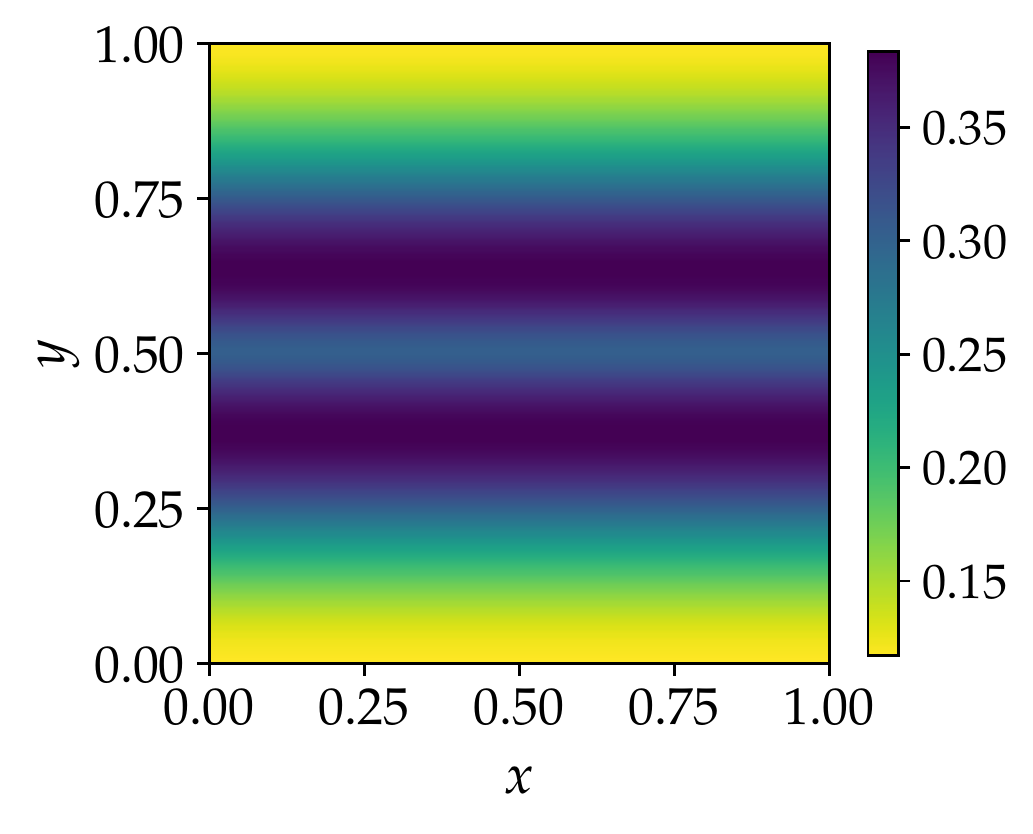}}
\hspace{0.2em}
\subfloat
{\includegraphics[width=0.35\textwidth]{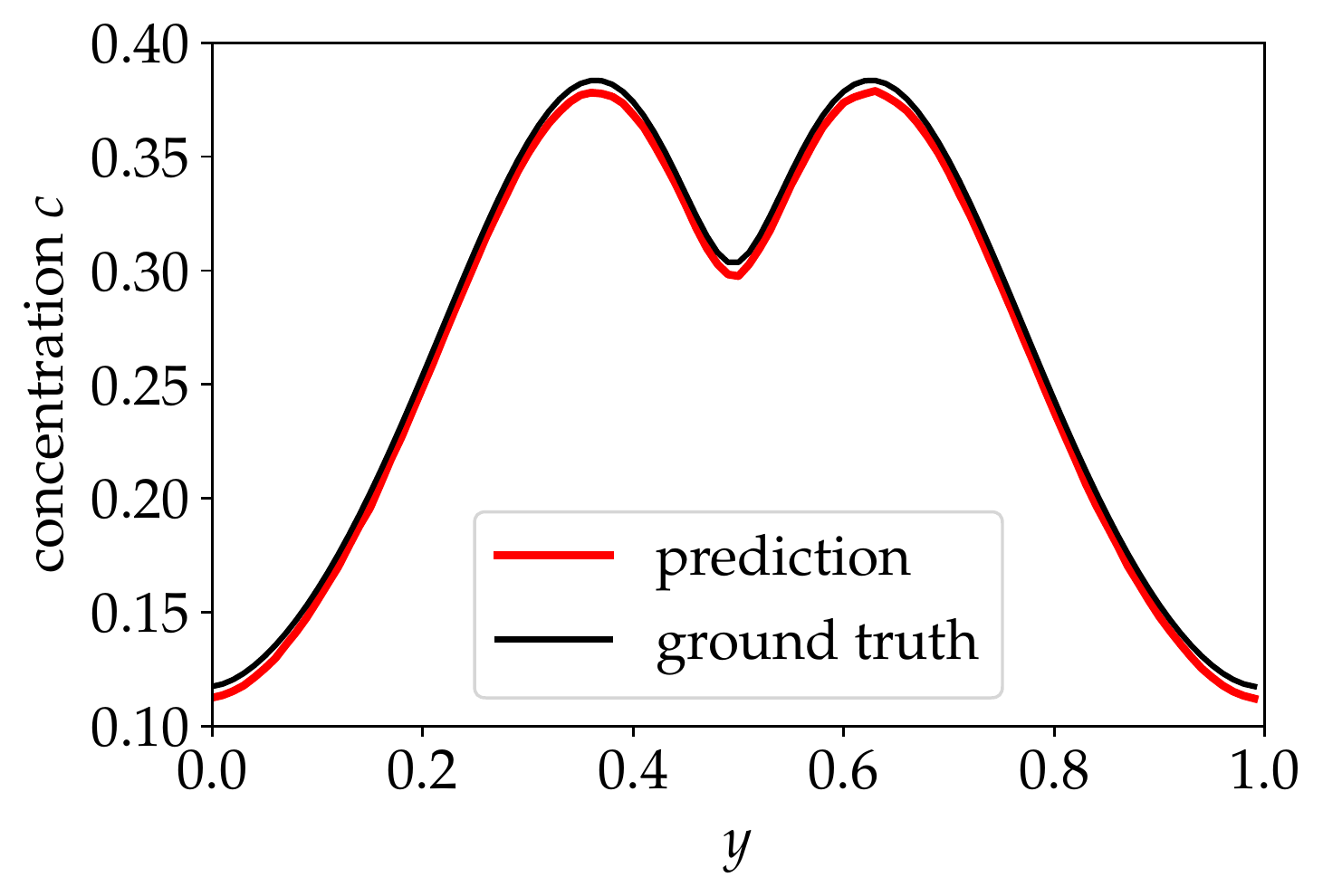}}\\
\addtocounter{subfigure}{-3}
\subfloat[ground truth of concentrations]
{\includegraphics[width=0.30\textwidth]{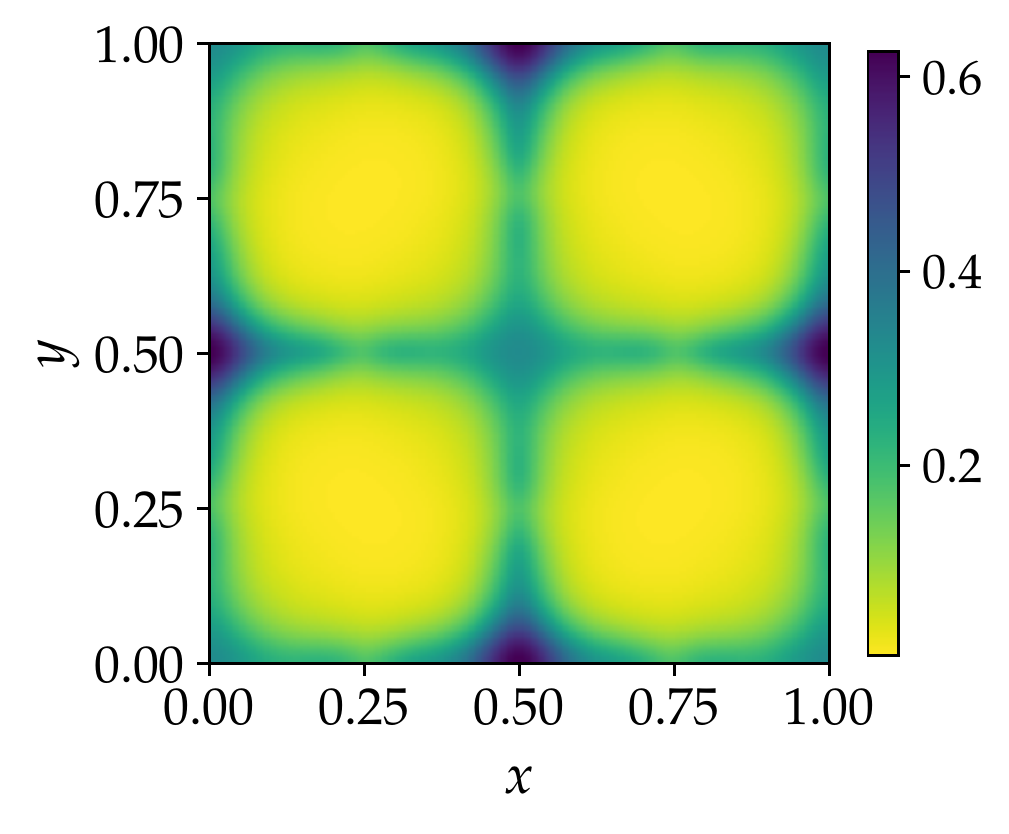}
}
\hspace{0.2em}
\subfloat[predicted concentrations]
{\includegraphics[width=0.30\textwidth]{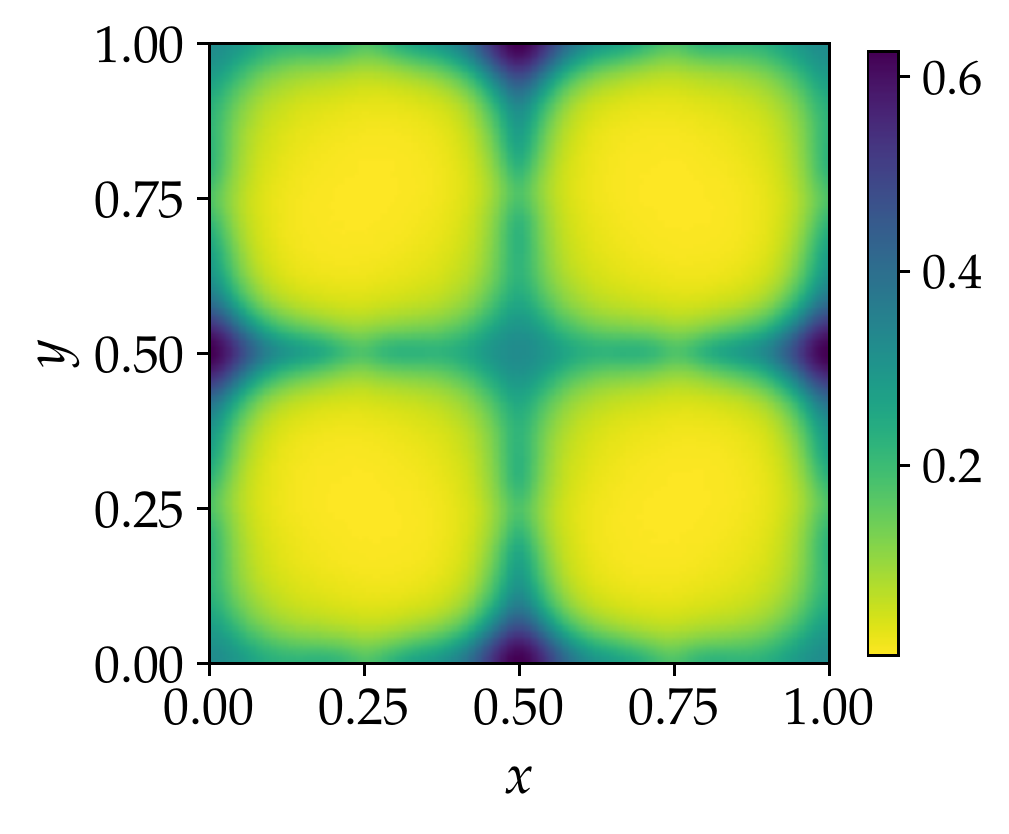}
}
\hspace{0.2em}
\subfloat[predict concentration $c$ at $x=0.5$]
{\includegraphics[width=0.35\textwidth]{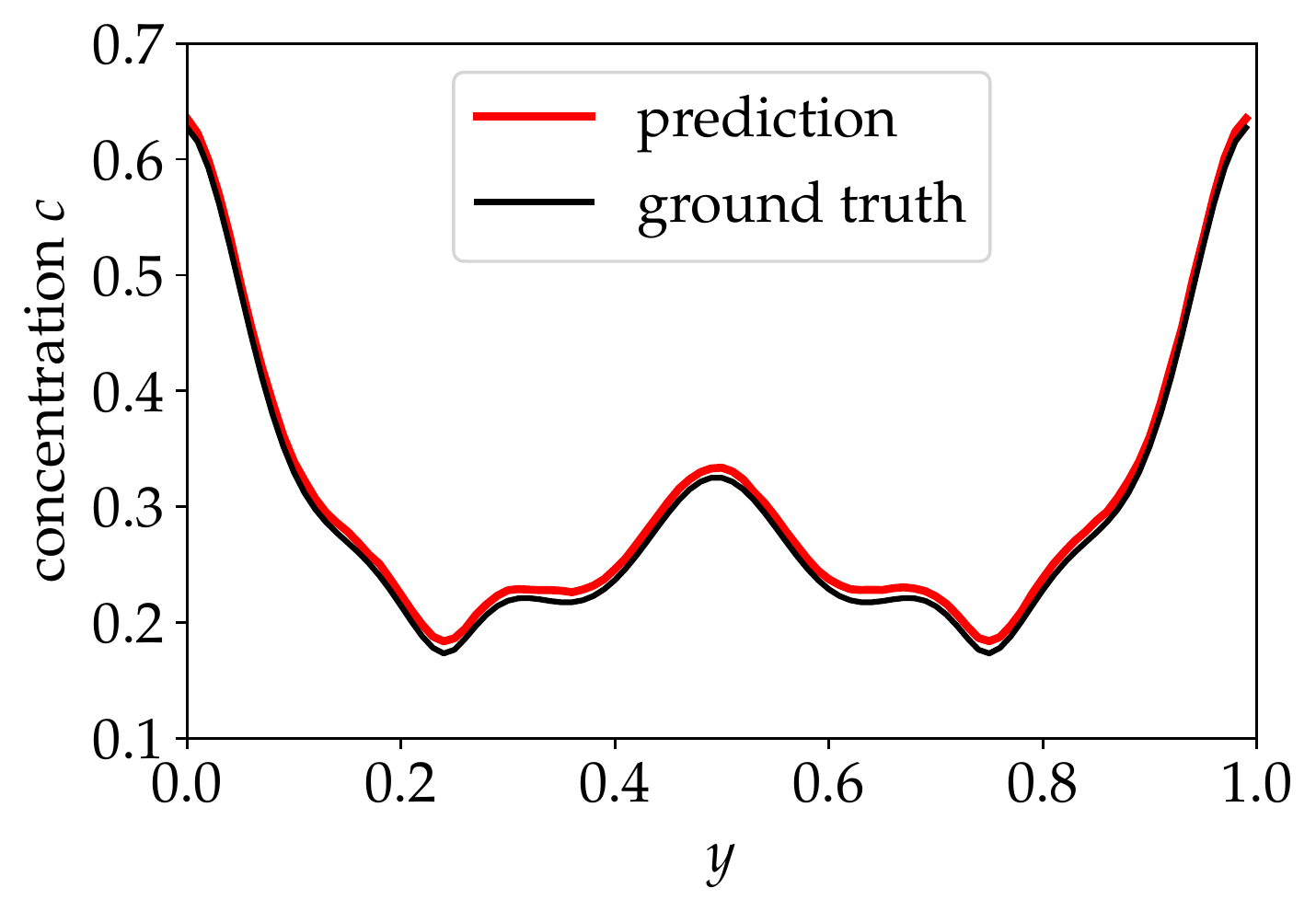}
}
  \caption{
  \label{fig:c-compare-case2}
  Comparison of the ground truth of the concentration fields (left column) and the corresponding predicted ones (middle column), and the concentration plots in two cross-sections (right column) in \emph{Case IIb} with a specified production term $\mathcal{P}(\mathbf{u}) = f(S)g(\Omega)$.
  The top row is for the plane Poiseuille flow and the bottom row is for the Taylor--Green vortex.
  }
\end{figure}

In \emph{Case IIb}, i.e., with production $\mathcal{P}(\mathbf{u}) = f(S)g(\Omega)$, the prediction error plots (omitted here for brevity; see \ref{sec:app-vp}) show excellent agreement between prediction and ground truth of the tracer concentration $c$. The normalized prediction errors for the Poiseuille flow and Taylor--Green vortex are $0.62\%$ and $1.05\%$, respectively.
A comparison of the predicted tracer concentration field and the ground truth for a representative test case (out of a total of five) is shown in Fig.~\ref{fig:c-compare-case2}, which suggests that the concentration fields are predicted very well. Compared to \emph{Case I} with a uniform velocity field, the predicted concentrations for both plane Poiseuille flow and Taylor--Green vortex are closer to the ground truths and much smoother, even though the flows are more complex. This is likely because the production field is smoother (compared to the production field drawn from the Gaussian processes).

\begin{figure}[!htb]
\centering
\subfloat
{\includegraphics[width=0.32\textwidth]{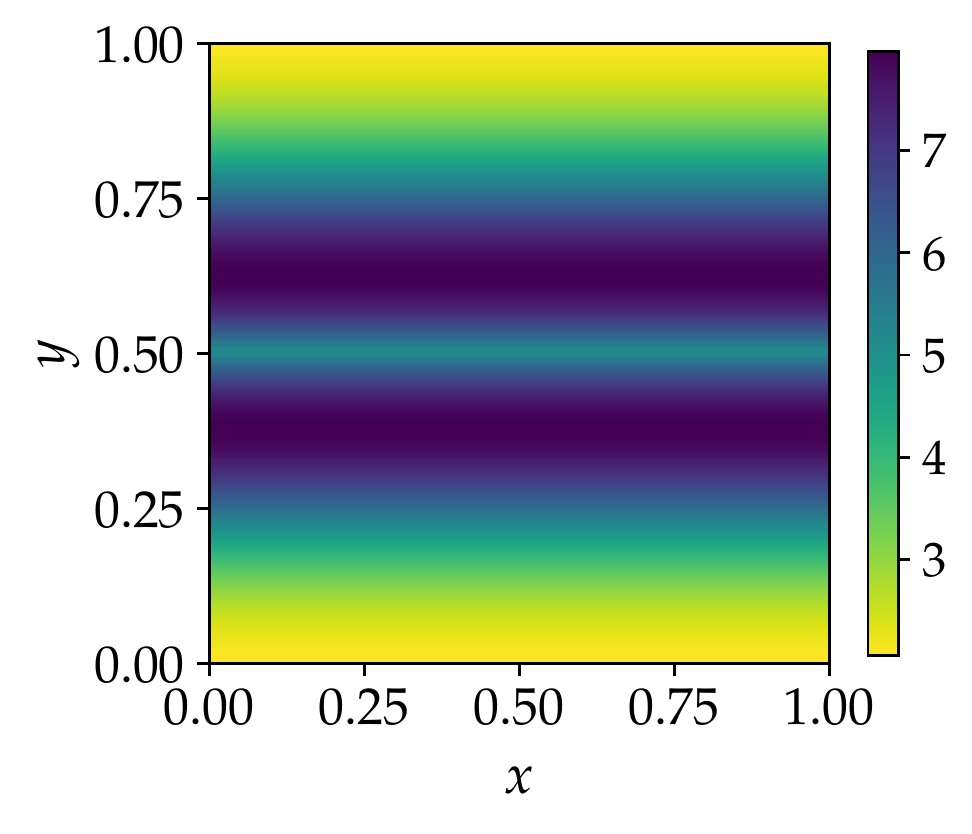}}
\hspace{0.5em}
\subfloat
{\includegraphics[width=0.32\textwidth]{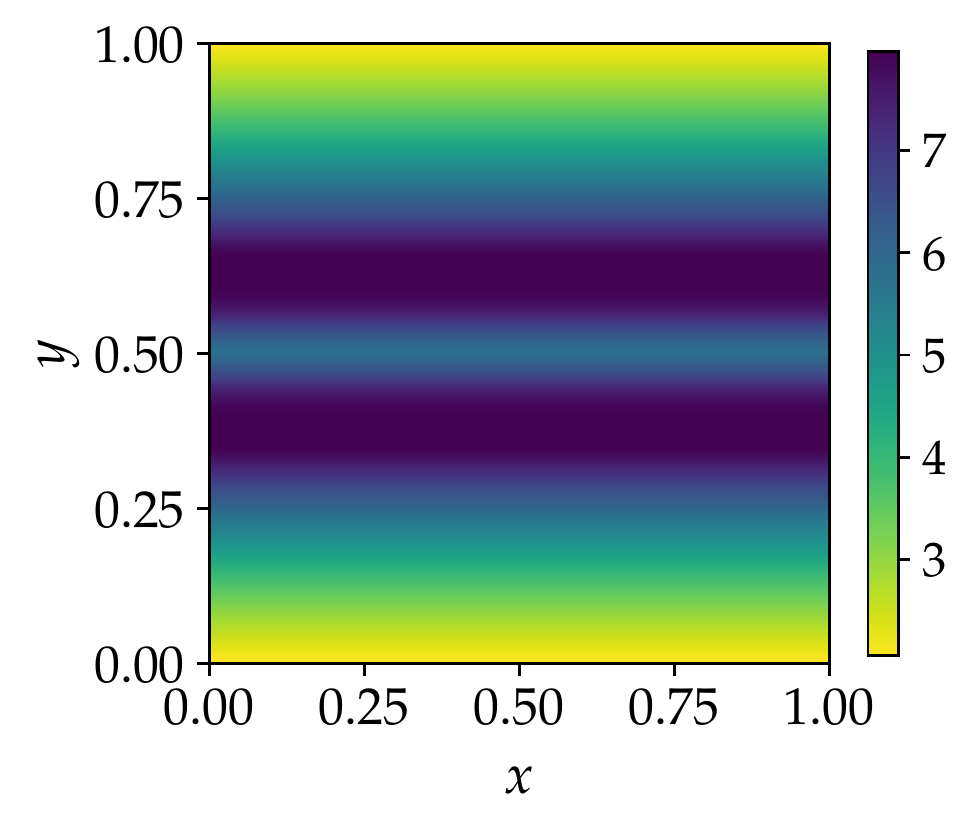}}
\quad
\addtocounter{subfigure}{-2}
\subfloat[ground truth of productions]
{\includegraphics[width=0.32\textwidth]{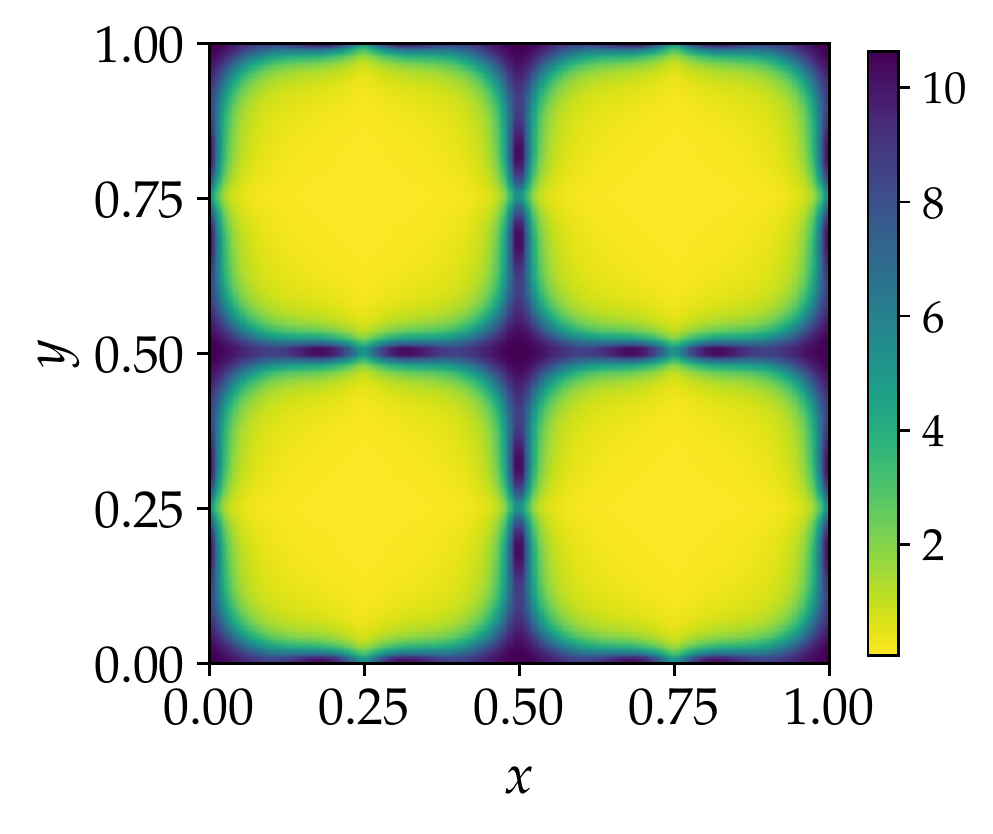}}
\hspace{0.5em}
\subfloat[predicted productions]
{\includegraphics[width=0.32\textwidth]{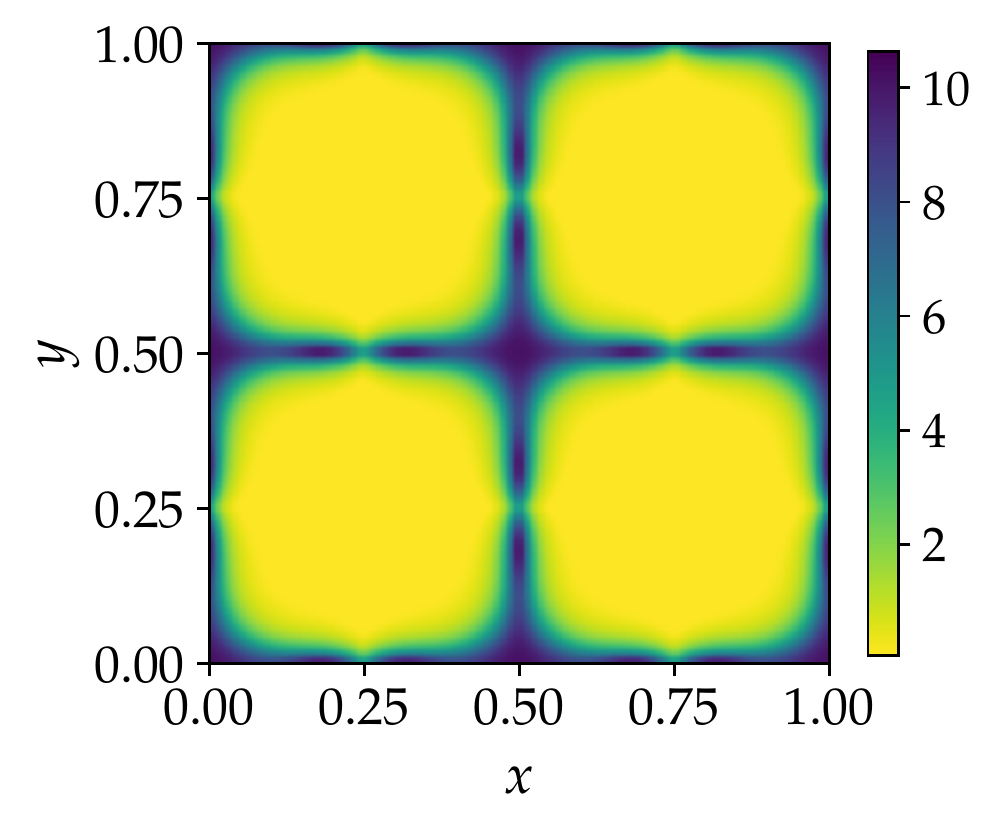}}
  \caption{
  \label{fig:production-compare-case2}
  {Comparison of the ground truth of the production fields (left column) and the corresponding predicted ones (right column) in \emph{Case IIb} with a specified production term $\mathcal{P}(\mathbf{u}) = f(S) g(\Omega)$. Top row: the plane Poiseuille flow; bottom row: Taylor--Green vortex.}
  }
\end{figure}

The network not only predicts the closure variable $c$ accurately, but also learns the underlying production. This is demonstrated in Fig.~\ref{fig:production-compare-case2}. 
As the solution $c(\mathbf{u}) = G(\mathbf{u}) \star P(\mathbf{u})$ is a convolution of the Green's kernel and the production term, we can only expect to learn both $G$ and $P$ up to a scaling constant. This is because the solution remains the same with respect to an arbitrary scaling $\beta$ to $P$ accompanied by a scaling of $1/\beta$ to $G$. 
Therefore, before the comparison we first multiplied the learned production field $\hat{P}$ by a scaling constant $\hat{\beta}$, i.e., $P = \hat{\beta} \hat{P}$. The scaling factor $\hat{\beta}$ is chosen such that the squared discrepancy between the learned production $\hat{P}$  and the ground truth $P^*$ is minimized, i.e., $\hat{\beta} = \argmin_\beta \| \beta \hat{P} - P^*\|^2$. 
With such a scaling, it can be seen that the predicted production fields in Fig.~\ref{fig:production-compare-case2} show similar patterns to that of the ground truths for both the plane Poiseuille flow and the Taylor--Green vortex.

\begin{figure}[!htb]
\centering
\includegraphics[width=0.5\textwidth]{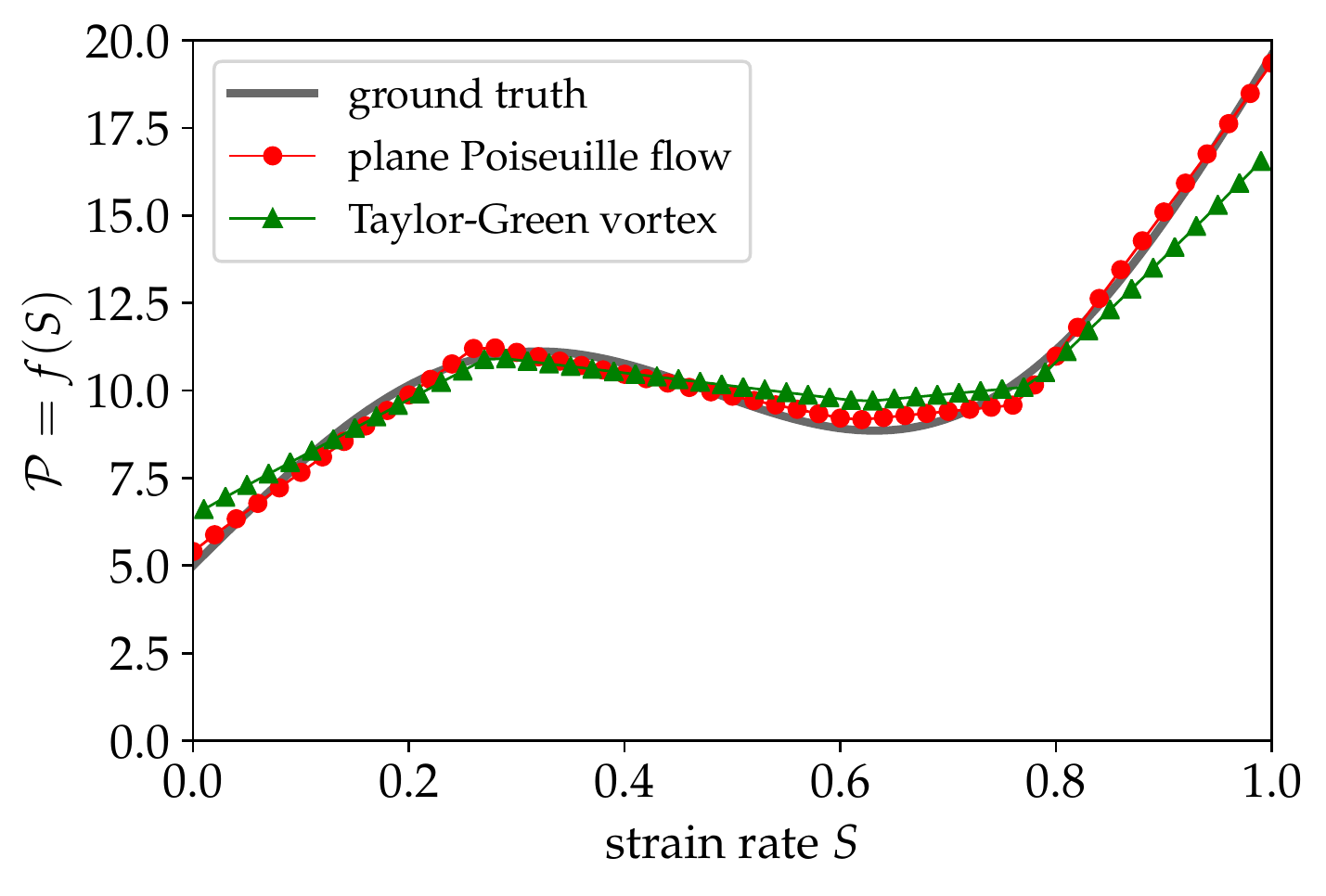}
  \caption{
  \label{fig:learning-Pfunction1}
  Learned production as functions of the strain rate $S$ in the plane Poiseuille flow and the Taylor--Green vortex compared to the ground truth (synthetic analytic function) in \emph{Case IIa}, i.e., $\mathcal{P}(\mathbf{u}) = f(S)$.
  }
\end{figure}

Given that the neural network predicts well the pattern of the production field, we further verify that the network is able to learn the underlying submodel of the production.
In \textit{Case IIa}, where $\mathcal{P}(\mathbf{u}) = f(S)$, 
we can see from Fig.~\ref{fig:learning-Pfunction1} that both the learned production functions for both the plane Poiseuille flows and Taylor--Green vortexes are close to the specified production function, suggesting that the neural network can learn the submodel $\mathcal{P}(\mathbf{u}) = f(S)$ accurately. Specifically, it learns the trends of the specified production function in three different regimes and two inflection points at the strain rate being around 0.3 and 0.68. A small shift of the learned inflection points may be caused by the uneven distribution of the training data, and the sharp change around the inflection points is likely due to the use of the rectified linear activation function (ReLU), i.e., $\text{ReLU}(x) := \max(0, x)$.

\begin{figure}[!htb]
\centering
\subfloat[plane Poiseuille flows]
{\includegraphics[width=0.42\textwidth]{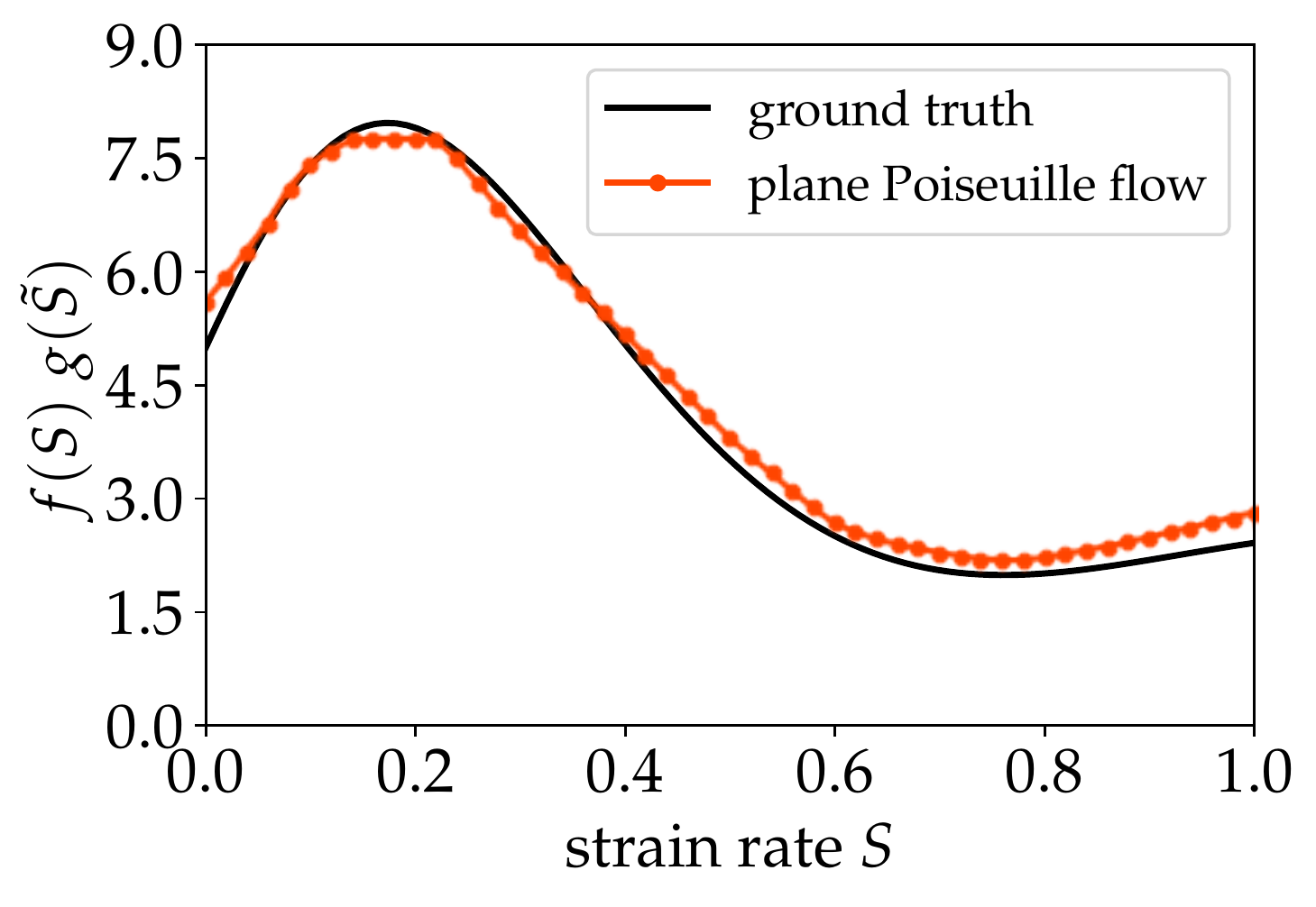}}
\hspace{0.2em} \\
\subfloat[Taylor--Green vortexes, $f(S)$]
{\includegraphics[width=0.42\textwidth]{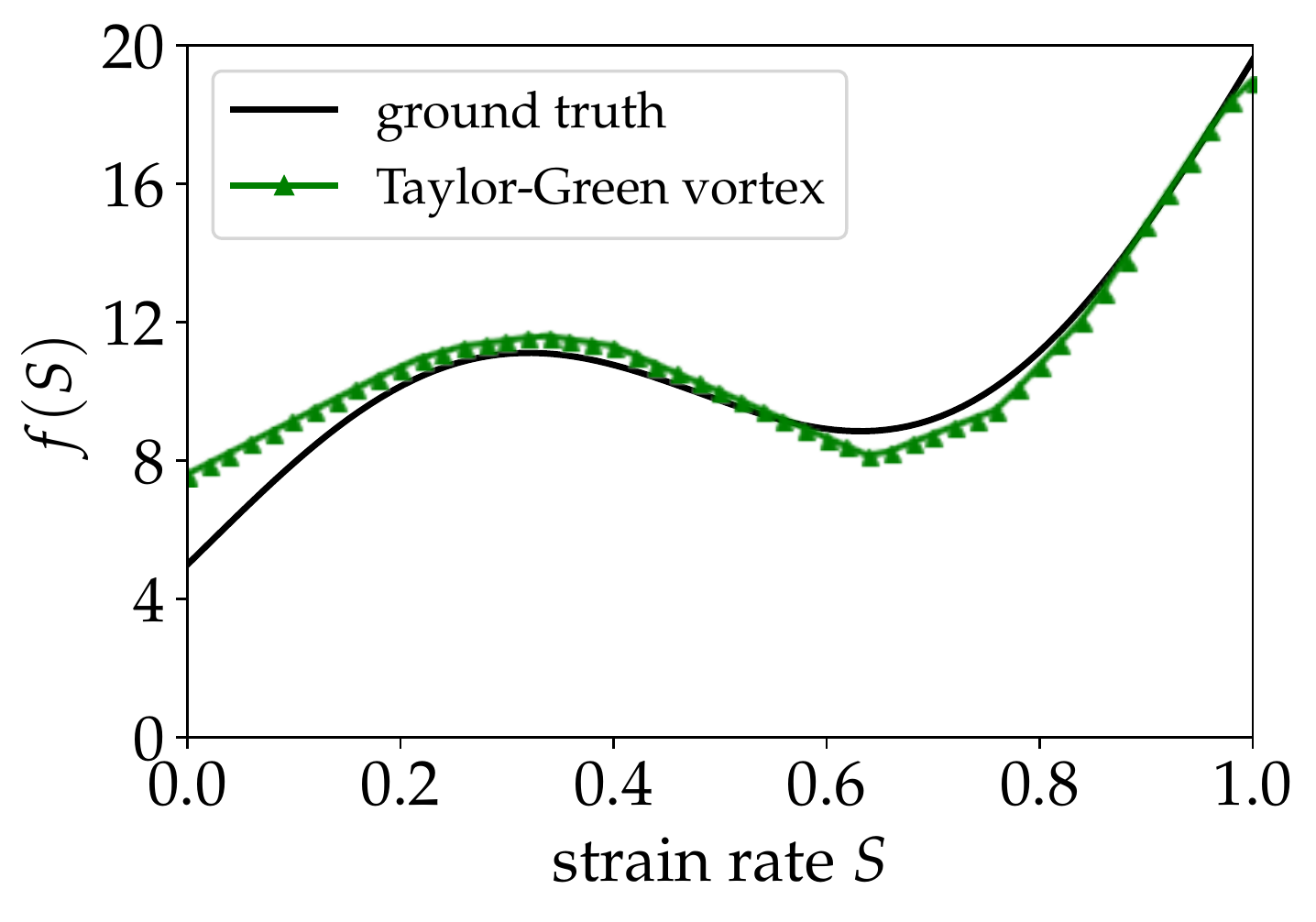}}
\hspace{0.2em}
\subfloat[Taylor--Green vortexes, $g(\Omega)$]
{\includegraphics[width=0.42\textwidth]{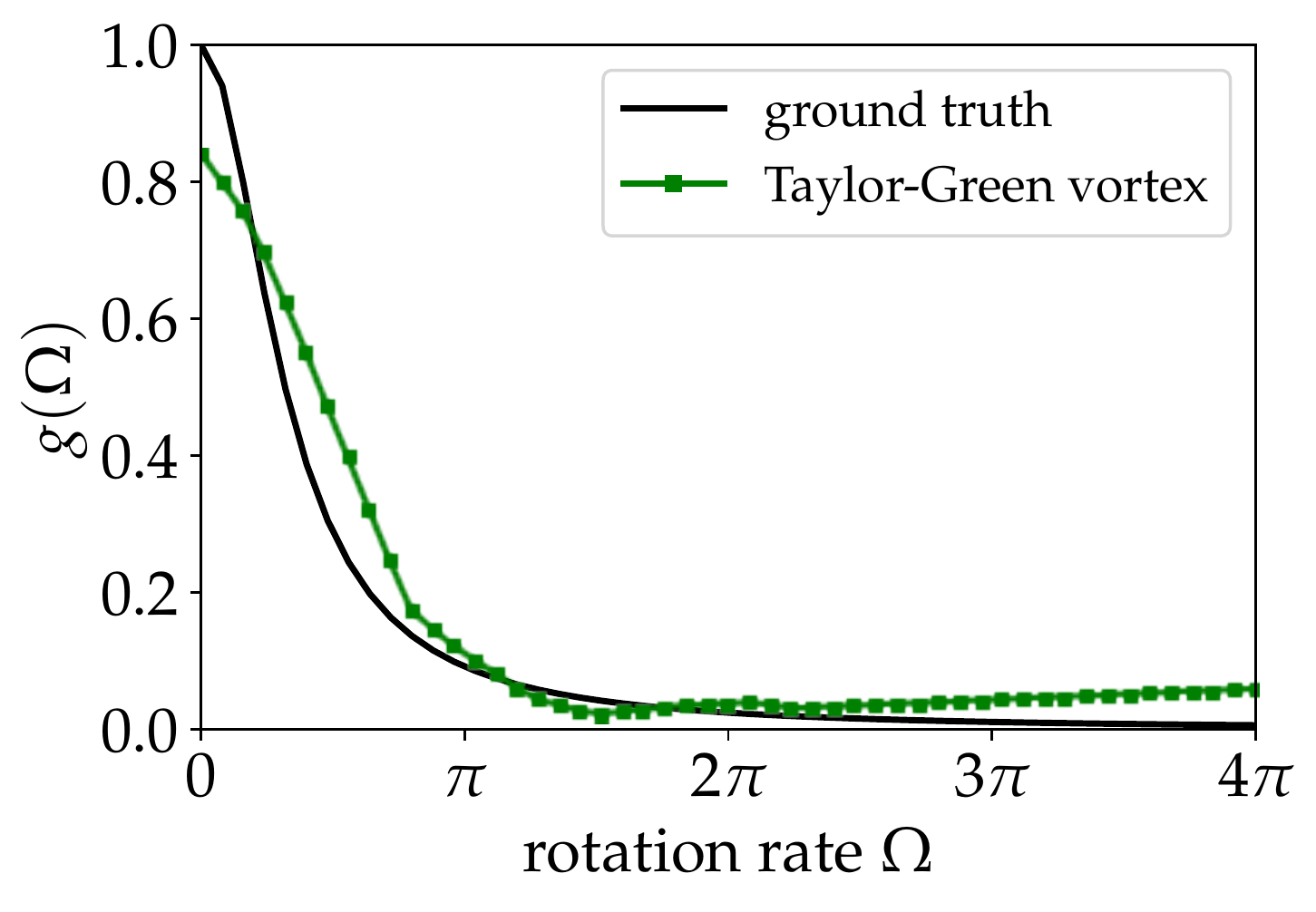}}
  \caption{
  \label{fig:learning-Pfunction2}
  Comparison of the learned production as functions of strain rate $S$ and rotation rate $\Omega$ in the plane Poiseuille flow and the Taylor--Green vortex with the corresponding ground truths (synthetic analytic functions) in \emph{Case IIb}, i.e., $\mathcal{P}(\mathbf{u}) = f(S)g(\Omega)$. 
  Due to the co-linearity of $S$ and $\Omega$ in the plane Poiseuille flows, we only present the learned production as a degenerate univaraite function of $S$ only in subfigure (a).
  }
\end{figure}

Even in the more challenging \textit{Case IIb}, where $\mathcal{P}(\mathbf{u}) = f(S)g(\Omega)$, the neural network is still able to learn the submodel of the production term, which is demonstrated in Fig.~\ref{fig:learning-Pfunction2}. However, for the plane Poiseuille flow, a caveat is that the strain rate and the rotation rate are co-linear. That is, $\tilde{S} = \tilde{\Omega} = |\frac{\partial u}{\partial y}|$ as $\frac{\partial v}{\partial x} = 0$, and thus $S$ and $\Omega$ only differ by a factor of the normalization constant. This coincidental co-linearity is due to the simplicity of the flow field (homogeneity in the stream direction). The co-linearity can also be confirmed from the scatter plot in Fig.~\ref{fig:cS-scatter} (for \emph{Case III}), as the projection of the points for the Poiseuille flow onto the $S$--$\Omega$ plane would form a straight line (not shown).
As a result, the bivariate production function $\mathcal{P}(S, \Omega)$ degenerates to a univariate function of $S$ only. In light of this observation, the neural network successfully learned the functional relationship between the production $\mathcal{P}(S, \Omega) = f(S) g(\Omega)$ and the strain rate $S$ (see Fig.~\ref{fig:learning-Pfunction2}a). For Taylor--Green vortex, the correlation between the strain rate and the rotation rate is much weaker, and thus the production $\mathcal{P}(S, \Omega)$ is determined jointly by both strain rate $S$ and rotation rate $\Omega$ (both normalized properly). Consequently, the neural network is able to learn both $f(S)$ and $g(\Omega)$ from the data, as shown in Fig.~\ref{fig:learning-Pfunction2}b and~\ref{fig:learning-Pfunction2}c, respectively. We note that, in all cases, there are discrepancies between the learned productions near the upper and lower bounds of the strain rate $S$ and rotation rate $\Omega$. This is mainly caused by the uneven distribution of strain rate $S$ and rotation rate $\Omega$ in the training data. Specifically, in plane Poiseuille flows, $S = \frac{1}{4}|\frac{\partial u}{\partial y}|=\frac{\alpha}{2} |y-0.5|$, where $\alpha$ is uniformly sampled in the interval between 1 and 4. In this way, the probability of $S$ being picked close to is 1 is relatively small only when $\alpha\approx 4$ and $y\approx 0 \text{ or }1$, which leads to the uneven data distribution. A similar situation happens to Taylor--Green vortexes in terms of both strain rate $S$ and rotation rate $\Omega$.

In~\ref{sec:appd_learn-h}, we study further a scenario (\textit{Case IIc}) showing the effect of the data distribution on the learning of submodel, especially when the data in the regime of interest is not well represented.
For both plane Poiseuille flows and Taylor--Green vortexes, the learned functions are close to the specified production functions. It is of significance that the neural network can learn the inflection points of the production function and their corresponding value of strain rate, especially for those where the submodel is complicated.

\subsection{Case III: Assessment of predictive performance for nonlinear PDEs}
\label{sec:res-all}
In this case, we again consider two types of flow fields: (1) plane Poiseuille flows and (2) Taylor--Green vortexes. The velocity fields of the flows are the same with those in \emph{Case II} as defined in Eq.~\eqref{eq:vp} and Eq.~\eqref{eq:vt}. The distribution of the flow fields and the constants of diffusion and dissipation are the same as those in \emph{Case II}.
Accordingly, the settings of stencil size and downsampling parameter are the same with those in \emph{Case II} such that the used region size matches approximately the suggested one.

\begin{figure}[!htb]
\centering
\includegraphics[width=0.55\textwidth]{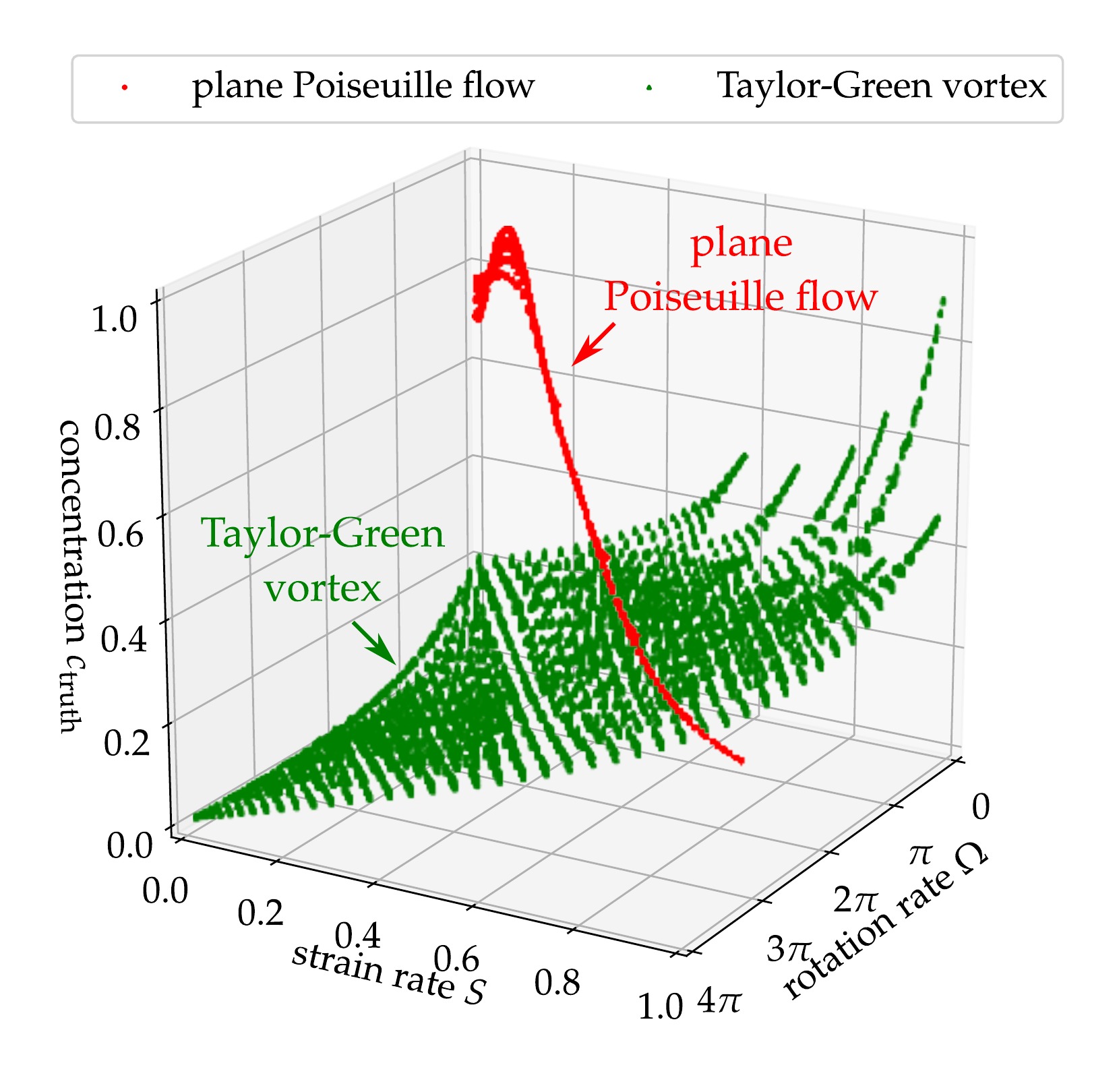}
  \caption{
  \label{fig:cS-scatter}
  Distribution of strain rate and concentrations at each point in \emph{Case III} (nonlinear PDE with a specified production field $\mathcal{P}(c, \mathbf{u}) = f(S) g(\Omega) (K\,h(c)+h_0)$), showing  strong nonlinearity between the strain rate, rotation rate and the concentrations on five different prediction flows of plane Poiseuille flows and Taylor--Green vortexes.
  }
\end{figure}

For each of the flows, we consider the production as a function of the (normalized) strain rate magnitude $S$, rotation rate magnitude $\Omega$ and concentration $c$, i.e., $\mathcal{P}(c,\mathbf{u}) = f(S)g(\Omega)(K\,h(c)+h_0)$ where $g(\Omega)$ and $h(c)$ are defined in Eq.~\eqref{eq:gomega} and Eq.~\eqref{eq:hc}.
The constants are chosen as $(K = 1.0, h_0=2.2)$ for plane Poiseuille flows and $(K=0.4, h_0=0.75)$ for Taylor--Green vortexes.
We generate training data by solving the general nonlinear transport PDE~\eqref{eq:cdr3} with the given velocity field and production. 
Strong nonlinearity in the resulting PDEs is observed from the scatter plot Fig.~\ref{fig:cS-scatter} between (normalized) strain rate $S$, rotation rate $\Omega$ and the concentration $c$.

\begin{figure}[!htb]
\centering
\subfloat
{\includegraphics[width=0.3\textwidth]{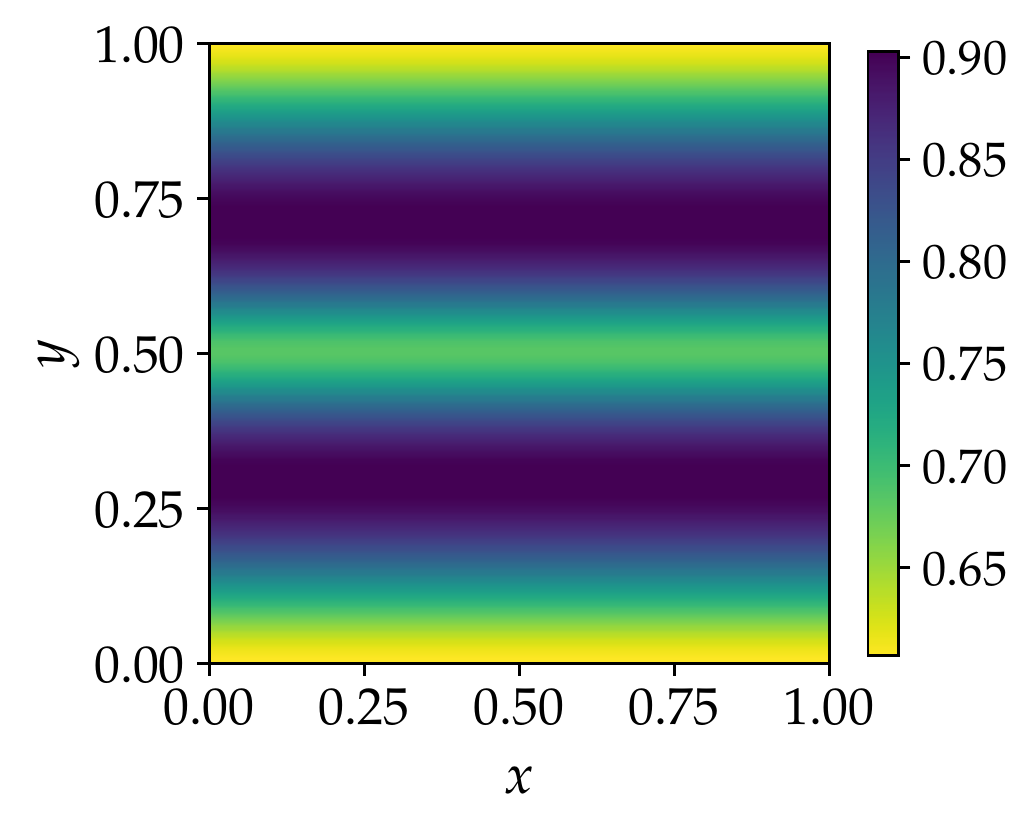}}
\hspace{0.2em}
\subfloat
{\includegraphics[width=0.3\textwidth]{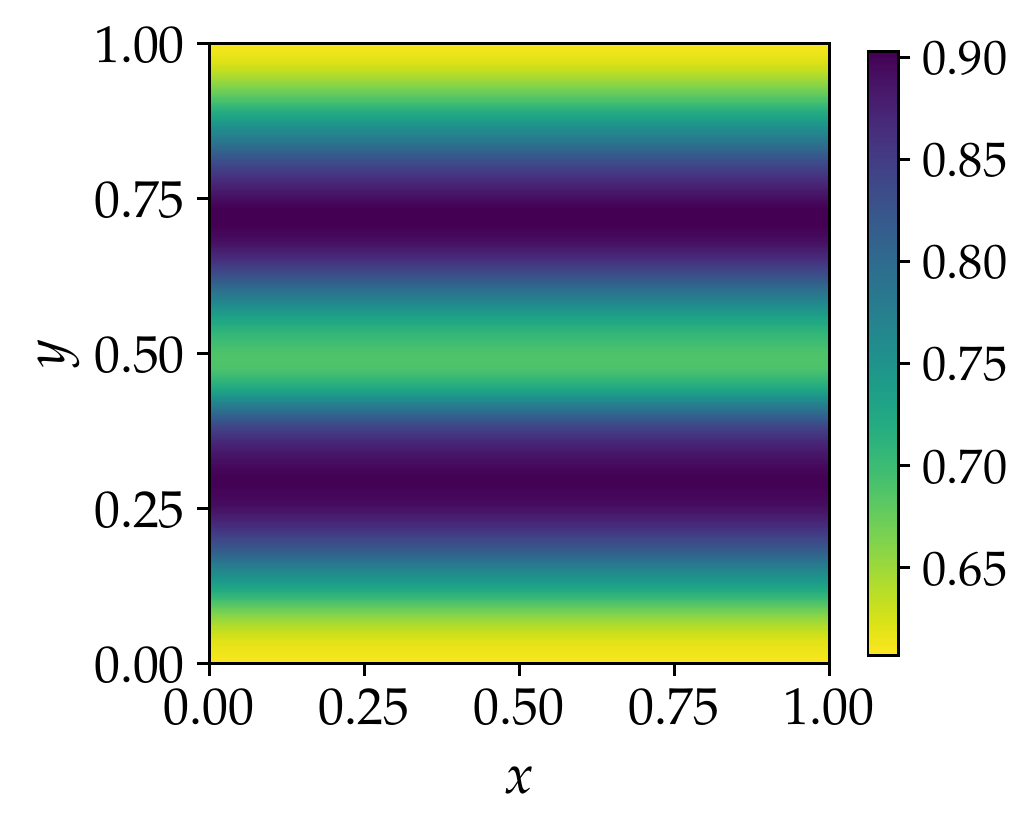}}
\hspace{0.2em}
\subfloat
{\includegraphics[width=0.35\textwidth]{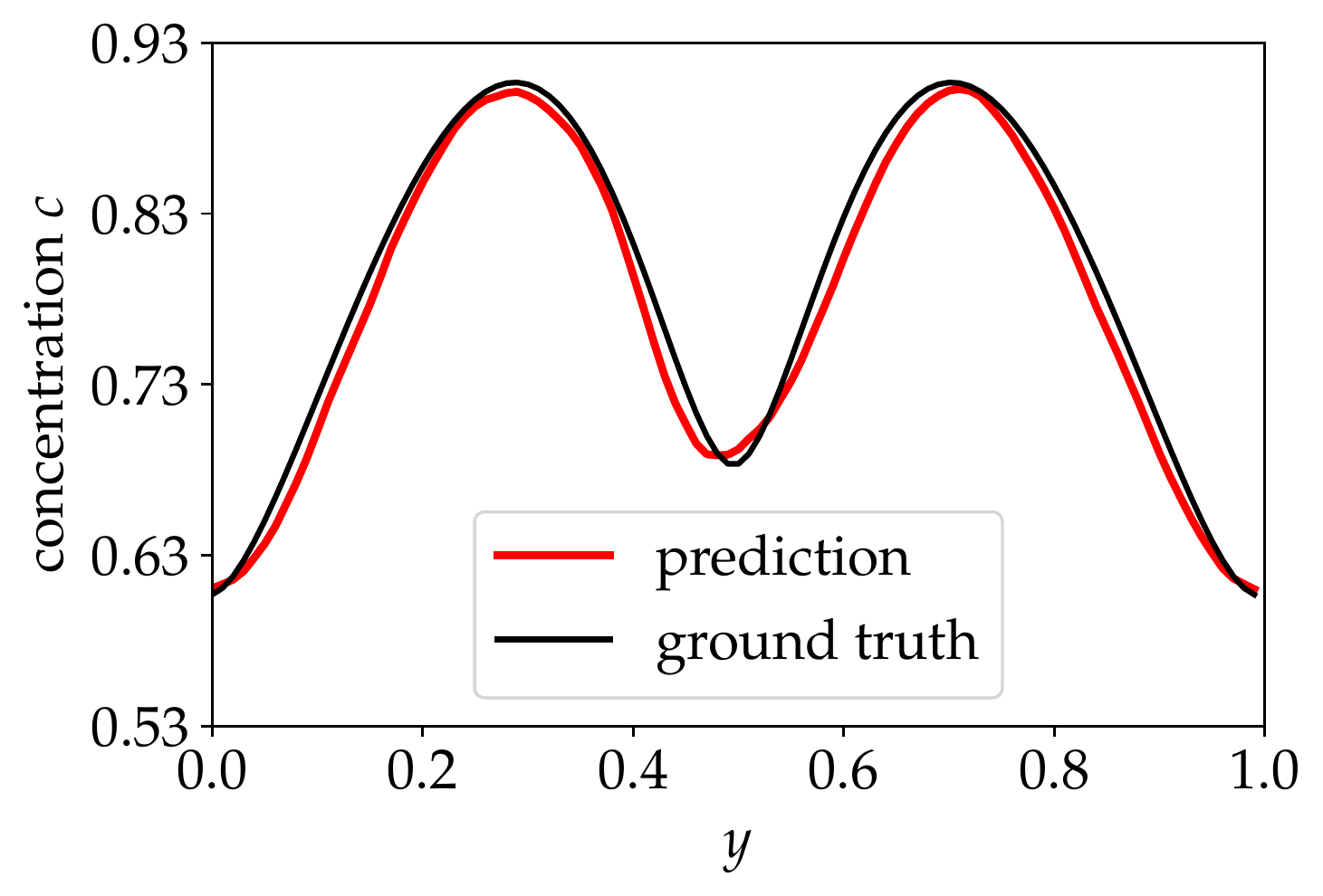}}
\quad
\addtocounter{subfigure}{-3}
\subfloat[ground truth of concentrations]
{\includegraphics[width=0.30\textwidth]{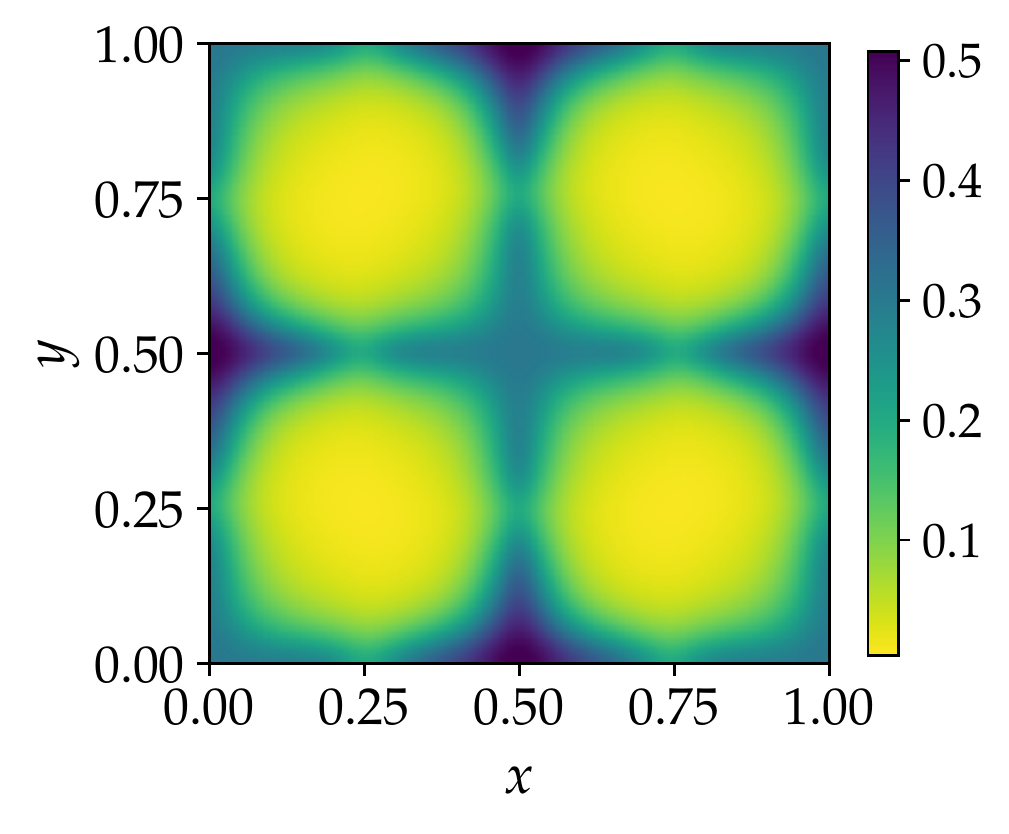}}
\hspace{0.2em}
\subfloat[predicted concentrations]
{\includegraphics[width=0.30\textwidth]{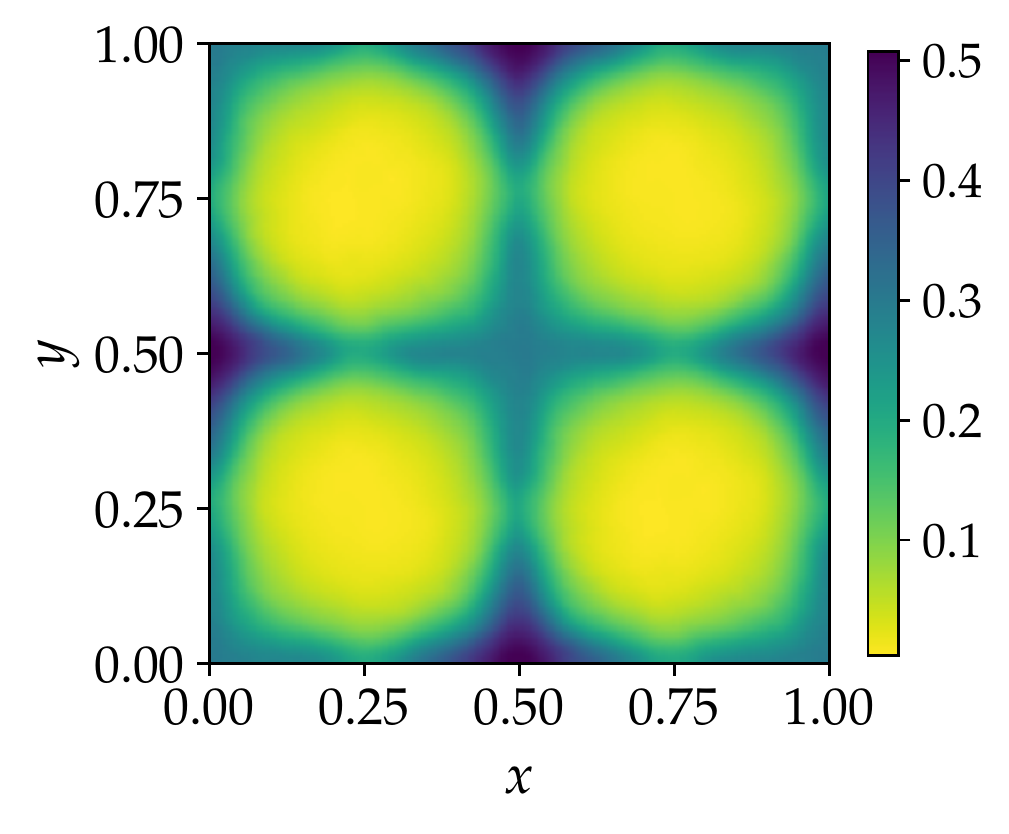}}
\hspace{0.2em}
\subfloat[predict concentration $c$ at $x=0.5$]
{\includegraphics[width=0.35\textwidth]{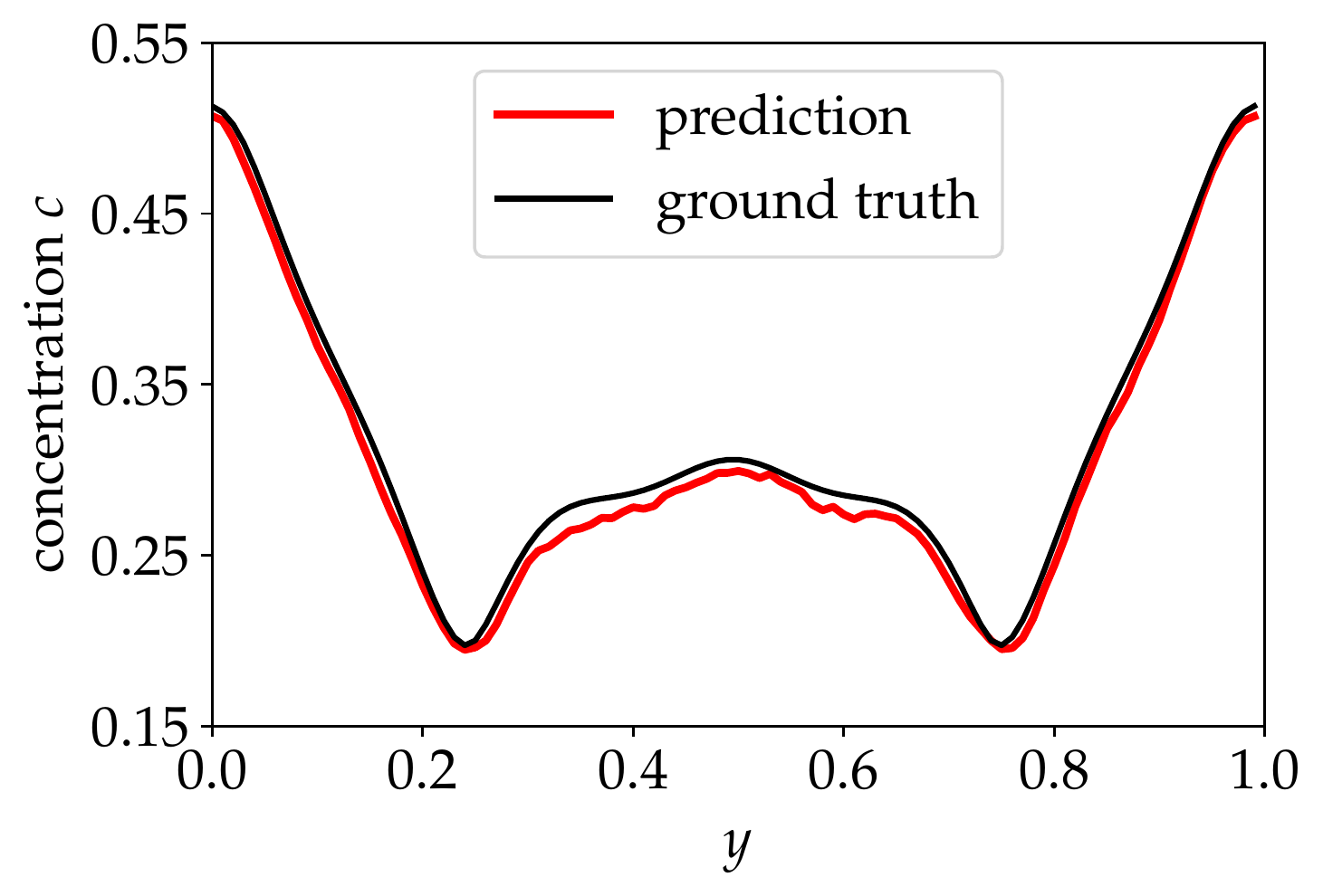}}
  \caption{
  Comparison of the ground truth of the concentration fields (left column) and the corresponding predicted ones (middle column) in \emph{Case III} with a specified production term $\mathcal{P}(c, \mathbf{u}) = f(S)g(\Omega)(K\,h(c)+h_0)$, showing the \textbf{interpolation capability} of the trained models. The concentration plots in two cross-sections (right column) are also presented.
  The top row is for the plane Poiseuille flow and the bottom row is for the Taylor--Green vortex.
  \label{fig:c-compare-case3}
  }
\end{figure}

\begin{figure}[!htb]
\centering
\subfloat
{\includegraphics[width=0.3\textwidth]{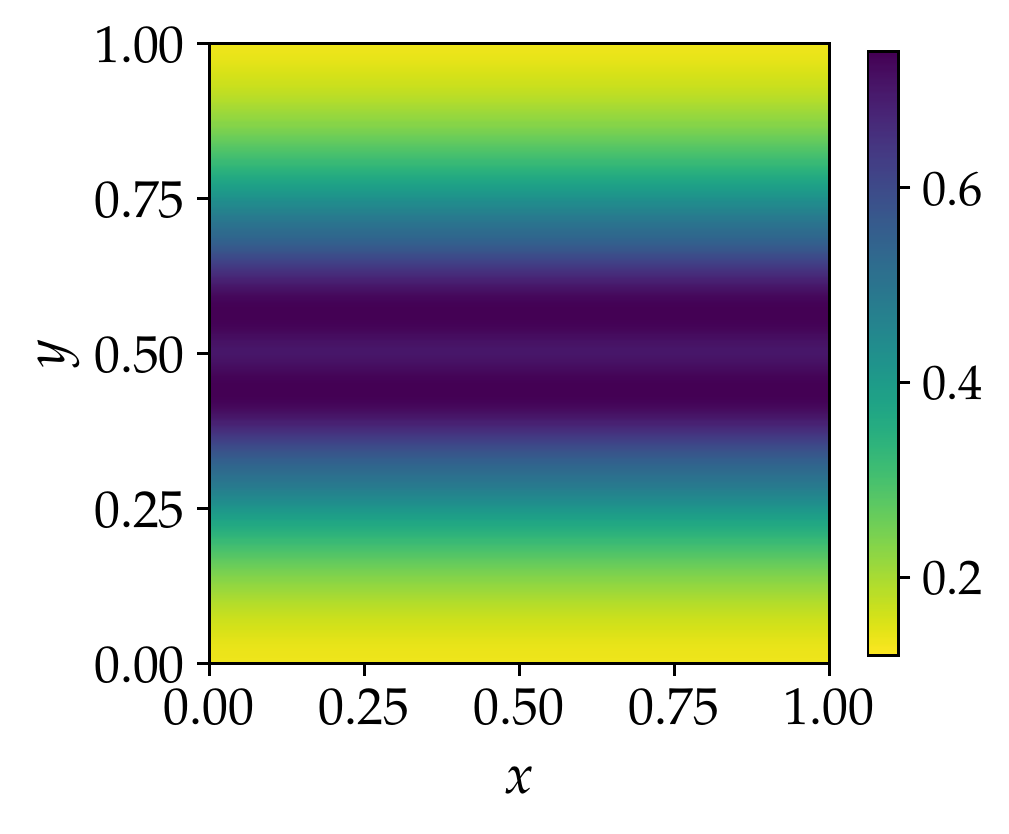}}
\hspace{0.2em}
\subfloat
{\includegraphics[width=0.3\textwidth]{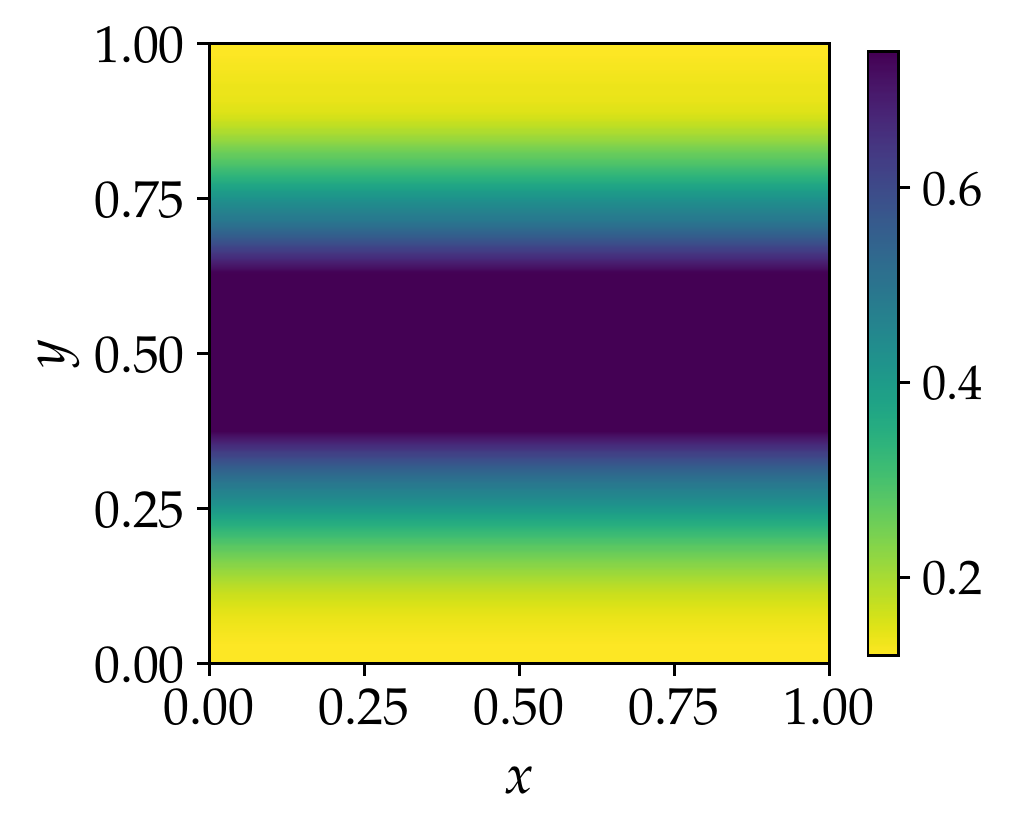}}
\hspace{0.2em}
\subfloat
{\includegraphics[width=0.35\textwidth]{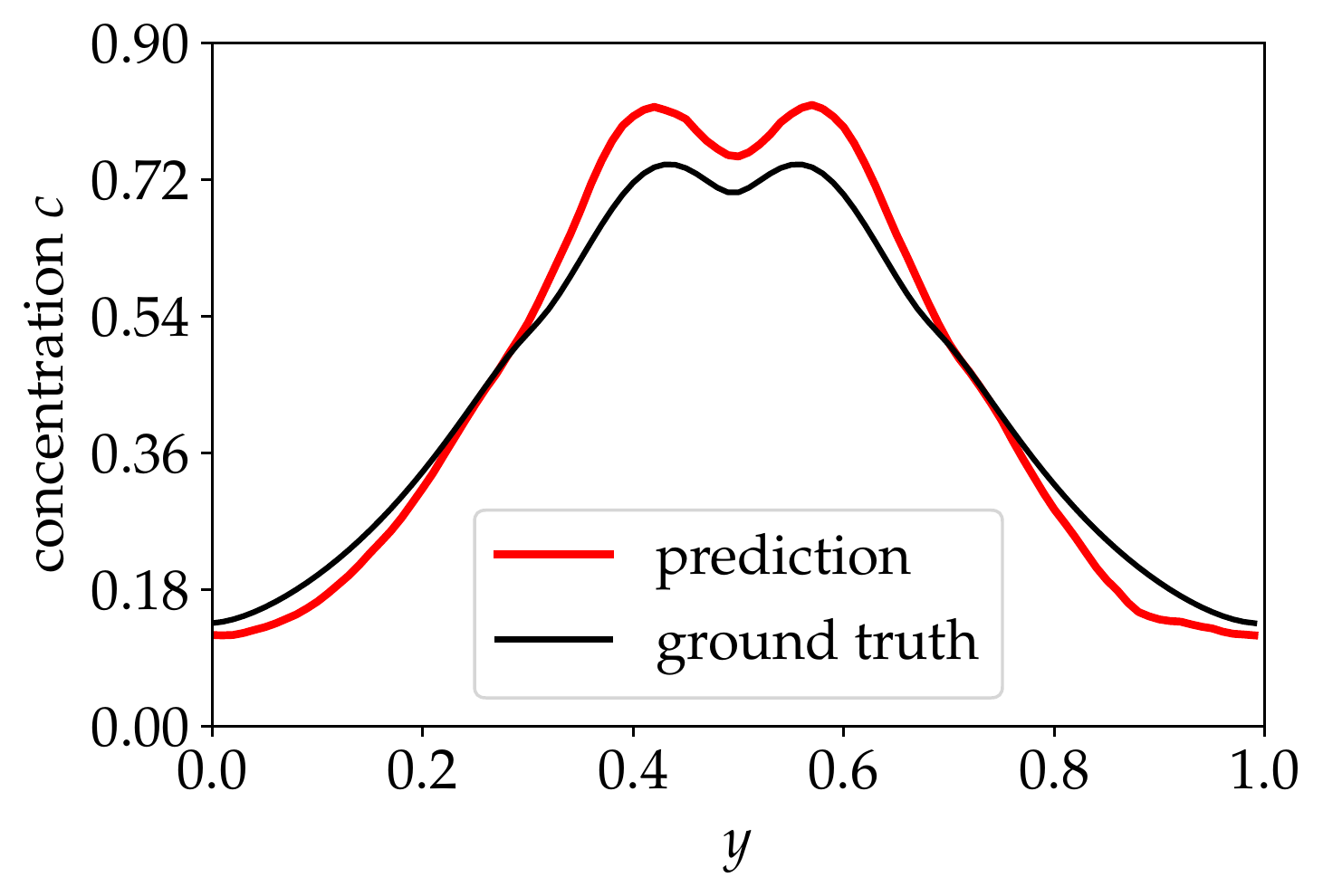}}
\quad
\addtocounter{subfigure}{-3}
\subfloat[ground truth of concentrations]
{\includegraphics[width=0.30\textwidth]{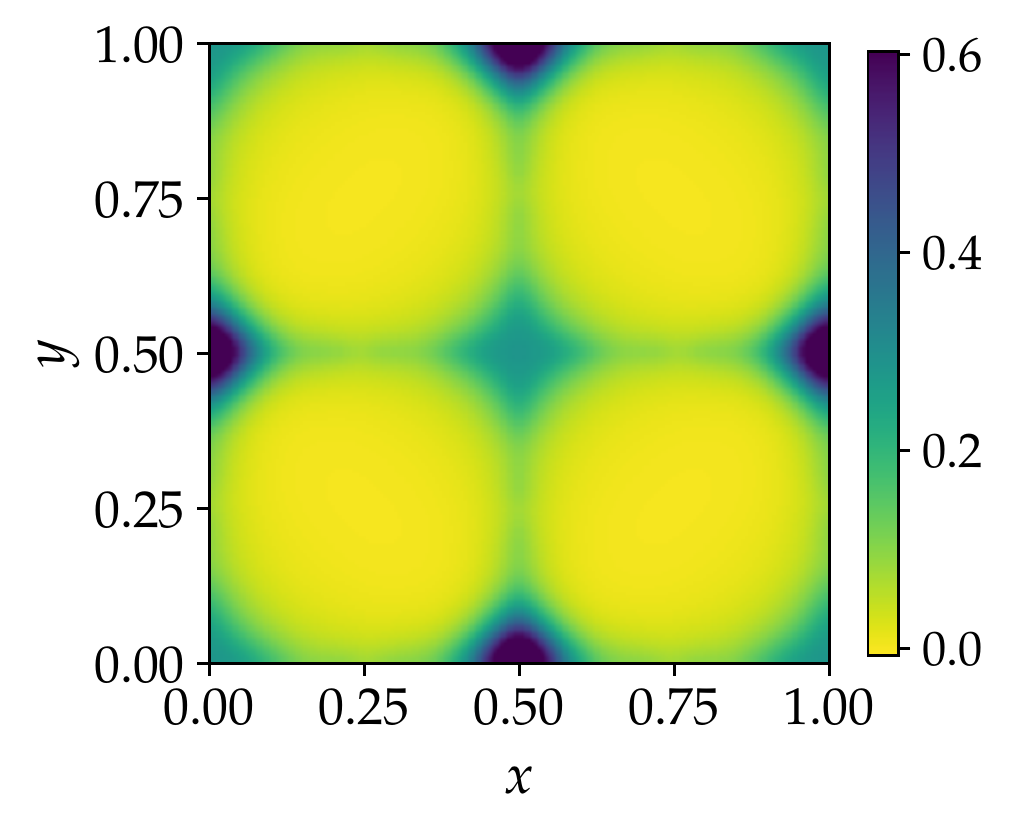}}
\hspace{0.2em}
\subfloat[predicted concentrations]
{\includegraphics[width=0.30\textwidth]{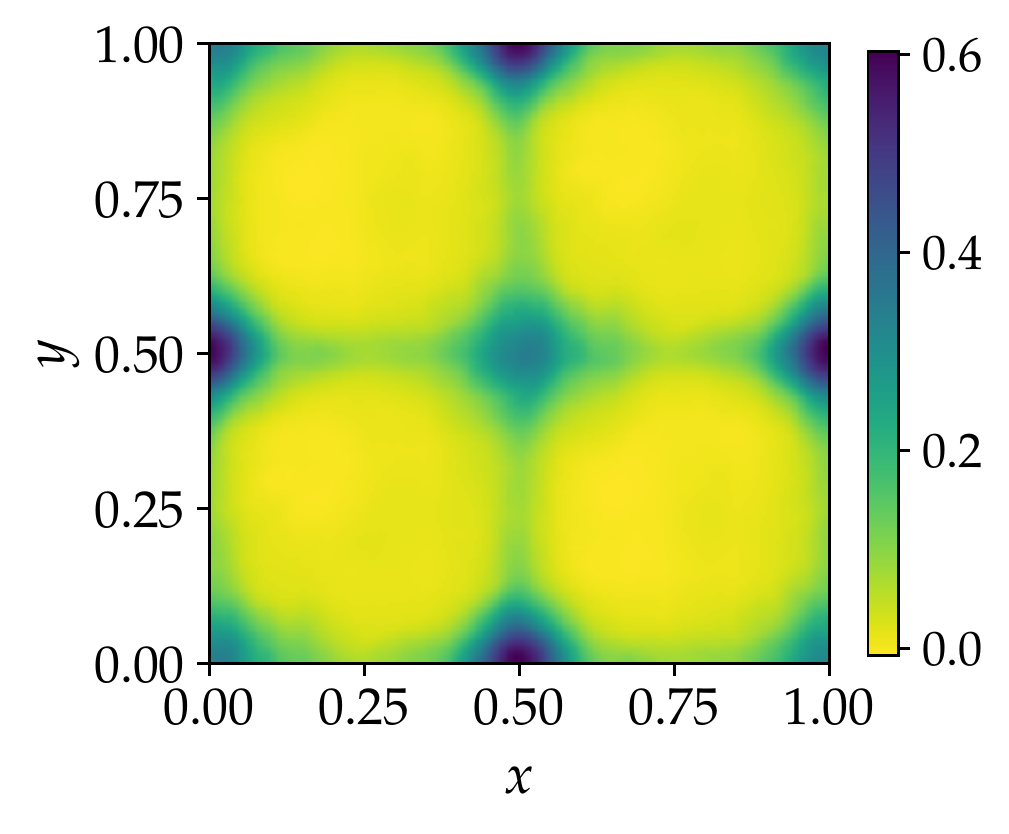}}
\hspace{0.2em}
\subfloat[predict concentration $c$ at $x=0.5$]
{\includegraphics[width=0.35\textwidth]{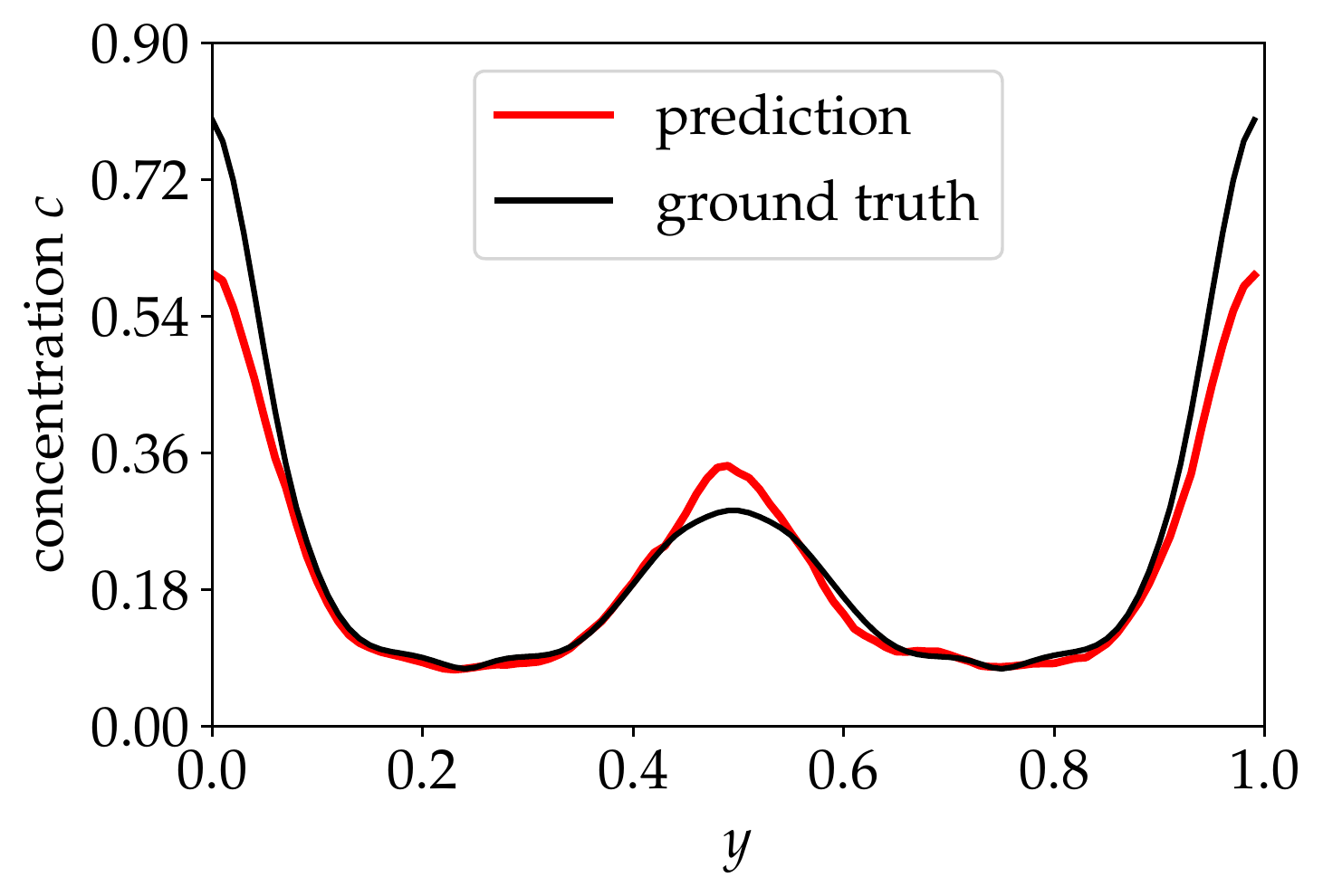}}
  \caption{
  Comparison of ground truth of the concentration fields (left column) and the corresponding predictions for flows in \emph{Case III} with conditions \emph{outside} the training range, demonstrating the \textbf{extrapolation capability} of the trained models. The figure shows the predicted concentration contours (middle column) and the concentration on cross-sections (right column).
  The top row is for the plane Poiseuille flow and the bottom row is for the Taylor--Green vortex.
  \label{fig:c-compare-case3-extrapolation}
  }
\end{figure}

The prediction error plots of plane Poiseuille flows and Taylor--Green vortexes (omitted here for brevity; see \ref{sec:app-vp}) show very well agreement between the predicted tracer concentration $c$ and the ground truth. The normalized prediction errors for the Poiseuille flows and the Taylor--Green vortexes are $1.54\%$ and $2.69\%$, respectively.
A comparison of the predicted tracer concentration field and the ground truth for a representative case is shown in Fig.~\ref{fig:c-compare-case3}.
For both flows, the predicted concentration field is similar to its corresponding ground truth, although there are slight differences seen: for plane Poiseuille flows, the region with low concentrations of prediction is larger than that of the ground truth; for Taylor--Green vortexes, there are minor fluctuations in prediction.
The overall good performance in predicting the concentration field showcases the ability of the proposed network architecture in closing the unknown nonlinear PDE with machine learning-based nonlocal submodels.

The aforementioned predictions in \emph{Case III} are both interpolations, i.e., the training and testing data have the same distributions. We also test the two trained models on extrapolation flows for both plane Poiseuille flows and Taylor--Green vortexes, respectively, to verify their generalizability. For plane Poiseuille flows, we change the velocity at the centerline from $u|_{y=0.5} = 1.0$ to $u|_{y=0.5} = 1.35$ and the pressure gradient parameter $\alpha$ to be in the interval between 2.4 and 5.4 such that the velocities at the top and bottom of the domain are still in the same range as that in the training data of \emph{Case III}; for Taylor--Green vortexes, the maximum velocities $U_0$ are changed to be in the range $[1, 1.35]$ instead of $[0, 1]$ in the training data. For each flow, we choose 5 different flow cases as the testing flow cases. 
The prediction error plots of plane Poiseuille flows and Taylor--Green vortexes (omitted here for brevity; see \ref{sec:app-vp}) show that the predicted tracer concentration $c$ and the ground truth are not in perfect agreement as the interpolation cases but still close to each other. The normalized prediction errors for the Poiseuille flows and the Taylor--Green vortexes are $5.19\%$ and $11.01\%$, respectively. A comparison of the predicted tracer concentration field and the ground truth for a representative extrapolation case is shown in Fig.~\ref{fig:c-compare-case3-extrapolation}. For both flows, we can still see a similar pattern of concentrations between the prediction and ground truth, albeit the distinctions near the boundary and the centerline.

\section{Conclusion}
\label{sec:conclude}
Many constitutive relations in computational mechanics, particularly those involving fluid flows and transport phenomena, are formulated as partial differential equations (PDEs) that describe the nonlocal mapping from primary variables to closure variables. These PDEs are often heuristically motivated and contain many submodels, which pose challenges for calibration or even for obtaining necessary data for their development. In this work we propose using neural networks to represent the region-to-point mappings in nonlocal constitutive relations. Such a representation leads to a constitutive neural network with a simple structure, which is inspired by the solution to linear convection-diffusion-reaction equations and is much easier to train than traditional models. We verify that the proposed network can learn the Green's kernel from the data generated by underlying linear PDEs. We further demonstrate that the data can be used to learn the submodels of a wide range of complexities without explicitly using the data from the latter level. Finally, we show the model has good predictive capability for general constitutive relations in data generated from nonlinear PDEs.

Despite the preliminary successes demonstrated here, we point out several directions of future study following the present work.
First and foremost, the performance of the learned closure model in solving the primary PDEs (e.g., the RANS momentum equation in turbulence modeling to solve for mean velocities and pressure) should be systematically examined in future work.
To this end, how to integrate the learned constitutive relation with other boundary conditions when their effects are nonnegligible is certainly worthy of future exploration.
Moreover, we assume a regular mesh to present the main idea of our data-driven approach to nonlocal constitutive modeling. Future work concerns extending the approach to unstructured meshes to facilitate the usage on more practical data coming from large-scale industrial CFD simulations and experiments.
Our follow-on work~\cite{zhou2021frame} has shown a lot of promise in this direction.
Finally, more efficient and adaptive sampling of data should be studied to boost the learning of more complicated flow fields.
The methodology developed here is demonstrated on 2D problems only. It can be extended to 3D quite straightforwardly and its numerical performance should be validated as well.

\afterpage{\clearpage}

\appendix

\section{Analysis of region size in 1D transport equation}
\label{sec:appd_green}
\subsection{Analytic solution}
The steady-state convection-diffusion-reaction equation in 1D with constant coefficients is written as:
\begin{equation}
     - \nu \frac{\ud^2}{\ud x^2}c(x) + u\frac{\ud}{\ud x}c(x) + \zeta c(x) = P(x), \quad \quad \quad x \in [a,b]
    \label{eq:1d_ss}
\end{equation}
with boundary condition $c(a)=c(b)=0$.
Here $\nu,u,\zeta$ denote diffusion, flow velocity and dissipation constants, respectively. $\nu, \zeta$ are both positive, and without loss of generality, we assume $u\geq 0$. 
If we define the corresponding differential operator $\mathcal{L}$ as 
\begin{equation}
    \mathcal{L}c \coloneqq - \nu \frac{\ud^2 c}{\ud x^2} + u\frac{\ud c}{\ud x} + \zeta c,
\end{equation}
the associated Green's function $G(x;x')$, with a fixed $x'\in[a,b]$ satisfies
\begin{equation}
    \mathcal{L}G(x;x')=\delta(x-x'),\quad\quad x\in[a,b]
\end{equation}
with $G(a;x')=G(b;x')=0$. 
Here $\delta(x)$ denotes the Dirac delta function.
With the Green's function, the solution to \eqref{eq:1d_ss} can be written as 
\begin{equation}
    c(x)=\int_{a}^b G(x;x')P(x')\ud x',
\end{equation}
in the same form as the first equality of \eqref{eq:green-sol}.

\begin{figure}[!htb]
\centering
  \includegraphics[width=0.99\textwidth]{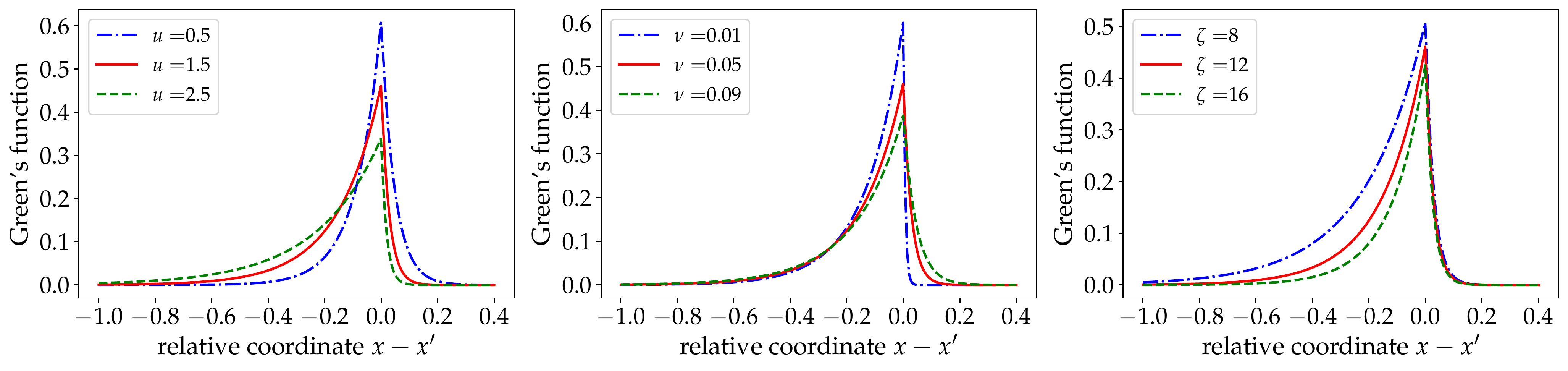}
  \caption{
  \label{fig:analytic_green}
  {Analytical Green's function under different $u, \nu, \zeta$. The default values are $u=1.5, \nu=0.05, \zeta=12$}.  
  }
\end{figure}

To eliminate the effect of boundary conditions and facilitate the analysis in the meantime, we further assume $a=-\infty, b=\infty$. In this case, it can be straightforwardly verified that
\begin{equation}
    G(x;x')=
    \begin{dcases}
    \frac{e^{\lambda_1(x-x')}}{\nu(\lambda_1-\lambda_2)}, & \quad (x\leq x')\\
    \frac{e^{\lambda_2(x-x')}}{\nu(\lambda_1-\lambda_2)}, & \quad (x>x')
    \end{dcases}
\end{equation}
where $\lambda_1,\lambda_2$ are two roots of the associated characteristic polynomial $-\nu y^2+uy+\zeta=0$:
\begin{equation}
    \lambda_1\coloneqq \frac{u+\sqrt{u^2+4\nu\zeta}}{2\nu} > 0 > \lambda_2 \coloneqq \frac{u-\sqrt{u^2+4\nu\zeta}}{2\nu}.
\end{equation}
The solution to \eqref{eq:1d_ss} becomes
\begin{equation}
\label{eq:1d_sol}
    c(x)=
    \int_{-\infty}^x \frac{e^{\lambda_2(x-x')}}{\nu(\lambda_1-\lambda_2)}P(x')\ud x' + 
    \int_{x}^{\infty}
    \frac{e^{\lambda_1(x-x')}}{\nu(\lambda_1-\lambda_2)}P(x')\ud x'.
\end{equation}
Notice that it is exactly the convolution form presented in Eq.~\eqref{eq:green-sol}, indicating the nonlocal dependence on the source. Specifically, recalling that we have assumed $u\geq0$, so we can interpret the two terms in Eq.~\eqref{eq:1d_sol} as the contribution from the upstream and downstream sources, respectively. Fig.~\ref{fig:analytic_green} illustrates the Green's function in terms of the relative coordinates under different choice of equation parameters. We can see that 
the influence region increases with larger diffusion coefficient $\nu$ and velocity magnitude $\|\mathbf{u}\|$ but decreases with larger dissipation coefficient $\zeta$. The quantitative relation between the influence region and these parameters will be derived below.

\begin{figure}[!htb]
\centering
  \includegraphics[width=0.5\textwidth]{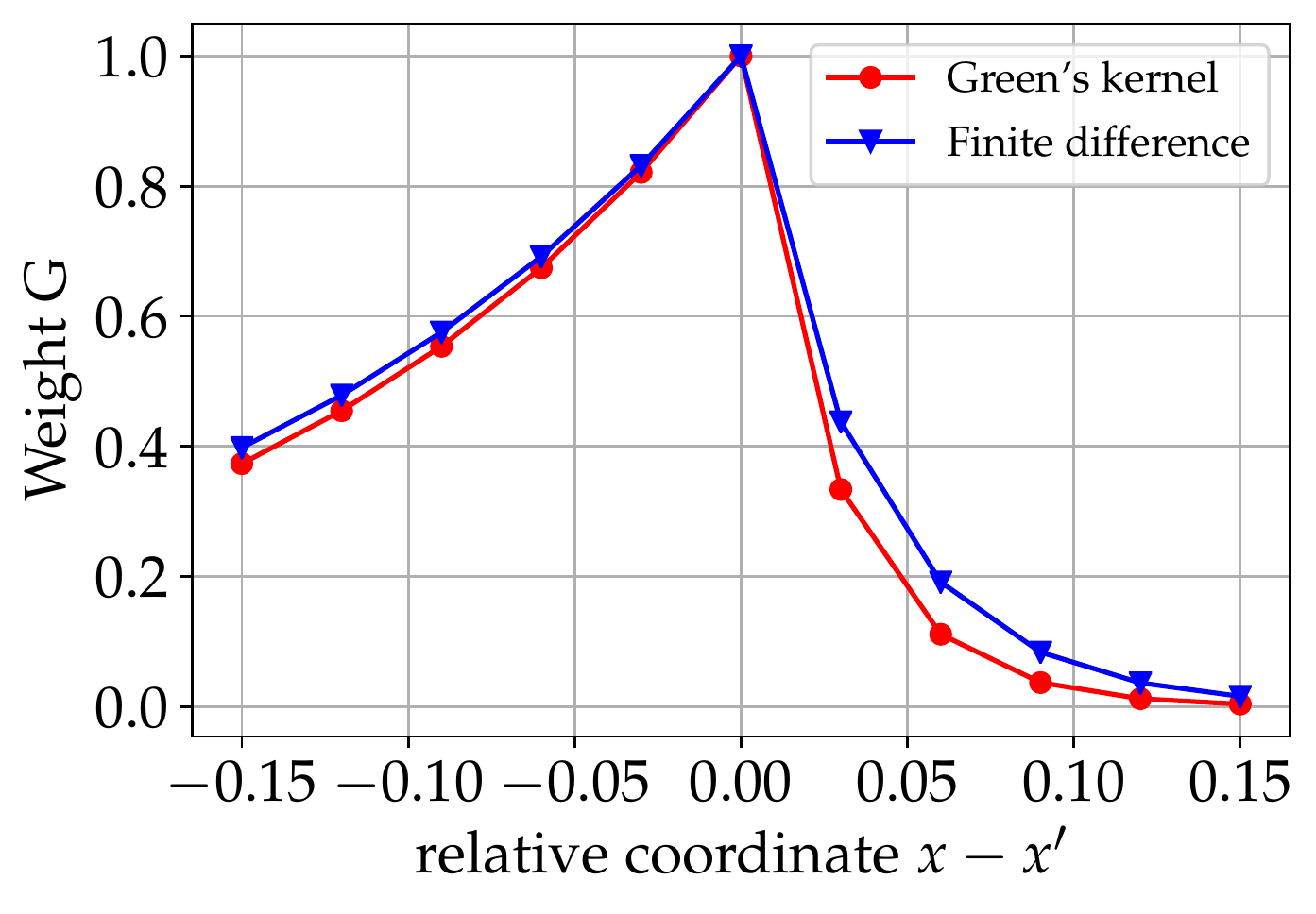}
  \caption{
  \label{fig:finite_difference}
  {The weights provided by the finite difference method approximates to the discrete value of the Green's function.}
  }
\end{figure}

In the discrete sense, the Green's function can also be recovered from the finite difference method.
Suppose we have a uniform mesh grid in 1D: $x_0<x_1<\cdots<x_N$ and denote the concentration and source at these grid points by $c_i$ and $P_i$, respectively.
The discretization (with upwind and central scheme for the convection and diffusion terms and zero boundary condition) of Eq.~\eqref{eq:1d_ss} leads to the following system of linear equations:
\begin{equation}
\begin{bmatrix}
\begin{array}{ccccc}
  a+2d+\zeta  &  -d  &      &       &     \\ 
 -(a+d)  &   a+2d+\zeta  & -d    &       &    \\ 
     &  -(a+d)  &  a+2d+\zeta    &  \ddots &     \\
     &      & \ddots  & \ddots & -d  \\
     &      &    &    -(a+d) & a+2d+\zeta
\end{array}
\end{bmatrix}
\begin{bmatrix}
c_1 \\
c_2\\
\vdots\\
c_{N-1}\\
c_N
\end{bmatrix}
=
\begin{bmatrix}
P_1 \\
P_2\\
\vdots\\
P_{N-1}\\
P_N
\end{bmatrix},
\label{eq:finite_difference}
\end{equation}
where coefficients $a = u/\Delta x$ and $d = \nu/\Delta x^2$ come from the discretization of the advection and the diffusion terms, respectively.
Denote the coefficient matrix in \eqref{eq:finite_difference} with , then the Green's kernel is closely related to $\mathsf{L}^{-1}$. In particular, the solution
\begin{align}
    c_i = \sum_{j=1}^{N}\mathsf{L}^{-1}_{ij}P_j,
\end{align}
indicates how the production terms at different location points contribute to the concentration through the weights~$\mathsf{L}^{-1}$. Indexing these weights according to the relative coordinates, we can see that they approximately equal to the discrete value of the Green's function, as shown in Fig.~\ref{fig:finite_difference}.

\subsection{Analysis of the size of influence region}
Equation~\eqref{eq:1d_sol} shows that the solution $c(x)$ depends on the source field $P$ in the whole domain, although the dependence has an exponential decay with distance from the point where the solution is sought. In practice, we would like to truncate the domain of influence to be finite for computation efficiency. If we denote the truncated size of the upstream and downstream region with $\ell^-$ and $\ell^+$, the approximation has the form
\begin{equation}
\label{eq:1d_sol_approx}
    c(x)=
    \int_{x-\ell^-}^x \frac{e^{\lambda_2(x-x')}}{D(\lambda_1-\lambda_2)}P(x')\ud x' + 
    \int_{x}^{x+\ell^+}
    \frac{e^{\lambda_1(x-x')}}{D(\lambda_1-\lambda_2)}P(x')\ud x'.
\end{equation}
To get more qualitative insight of the relationship between the size of the truncated region and the corresponding accuracy, we consider a simplified setting. Suppose the production term $P(x)$ varies a little such that it is roughly a constant. In this situation, in order to achieve a relative error tolerance $\epsilon$ for the contribution in the upstream region (the first term in Eq.~\eqref{eq:1d_sol_approx}), we require
\begin{equation}
    \frac{\displaystyle{
    \int_{x-\ell^-}^{x}e^{\lambda_2(x-x')}\ud x'}}
    {\displaystyle{
    \int_{-\infty}^{x}e^{\lambda_2(x-x')}\ud x'}}
    \simeq 1-\epsilon,
\end{equation}
which implies
\begin{equation}
    \ell^- \simeq \bigg|\frac{\log\epsilon}{\lambda_2}\bigg|.
\end{equation}
Similarly, the analysis of the downstream region gives us
\begin{equation}
    \ell^+ \simeq \bigg|\frac{\log\epsilon}{\lambda_1
    }\bigg|.
\end{equation}
Noticing that $|\lambda_2|<|\lambda_1|$, we see the influence region of the upstream is larger than that of the downstream. However, in practical 2D and 3D applications, we may not know the direction of flow in advance, so for simplicity we shall choose a conservative patch size $2\ell^-$ to guarantee the small error. Furthermore, in the 2D case, we can assume the truncated domain is square such that we can reuse the result in the 1D case, in which $u$ is replaced by the magnitude $\|\mathbf{u}\|$ as a conservative choice. Taking the above simplifications into consideration, we arrive at the final half-length of the patch
\begin{equation}
    \ell = \bigg|\frac{2\nu\log\epsilon}{\sqrt{\|\mathbf{u}\|^2+4\nu\zeta}-\|\mathbf{u}\|}\bigg|.
\end{equation}

\section{Details of network architecture and training parameters}
\label{sec:appd_parameter}

The details of the employed network architecture are listed in Table~\ref{tab:nn-details}. The entire neural network consists of two sub-networks: (1) a convolutional neural network and (2) a fully connected network. 

The convolutional neural network learns a nonlinear region-to-region mapping $\region{\mathbf{u}}_{i, j} \mapsto \region{G}_{i, j}$. It consists of six convolutional layers without pooling layers. The number of channels in each layer is chosen to be the powers of two empirically. The convolutional kernels in all layers are of size $3\times3$, which proves to be very effective in image recognition~\cite{simonyan2014very,he2016deep}. The amount of zero padding is set as one to ensure the output and input of each layer is of the same size. Other padding modes including reflect and replicate padding modes are also tested when learning the Green's kernel in \emph{Cass II}, with only slight differences. Batch normalization~\cite{ioffe2015batch} is used in each layer except the last layer to stabilize the learning process and accelerate the training. 

The fully connected network learns a pointwise mapping $\mathbf{u}_{i, j} \mapsto P_{i, j}$, applicable for every stencil point in the influence region $\region{\mathbf{u}}_{i, j}$. In fact, this mapping consists of two steps: $\mathbf{u}_{i, j} \mapsto (\boldsymbol{S}_{i, j}, \mathbf{\Omega}_{i, j})$ and $(\boldsymbol{S}_{i, j}, \mathbf{\Omega}_{i, j}) \mapsto P_{i, j}$. The first step calculates the physical properties (strain rate and rotation rate) of the flow without any trainable parameters. The second step corresponds to this fully connected network. We set the number of layers as five by default. When the source function is $\mathcal{P} = \mathcal{P}(c; \mathbf{u})$ (\emph{Case III}) or $\mathcal{P} = h(S)$ (\ref{sec:appd_learn-h}), we add one more layer ($32\times32$) between layer 1 and 2 to ensure enough complexity of the fully connected network. We select the rectified linear activation function (ReLU) as the activation function because it is easier to train and often achieves good performance.

We adopt the Adam optimizer~\cite{kingma2015adam} to optimize the parameters in neural networks. In all the cases the training takes 500 epochs with constant learning rate 0.001. The only exception is \emph{Case I} where the random production generated by the Gaussian process makes it harder to train the neural network.
In \emph{Case I}, the training takes 2000 epochs, and the learning rate is 0.001 at the beginning and then multiplied by 0.7 every 600 epochs.

\begin{table}[htbp]
\caption{Detailed architectures of the convolutional neural network ($\region{\mathbf{u}}_{i, j} \mapsto \region{G}_{i, j}$) and the fully connected networks ($\mathbf{u}_{i, j} \mapsto P_{i, j}$). Acronyms used: C, number of filters (channels); FS, filter size; S, stride; Pad, the amount of padding; BN, batch normalization; N, neurons.}
\centering
\begin{tabular}{P{3cm} P{4.7cm} P{4.0cm}}
\toprule[1pt]
 & Convolutional Neural Network  ($\region{\mathbf{u}}_{i, j} \mapsto \region{G}_{i, j}$) & Fully Connected Network ($\mathbf{u}_{i, j} \mapsto P_{i, j}$)\\
\hline \hline
Layer 1 & C: 32, FS: $3\times3$, S: 1, Pad: 1, BN, ReLU & N : $2\times32$, ReLU  \\
\hline
Layer 2 & C: 32, FS: $3\times3$, S: 1, Pad: 1, BN, ReLU & N : $32\times16$, ReLU  \\
\hline
Layer 3 & C: 16, FS: $3\times3$, S: 1, Pad: 1, BN, ReLU  & N : $16\times8$, ReLU  \\
\hline
Layer 4 & C: 8, FS: $3\times3$, S: 1, Pad: 1, BN, ReLU  & N : $8\times4$, ReLU  \\
\hline
Layer 5 & C: 4, FS: $3\times3$, S: 1, Pad: 1, BN, ReLU  & N : $4\times1$  \\
\hline
Layer 6 & C: 1, FS: $3\times3$, S: 1, Pad: 1  &   \\
\bottomrule[1pt]
\end{tabular}
\label{tab:nn-details}
\end{table}

\section{Learning of a more challenging submodel function}
\label{sec:appd_learn-h}

Learning submodels (e.g., $\mathcal{P(S)}$) from closure variable data (e.g., $\{\region{\mathbf{u}}, c\}$) is of great practical importance as closure models in computational physics often have separate embedded models to account for different unresolved physical processes. While we have demonstrated that the proposed framework has successfully learned univariate production submodel $\mathcal{P}(S) = 4\sin{(2\pi S)} + 6 {S}^2 + 5 e^{S}$, we further evaluate our framework in learning a more challenging production model. Specifically, in this assessment the submodel $\mathcal{P}(S)$ has three distinct regimes with steep gradient in the middle regime as shown in Eq.~\eqref{eq:hc} (and depicted in Fig.~\ref{fig:learning-Pfunction1}):
\begin{equation}
\label{eq:hs}
    \mathcal{P}(S) = h(S)  = 
    \begin{cases}
    e^{{-3S(1-S)}} & S \in [0, 0.5] \\
    e^{{-10(S-0.5)}} + h_{12} & S \in [0.5, 0.55] \\
    0.8 \ln S + h_{23} & S \in [0.55, 1] 
    \end{cases}
\end{equation}
with constants $h_{12}$ and $h_{23}$ chosen to be the same values as in Eq.~\eqref{eq:hc} to ensure the continuity of the function.
The function consists of a positive gradient section $S \in [0, 0.5]$ and a negative gradient section $S \in [0.55, 1]$, which are connected by a sharp transition section $S \in [0.5, 0.55]$. 
The tests are performed on plane Poiseuille flows with the parameters settings as those in \emph{Case II} (Section \ref{sec:res-P}).

\begin{figure}[!htb]
\centering
{\includegraphics[width=0.6\textwidth]{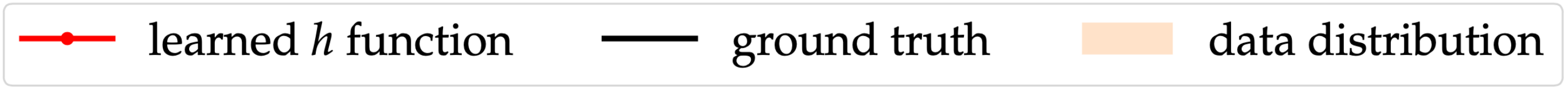}\vspace{-8pt}}\\
\subfloat[$h_1$]
{\includegraphics[width=0.33\textwidth]{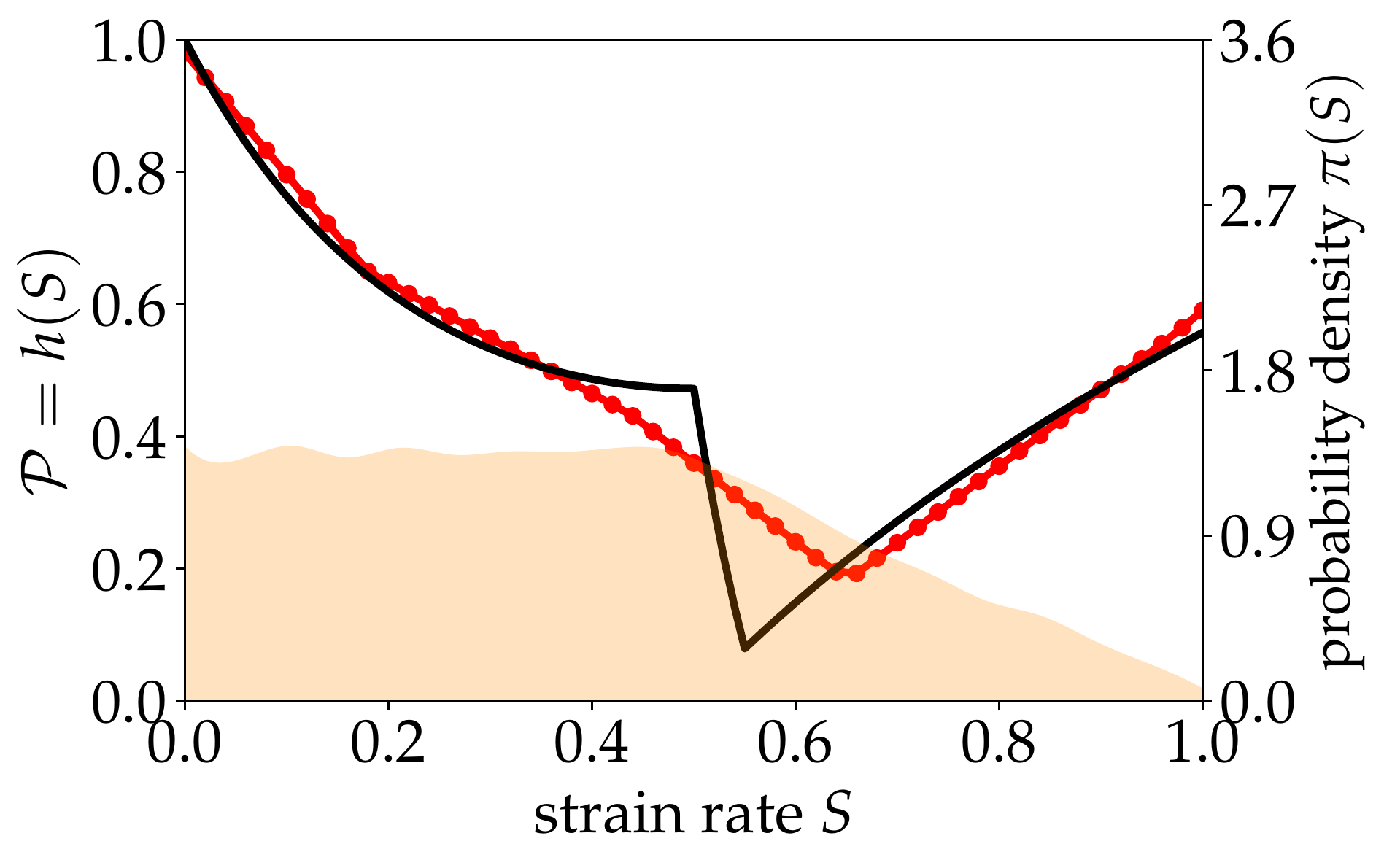}}
\subfloat[$h_2$]
{\includegraphics[width=0.33\textwidth]{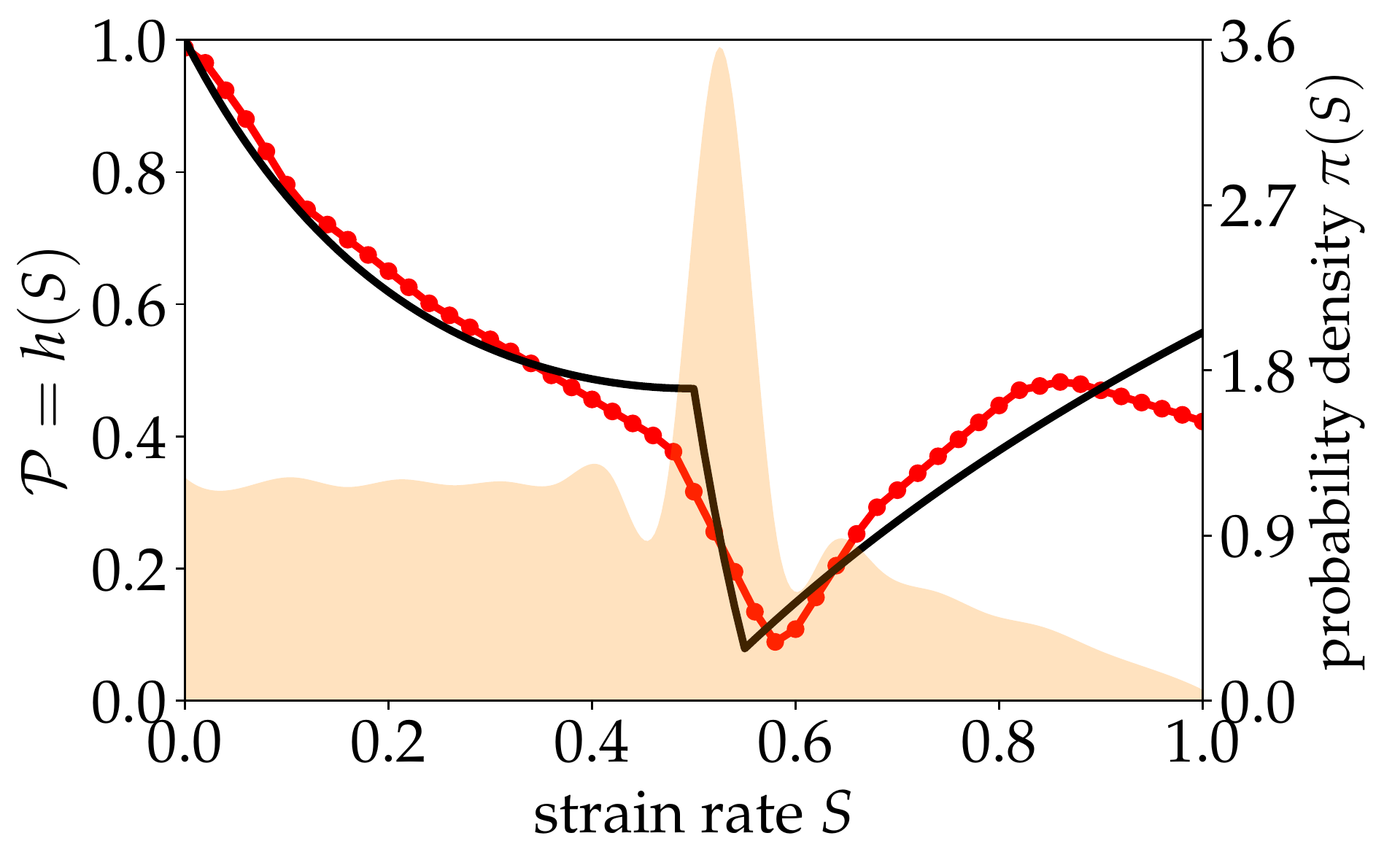}}
\subfloat[$h_3$]
{\includegraphics[width=0.33\textwidth]{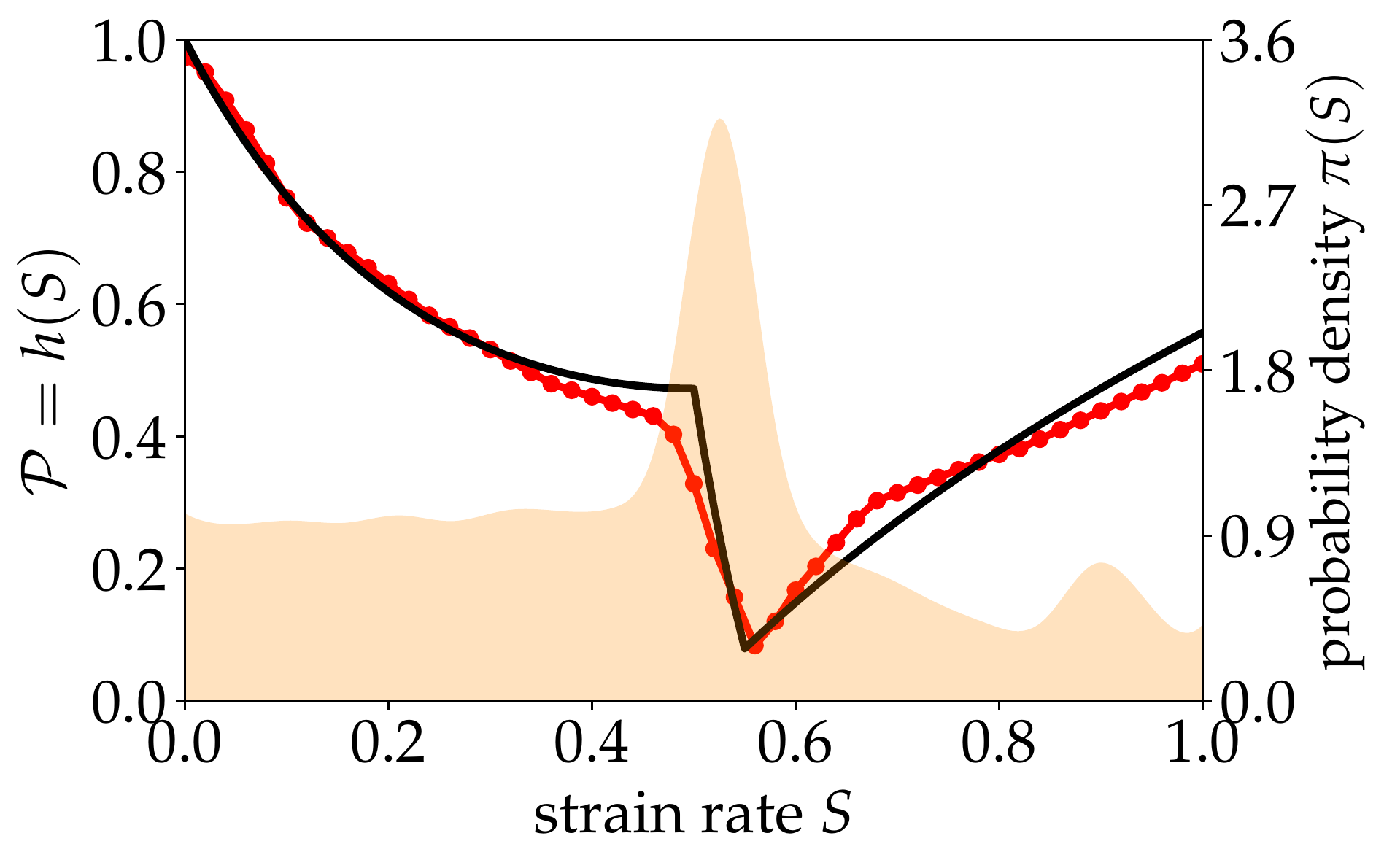}}
  \caption{
  \label{fig:learning-hs}
  Comparison of learned submodels with the ground truth under a specified production term $\mathcal{P}(\mathbf{u}) = h(S)$ and three different training data distributions, showing data distribution has a great influence on the learning performance of the neural network.
  }
\end{figure}

The overall trend of the submodel function $h(S)$ can be learned correctly, as shown in Fig.~\ref{fig:learning-hs}a. However, the learned model failed to capture the sharp transition in the region $S \in [0.5, 0.55]$. Rather, the learned function has a relatively mild transition connecting the descending and ascending regimes. This is caused by the distribution of the training data used to learn the function. 
Specifically, we generated the training data under the plane Poiseuille flows, in which the velocity along the central axis is fixed and the pressure gradient parameter $\alpha$ is sampled from a uniform distribution on the interval $[1,4]$. Therefore, the distribution of velocity gradient (and thus strain rate magnitude $S$) is not explicitly controlled but implicitly determined by the flow field. The probability distribution of $S$ is presented as overlaying shades in Fig.~\ref{fig:learning-hs}. It can be seen that the percentage of the data in the transition regime of $h(S)$ is rather small (31,380 or 6.3\% of the 500,000 data points in total) in this regime. As a result, they are not sufficiently weighted in the training process for learning the steep gradient in the middle regime.

The learning of submodels can be improved by augmenting the data in the region of interest (the middle regime $S \in [0.5, 0.55]$ in this example). 
In some applications this is well justified. For example, consider the submodel $C_d(Re)$ of drag coefficient $C_d$ on a spherical particle as a function of the particle Reynolds number $Re$. The range of $Re$ in the transition regime (which connects the laminar regime at low $Re$ and the turbulent regime at high $Re$) is known \textit{a priori} from experimental evidence and the physics of transition. Hence, the augmentation of data in this regime can be performed. In fact, such augmentation shares a similar spirit with stratified sampling and importance sampling, which are commonly used techniques in statistics and machine learning~\cite{zhu2016gradient,wang2018optimal,ting2018optimal,katharopoulos2018not,kuchnik2018efficient}.
After we augment the training data by increasing the number of points in the transition regime (which could have been achieved equally well by increasing sample weights in this region instead), the learned function agrees much better in the transition regime, as shown in Fig.~\ref{fig:learning-hs}b. Unfortunately, the agreement in the high strain-rate regime ($S \in [0.7, 1.0]$) deteriorated as compared to Fig.~\ref{fig:learning-hs}a, as the relative weights in this regime are too small. When we increase the percentage of samples in this regime in the training data, the submodel learning can be improved. The best result in this experiment in terms of the learned submodel is shown in Fig.~\ref{fig:learning-hs}c, which is obtained by the data distribution shown in the corresponding background shade. 

In summary, the results here demonstrate that the learning of the submodel has achieved overall success, but the performance is affected by the data distribution of the submodel variable $S$. If the overall behavior of the submodel is known, we can use such prior knowledge to guide the sampling in the training data and thus to improve the learning of the submodel. This idea is consistent with the well-known importance sampling techniques in statistics. Otherwise, some adaptive sampling techniques~\cite{zhu2016gradient,wang2018optimal,ting2018optimal,katharopoulos2018not,kuchnik2018efficient} may also help to improve the learning performance.

\section{Additional plots in results}
\label{sec:app-vp}

\begin{figure}[!htb]
\centering
\subfloat[plane Poiseuille flows]
{\includegraphics[width=0.4\textwidth]{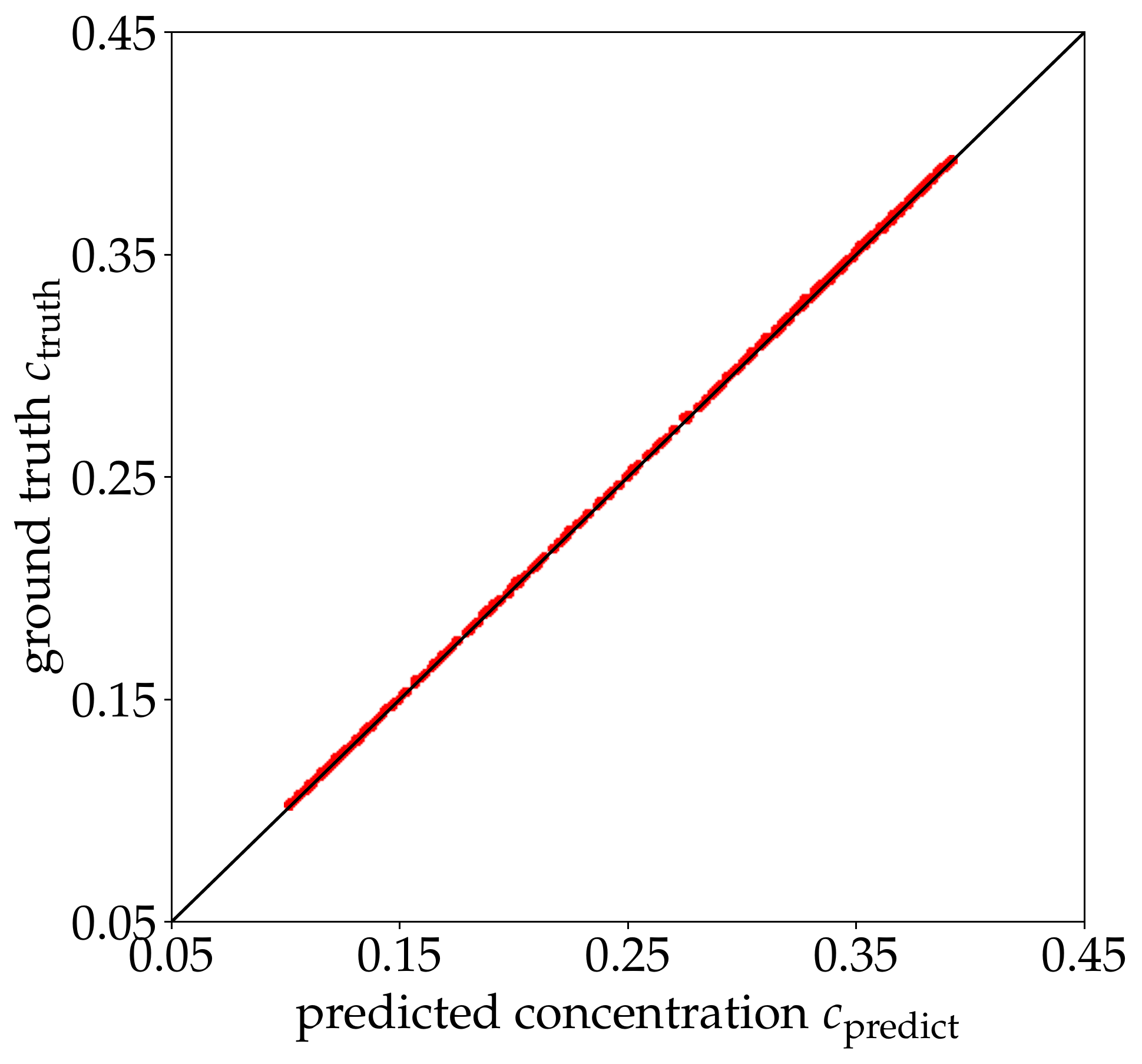}}
\hspace{0.5em}
\subfloat[Taylor--Green vortexes]
{\includegraphics[width=0.4\textwidth]{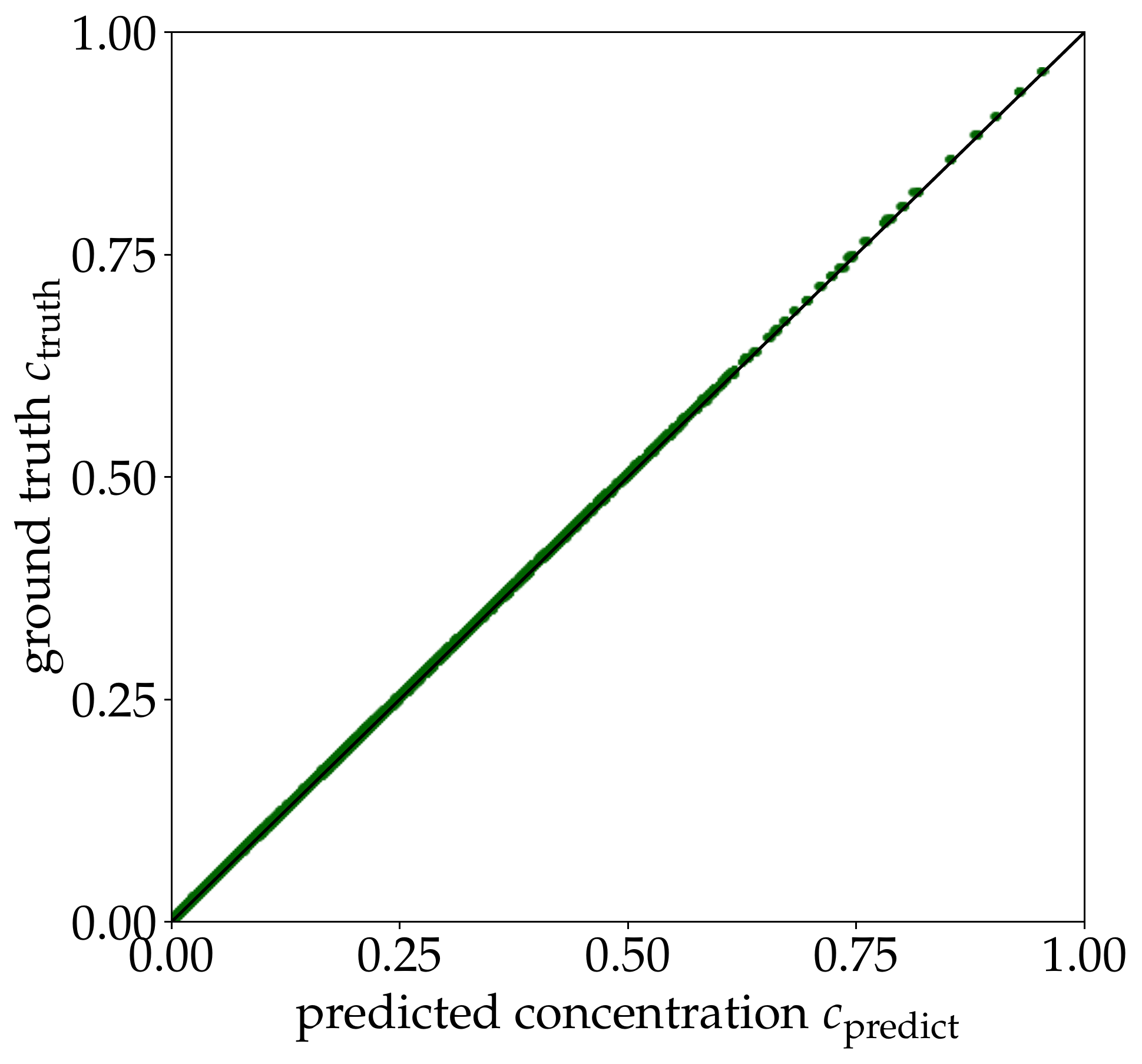}}
  \caption{
  \label{fig:vp-case2}
  Prediction error plots for \emph{Case IIb} (linear PDE with specified source field $P(S,\Omega)$), showing agreement between the predicted tracer concentrations and their ground truths on five different test flows of plane Poiseuille flows and Taylor--Green vortexes.)
  }
\end{figure}

\begin{figure}[!htb]
\centering
\subfloat[plane Poiseuille flows]
{\includegraphics[width=0.4\textwidth]{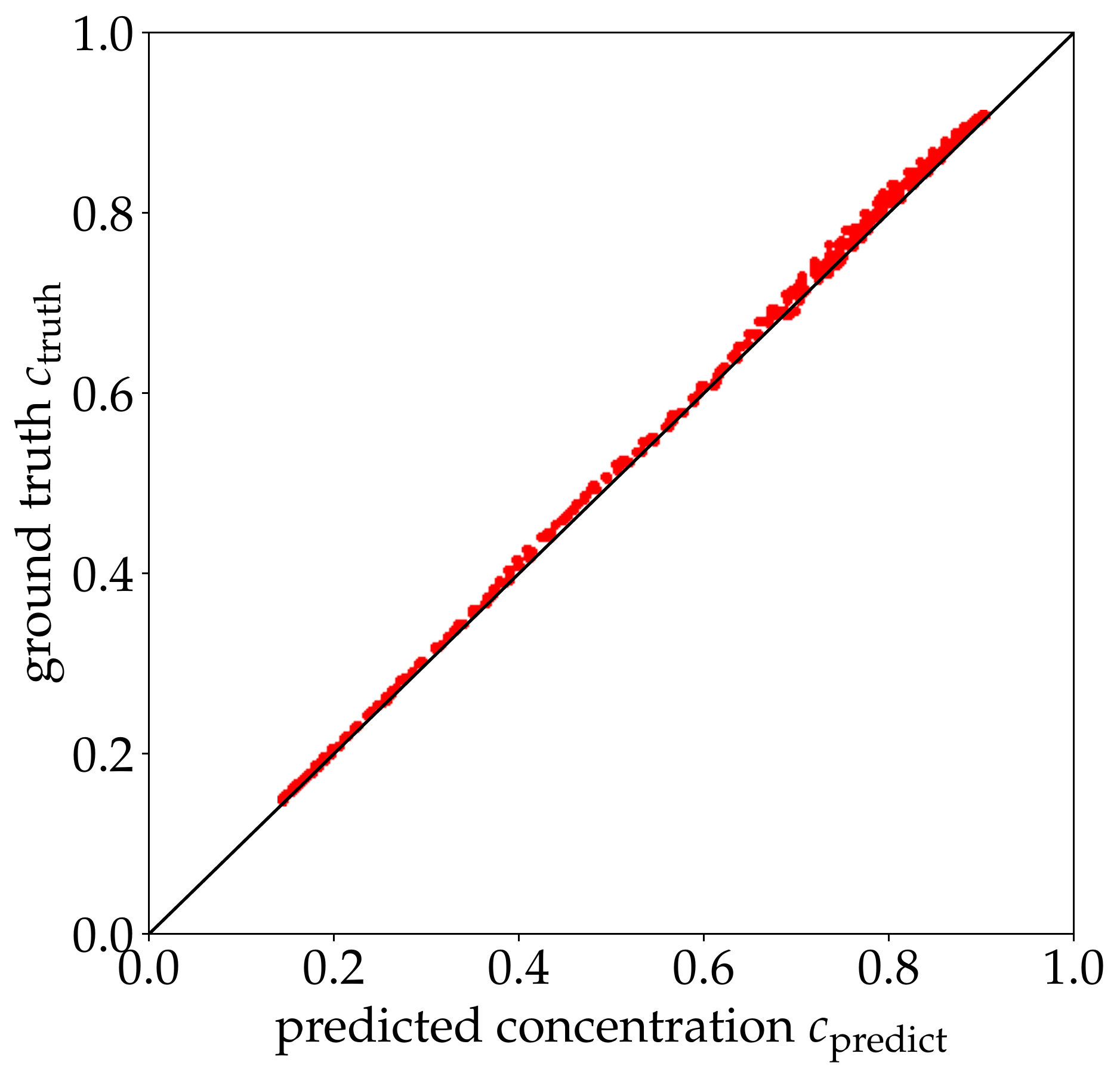}}
\hspace{0.5em}
\subfloat[Taylor--Green vortexes]
{\includegraphics[width=0.4\textwidth]{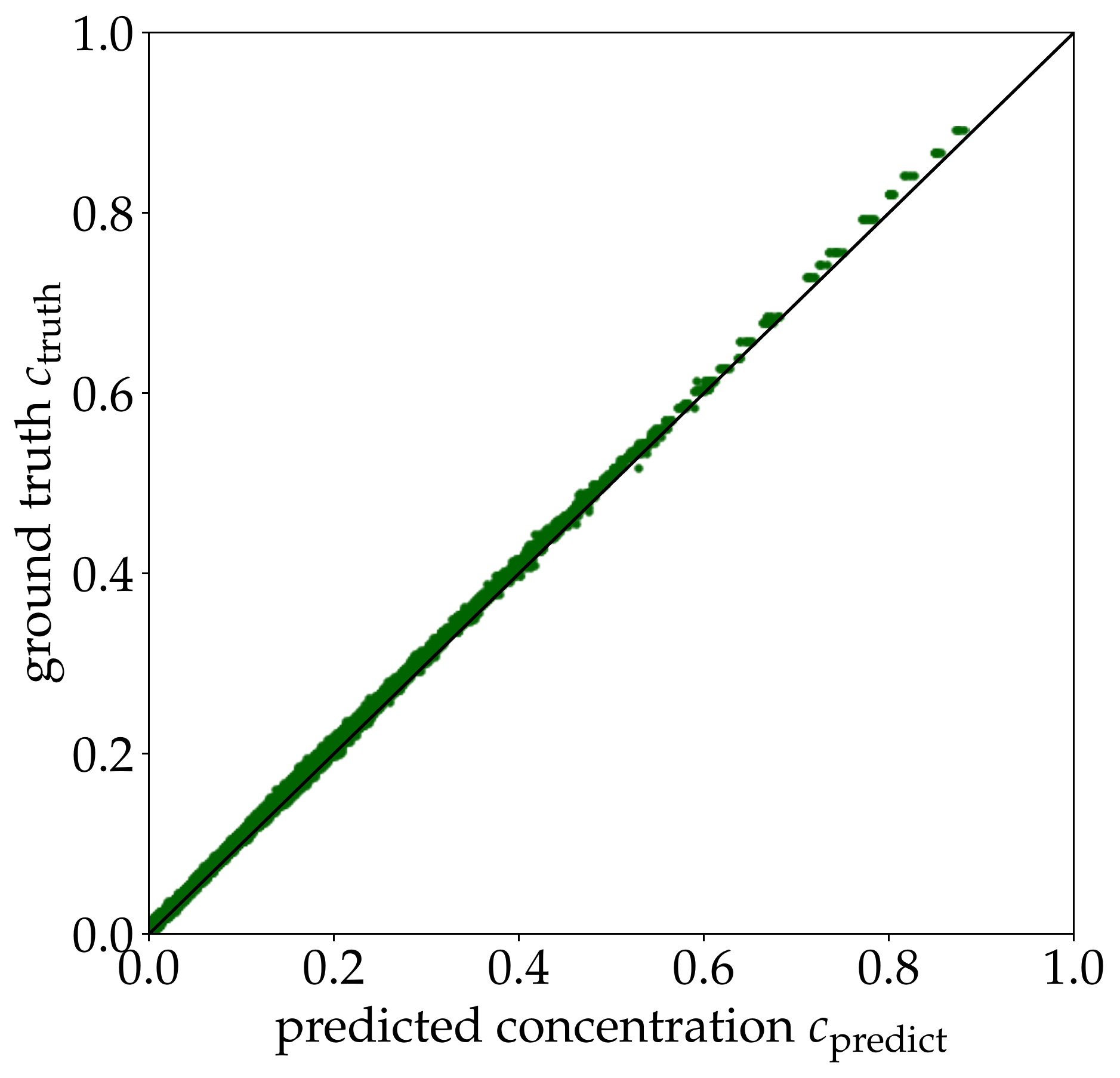}}
  \caption{
  \label{fig:vp-case3}
  Prediction error plots plot for \emph{Case III} (nonlinear PDE with $\mathcal{P}(c, \mathbf{u}) = f(S)g(\Omega)(K\,h(c)+b)$), showing agreement between the predicted tracer concentrations and their ground truths on five different test flows of plane Poiseuille flows and Taylor--Green vortexes.)
  }
\end{figure}

The prediction error plots for \textit{Cases IIb} and \textit{III} are presented in Figs.~\ref{fig:vp-case2} and ~\ref{fig:vp-case3}, respectively.

\begin{figure}[!htb]
\centering
\subfloat[plane Poiseuille flows]
{\includegraphics[width=0.4\textwidth]{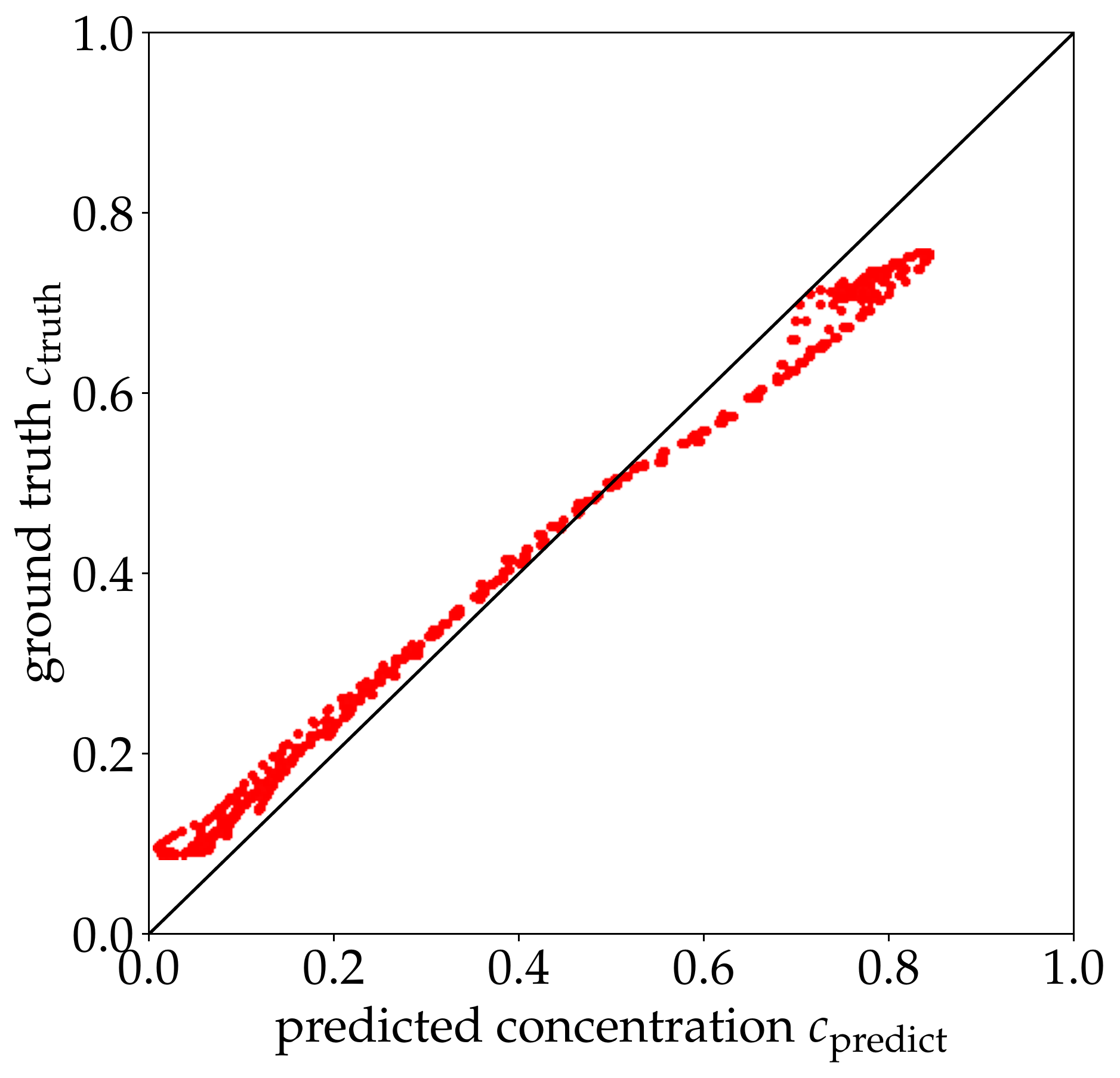}}
\hspace{0.5em}
\subfloat[Taylor--Green vortexes]
{\includegraphics[width=0.4\textwidth]{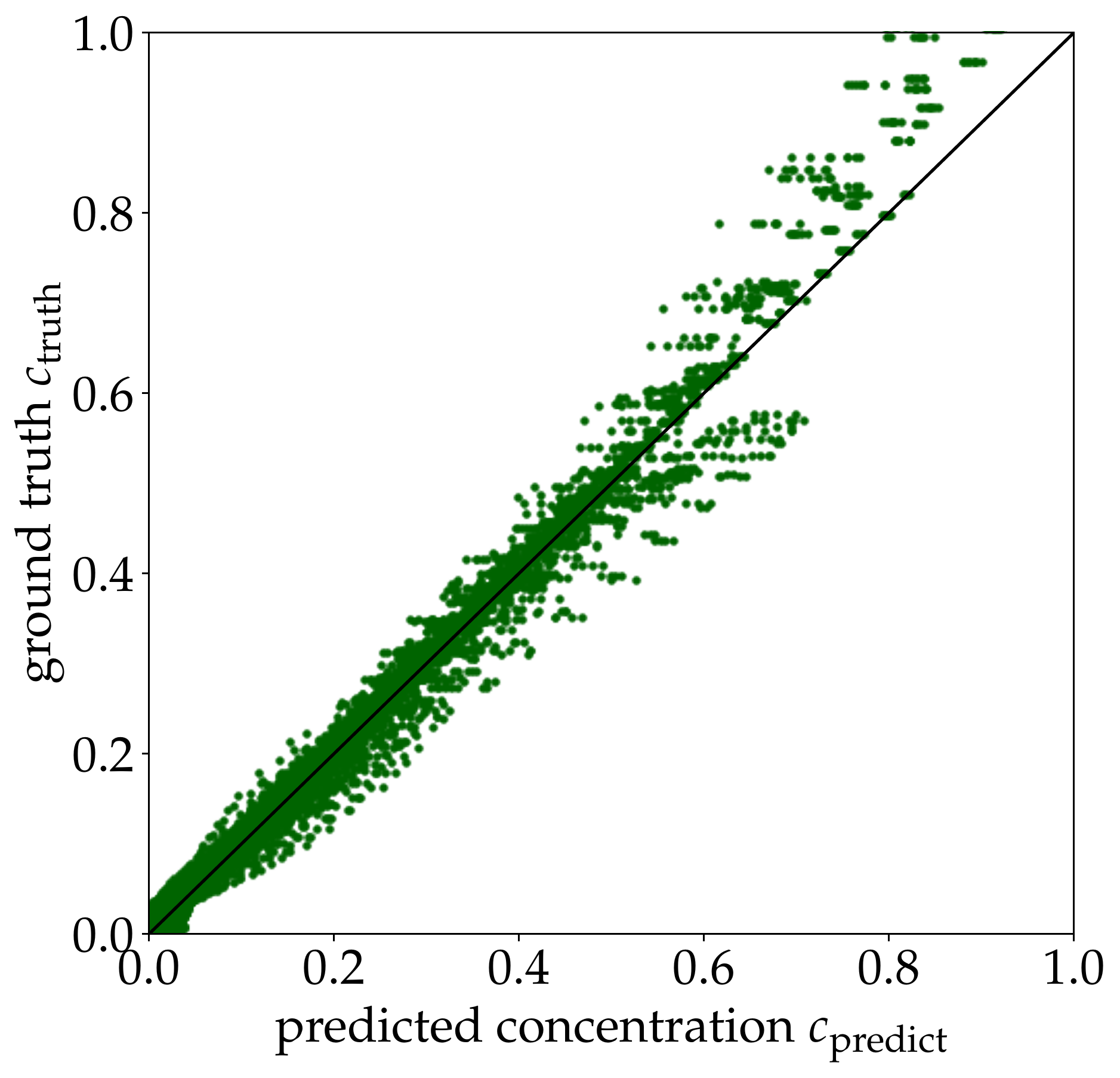}}
  \caption{
  \label{fig:vp-case3-extrapolation}
  Prediction error plots for \emph{Case III} (nonlinear PDE with $\mathcal{P}(c, \mathbf{u}) = f(S)g(\Omega)(K\,h(c)+b)$), showing agreement between the predicted tracer concentrations and their ground truths on five different test flows of plane Poiseuille flows and Taylor--Green vortexes.)
  }
\end{figure}

\section*{Acknowledgments}
The computational resources used for this project were provided by the Advanced Research Computing (ARC) of Virginia Tech, which is gratefully acknowledged. HX would like to thank Dr.~Jin-Long Wu of Caltech for his valuable discussions in the conceptual design of this work when Dr.~Wu was a Ph.D. student at Virginia Tech. The authors appreciate the contributions of code by Muhammad Irfan Zafar during the early stage of this research.

\thenomenclature
\nomgroup{A}
\item [{$\nabla$}]\begingroup spatial derivatives
\item [{$\bar{\boxed{}}$}]\begingroup mean
\item [{$\hat{\boxed{}}$}]\begingroup prediction
\item [{$\star$}]\begingroup convolution
\item [{$|{\cdot}|$}]\begingroup absolute value
\item [{$\|{\cdot}\|$}]\begingroup magnitude
\item [{$\region{\cdot}$}]\begingroup variables in a region (patch)
\item [{$\boxed{}_{\rm{max}}$}]\begingroup maximum
\item [{$\boxed{}^\top$}]\begingroup matrix transpose
\item [{$\boxed{}^{-1}$}]\begingroup matrix inverse
\item [{$\boxed{}^{+}$}]\begingroup downstream
\item [{$\boxed{}^{-}$}]\begingroup upstream
\nomgroup{B}
\item [{$\mathbf{u}$}]\begingroup mean velocity field
\item [{$\mathbf{v}$}]\begingroup velocity
\item [{$\mathbf{x}$}]\begingroup spatial coordinate
\item [{$c$}]\begingroup closure variable (e.g. concentration)
\item [{C}]\begingroup number of channels
\item [{$D$}]\begingroup selected region size
\item [{$e$}]\begingroup Euler's number
\item [{$\epsilon$}]\begingroup error tolerance
\item [{$\mathcal{E}$}]\begingroup turbulence dissipation
\item [{$f$}]\begingroup generic function
\item [{$F$}]\begingroup point--to--point mapping
\item [{$\mathcal{F}$}]\begingroup region--to--point mapping
\item [{$g$}]\begingroup generic function
\item [{$G$}]\begingroup Green's kernel
\item [{$h$}]\begingroup generic function
\item [{$h_0$}]\begingroup constant
\item [{$h_{12}$}]\begingroup constant
\item [{$h_{23}$}]\begingroup constant
\item [{$k$}]\begingroup stencil parameter (stencil size: $(2k+1) \times (2k+1)$)
\item [{$K$}]\begingroup constant
\item [{$\mathsf{K}$}]\begingroup covariance function in Gaussian process
\item [{$\ell$}]\begingroup half-length of the region size (determined by PDE)
\item [{$\mathcal{L}$}]\begingroup linear differential operator
\item [{$n_0$}]\begingroup downsampling parameter
\item [{N}]\begingroup neurons in fully connected network
\item [{$\mathcal{N}$}]\begingroup differential operator 
\item [{$p$}]\begingroup pressure
\item [{$\mathcal{P}$}]\begingroup production function
\item [{$P$}]\begingroup production term
\item [{$R_\mathbf{x}$}]\begingroup influence region centered at $\mathbf{x}$
\item [{S}]\begingroup stride in convolutional neural network
\item [{$S$}]\begingroup magnitude of normalized strain rate
\item [{$\tilde{S}$}]\begingroup magnitude of strain rate
\item [{$t$}]\begingroup temporal coordinate
\item [{$T$}]\begingroup flux in Reynolds stress transport
\item [{$U_0$}]\begingroup maximum velocity in Taylor--Green vortexes
\nomgroup{C}
\item [{$\alpha$}]\begingroup pressure gradient parameter
\item [{$\beta$}]\begingroup scaling parameter
\item [{$\delta(x)$}]\begingroup Dirac delta function
\item [{$\zeta$}]\begingroup dissipation coefficient
\item [{$\nu$}]\begingroup kinematic viscosity (diffusion coefficient)
\item [{$\mu$}]\begingroup dynamic viscosity
\item [{$\bm{\xi}$}]\begingroup spatial offset
\item [{$\Pi$}]\begingroup pressure--strain-rate tensor
\item [{$\rho$}]\begingroup density
\item [{$\sigma$}]\begingroup variance in covariance function
\item [{$\bm{\tau}$}]\begingroup Reynolds stress
\item [{$\phi$}]\begingroup closure variable
\item [{$\Omega$}]\begingroup magnitude of normalized rotation rate
\item [{$\tilde{\Omega}$}]\begingroup magnitude of rotation rate

\end{document}